\documentclass{fundam}

\usepackage{lineno}
\usepackage{todonotes}
\usepackage{needspace}
\usepackage[T1]{fontenc}
\usepackage[utf8]{inputenc}
\usepackage{url}
\usepackage{hyperref}

\usepackage{amssymb}
\usepackage{enumerate}
\usepackage{ifthen}

\usepackage{graphicx}
\DeclareGraphicsExtensions{.pdf,.png,.jpg}
\usepackage[misc,geometry]{ifsym} 
\usepackage{epsf}
\usepackage{latexsym}
\usepackage{amsfonts}
\usepackage{url}
\usepackage{rotating}
\usepackage{mathtools}
\usepackage{tikz}
\usetikzlibrary{decorations,decorations.pathreplacing,decorations.pathmorphing,decorations.markings}
\usetikzlibrary{calc}
\usetikzlibrary{arrows}
\usetikzlibrary{arrows.meta}
\usetikzlibrary{shapes.geometric}
\usetikzlibrary{shapes.symbols}
\usepackage{enumitem}
\setlist{nosep}
\usepackage{scalefnt}
\usepackage{todonotes}

\newcommand{\PT}{\mathit{PT}}
\newcommand{\lts}{LTS}
\newcommand{\drop}[1]{}

\newcommand{\es}{\emptyset}

\newcommand{\dt}{\bullet}

\newcommand{\pre}[1]{{}^\bullet{#1}}
\newcommand{\post}[1]{{#1}^\bullet}
\newcommand{\leer}{\varepsilon}
\newcommand{\nsymbol}{\mathbb{N}}

\newcommand{\minus}{\setminus} 
\newcommand{\goesto}{\rightarrow}
\renewcommand{\goesto}[1]{\stackrel{#1}{\longrightarrow}}

\newcommand{\fire}{\goesto}
\newcommand{\xfire}[1]{\xrightarrow{{#1}}}
\newcommand{\rset}[1]{[#1\rangle}
\newcommand{\emptyseq}{\varepsilon}
\newcommand{\impl}{\Rightarrow}
\newcommand{\Parikh}{\Psi}

\newcommand{\black}[1]{{\color{black} #1}}

\newcommand{\BX}[1]{{\unskip\nobreak\hfil\penalty50
                    \hskip2em\hbox{}\hfil
$\square$\/ {{\rm #1}}
                    \parfillskip=0pt \finalhyphendemerits=0 \par
                   }}

\newcommand{\EXA}[2]{\goodbreak\begin{example}
                     \label{#1}\begin{rm}{\sc #2}

                    }
\newcommand{\ENDEXAE}[1]{\BX{E\ref{#1}}
                        \end{rm}\end{example}
                       }
\newcommand{\ENDEXA}[1]{\BX{\ref{#1}}
                        \end{rm}\end{example}
                       }

\newcommand{\DEF}[2]{\goodbreak\begin{definition}
                     \label{#1}\begin{rm}{\sc #2}

                    }
\newcommand{\ENDDEFD}[1]{\BX{D\ref{#1}}
                        \end{rm}\end{definition}
                       }
\newcommand{\ENDDEF}[1]{\BX{\ref{#1}}
                        \end{rm}\end{definition}
                       }

\newcommand{\ENXDEF}{\end{rm}\end{definition}}

\newcommand{\KRYPT}[2]{\goodbreak\begin{kryptosystem}
                     \label{#1}\begin{rm}{\sc #2}

                    }
\newcommand{\ENDKRYPT}[1]{\BX{\ref{#1}}
                        \end{rm}\end{kryptosystem}
                       }
\newcommand{\CONJ}[2]{\goodbreak{\bf Conjecture:}
                     \label{#1}{\sc #2}
                    }
\newcommand{\ENDCONJ}[1]{\BX{}
                       }
\newcommand{\ENXCONJ}{
\end{conjecture}}
\newcommand{\KOR}[2]{\goodbreak\begin{corollary}
                     \label{#1}{\sc #2}

                    }
\newcommand{\ENDKORC}[1]{\BX{C\ref{#1}}
\end{corollary}
                       }
\newcommand{\ENDKOR}[1]{\BX{\ref{#1}}
\end{corollary}
                       }
\newcommand{\ENXKOR}{
\end{corollary}}
\newcommand{\PROP}[2]{\goodbreak\begin{proposition}
                      \label{#1}{\sc #2}

                     }
\newcommand{\ENDPROP}{
\end{proposition}}
\newcommand{\ENXPROP}[1]{\BX{\ref{#1}}
\end{proposition}
                       }
\newcommand{\ENXPROPP}[1]{\BX{\ref{#1}}
\end{proposition}
                       }
\newcommand{\THEO}[2]{\goodbreak\begin{theorem}
                     \label{#1}{\sc #2}

                    }
\newcommand{\ENDTHEO}{
\end{theorem}}
\newcommand{\SATZ}[2]{\goodbreak\begin{theorem}
                     \label{#1}{\sc #2}
 
                    }
\newcommand{\ENDSATZ}{
\end{theorem}}
\newcommand{\ENXSATZ}[1]{\BX{\ref{#1}}
\end{theorem}}
\newcommand{\LEM}[2]{\goodbreak\begin{lemma}
                     \label{#1}{\sc #2}

                    }
\newcommand{\ENDLEM}{
\end{lemma}}
\newcommand{\ENXLEM}[1]{\BX{\ref{#1}}
\end{lemma}}
\newcommand{\BEW}{\goodbreak
{\bf Proof:}
                 }

\newcommand{\ENDBEW}[1]{\BX{\ref{#1}}
                       }

\newcommand{\ENXBEW}{}

\newcommand{\NOT}[2]{\goodbreak\begin{notation}
                     \label{#1}\begin{rm}{\sc #2}

                    }
\newcommand{\ENDNOT}[1]{\BX{\ref{#1}}
                        \end{rm}\end{notation}

                       }
\newcommand{\ENXNOT}{\end{rm}\end{notation}}
\newcommand{\REM}[2]{\goodbreak\begin{remark}
                     \label{#1}\begin{rm}{\sc #2}

                    }
\newcommand{\ENDREMR}[1]{\BX{R\ref{#1}}
                        \end{rm}\end{remark}
                       }
\newcommand{\ENDREM}[1]{\BX{\ref{#1}}
                        \end{rm}\end{remark}
                       }
\newcommand{\ENXREM}{\end{rm}\end{remark}}
\newcommand{\BSP}[2]{\goodbreak\begin{beispiel}
                     \label{#1}\begin{rm}{\sc #2}

                    }
\newcommand{\ENDBSP}[1]{\BX{\ref{#1}}
                        \end{rm}\end{bespiel}
                       }

\newcommand{\TS}{\mathit{TS}}

\newcommand{\PN}{\mathit{N}}
\newcounter{examplePNcounter}
\newcommand{\PNref}[1]{\PN_{\ref{#1}}}

\newcounter{exampleTScounter}
\newcommand{\TSref}[1]{\TS_{\ref{#1}}}

\begin{document}


\title{Persistent Permutability in Choice Petri Nets} 

\author{Eike Best\\
Department of Computer Science, Carl von Ossietzky Universit\"at Oldenburg,\\
D-26111 Oldenburg, Germany
\and Raymond Devillers\\
D\'epartement d'Informatique, Universit\'e Libre de Bruxelles,\\
B-1050 Brussels, Belgium
}

\maketitle

\runninghead{Best, Devillers}{Persistent Permutability in Choice Petri Nets}

\begin{abstract}
Persistence is a strong, global, behavioural property of a Petri net,
meaning that no activity (represented by a net's transition) can disable a different activity.
Persistent permutability is a weaker property,
pertaining to the individual interleavings of a Petri net
and stating that a non-persistent sequence can be permuted into a persistent one.
This paper shows that in the classes of
equal-conflict Petri nets and pure dissymmetric choice Petri nets,
persistent permutability already suffices to imply overall persistence.
These classes generalise free-choice Petri nets
and are related to Petri's concept of ``confusion'',
while they are distinguished from each other by diverse 
restrictions on the choice structure of a net.
\end{abstract}

\begin{keywords}
Asymmetric choice, dissymmetric choice, embedding, free-choice, equal conflict,
Parikh equivalence, permutations, permutation equivalence, persistence, Petri nets.
\end{keywords}

\section{Introduction}
\label{intro.sct}

A marked Petri net is {\it persistent} if, at any reachable marking, the occurrence of a transition
cannot switch a {\it different} transition from being enabled to being disabled.
The notion of persistence has been well-established in Petri net theory 
since \cite{DBLP:journals/jcss/KarpM69} (Karp and Miller),
\cite{DBLP:conf/sagamore/Keller74} (Keller),
and especially \cite{DBLP:journals/jacm/LandweberR78}
(where Landweber and Robertson
proved that persistent
Petri nets have semilinear reachability sets).

Persistence is usually viewed as a property of a marked Petri net as a whole,
but in \cite{och}, the concept was widened to apply to individual interleavings
(i.e., firing sequences) of a net.
In \cite{och,folco-2014}, Ochma\'nski defined a {\it persistent permutability} property,
which states that one can deduce, from the existence of a given \emph{non-persistent}
interleaving,
the existence of an equivalent (in terms of available concurrent permutations),
\emph{persistent} interleaving, into which the given one can be permuted.
He also conjectured that under certain assumptions, the property of permutability is satisfied.
This conjecture is difficult to prove because it concerns the permutability of infinite firing sequences,
a property which is not easy to characterise.

If a Petri net is persistent, then it is also persistently permutable,
for the simple reason that the identity permutation can be applied to any firing sequence.
Our paper addresses the converse question:
\begin{equation}\label{q.eq}
\text{Does persistent permutability of finite sequences (called SPE) imply persistence?}
\end{equation}
Here, ``SPE'' is a shorthand for ``short permutation equivalence'',
where ``short'' refers to ``finite''.
Once such a result is established for a class of nets, then the permutability of infinite sequences
(and in its wake, also the validity of Ochma\'nski's conjecture) follows for this class of nets.

For general Petri nets, the answer to (\ref{q.eq}) is negative.
Nevertheless, in a precursor of this paper \cite{DBLP:conf/apn/BestD25},
we examined various Petri net classes and were able to show
that the answer to (\ref{q.eq}) is ``yes''
in the class of pure and safe dissymmetric choice (DC) nets
and in the class of safe free-choice (FC)
Petri nets \cite{de95,hack72}.\footnote{In \cite{DBLP:conf/apn/BestD25},
DC nets were called asymmetric choice.
Appendix~\ref{history.app} clarifies and disambiguates
this terminology.}
DC nets generalise FC nets and exclude a class of asymmetrically confused
nets.\footnote{This terminology is due to Petri \cite{petri-gnt-1976}.}
The results contained in the present paper extend both results of \cite{DBLP:conf/apn/BestD25} in the following way:
\begin{itemize}
\item[(i)]
SPE$\impl$persistence also holds in the entire class of pure (not necessarily safe)
DC Petri nets, which properly includes the class of pure FC nets (but not the class of safe FC nets), and
\item[(ii)]
SPE$\impl$persistence holds in the full class of (bounded or unbounded) FC nets, and even in equal conflict (EC)  nets. 
\end{itemize}
The first result is also in \cite{conf/apn/bestd26}, while the second result is new.

Ochma\'nski's conjecture has a straightforward theoretical appeal.
Its practical interest is perhaps less obvious.
On the one hand, one can guess that the existence of persistently equivalent executions
of a Petri net might give rise to feasible concurrent scheduling strategies that minimise arbitrary choices.
On the other hand, if
the sequential behaviour of a Petri net in general is limited to persistent sequences,
then it even becomes Turing powerful
and zero tests could be simulated by such
strategies~\cite{DBLP:journals/iandc/BarylskaMO13}. 

This paper is organised as follows.
In Sections \ref{ts.sct} and \ref{pn.sct}, all necessary basic concepts, in particular
plainness, pureness, safeness, FC nets, DC nets, EC nets,
and persistence, are introduced.
Section~\ref{pt.sct} contains the definition of a pattern and how
a pattern can be embedded in a directed graph.
In Section~\ref{perm1.sct}, persistent firing sequences are defined,
while in Section~\ref{perm2.sct}, persistent permutations are introduced.
Section~\ref{och.sct} explains the notion of fairness employed in this paper
and introduces permutation equivalence, along with Ochma\'nski's conjecture,
originally limited to elementary nets \cite{DBLP:conf/ac/RozenbergE96}.
In Section~\ref{main.sct}, the two results described above
are stated and provided with proofs.
The proofs are indirect, and one of them makes use of patterns as described in Section \ref{pt.sct}.
Section~\ref{counter.sct} discusses a canonical example
for which the implication SPE$\impl$persistence does \emph{not} hold
and for which Ochma\'nski's conjecture is invalid.
This example is 2-bounded (rather than safe),
and its reachability graph has provably no safe and no dissymmetric choice Petri net solution.
In Section~\ref{fair.sct}, a spectrum of fairness and persistent permutability notions is investigated,
and it is discussed how they are related to the ones employed in the main part of the paper.
Section \ref{concl.sct} concludes the paper with a few remarks on context
and on open problems.
Appendix~\ref{history.app} explains our naming discipline and also contain a brief historical account
of the roots of the classes of dissymmetric (and other) choice Petri nets.

\section{Labelled transition systems and their properties} 
\label{ts.sct}

Starting, perhaps, with Keller's paper \cite{DBLP:conf/sagamore/Keller74},
but also influenced by Kripke's earlier paper \cite{kripke-1963},
labelled transition systems (\lts) have served very widely as a basic
semantic model of asynchronous or nondeterministic systems.
They are essentially directed graphs whose edges are labelled by transitions
(or more complex objects, for instance for the explicit representation of concurrency)
and whose nodes may be labelled in some suitable way.
In this paper, we shall use labelled transition systems in order to represent
the reachability graphs of Petri nets, using the transitions
of such a net as their edge labels.

\DEF{lts.def}{Labelled transition systems}
A \emph{labelled transition system} with initial state, abbreviated \lts{},
is a quadruple $\TS=(S,\to,T,s_0)$
where $S$ is a set of \emph{states},
$T$ is a set of \emph{labels} (or \emph{transitions}),
$\to\;\subseteq(S\times T\times S)$ is the \emph{transition relation},
and $s_0\in S$ is an \emph{initial state} (so that $S$ may not be empty).
A label $t$ is \emph{enabled} in a state $s\in S$,
denoted by $s\xfire{t}$, if there is some state $s'\in S$ such that $(s,t,s')\in\to$,
and \emph{disabled at $s$} if there is \emph{no} state $s'\in S$ such that $(s,t,s')\in\to$.
For $t\in T$, we write $s\xfire{t}s'$ if{}f $(s,t,s')\in\to$,
meaning that $s'$ is \emph{reachable} from $s$ through the execution of $t$.
These definitions are extended to finite sequences
$\sigma\in T^*$ as follows:\footnote{We
use the shorthands $T^*$ and $T^\infty$ to denote the set of
finite and infinite sequences of transitions, respectively.
The length of a sequence $\tau\in T^*$ is denoted by $|\tau|$.
By $\leer$, we denote the empty sequence (of length $0$).} 
\[\begin{array}{l}
s\xfire{\emptyseq}\text{ and }s\xfire{\emptyseq}s\text{ are always true, by definition, and} \\
\text{$s\xfire{\sigma t}\;$ (or $s\xfire{\sigma t} s'$)$\;$ if $\;\exists q\in S$
with $s\xfire{\sigma}q\xfire{t}\;$ (with $s\xfire{\sigma}q\xfire{t} s'$, respectively).}
\end{array}
\]
Enabledness may be extended to infinite sequences $\sigma\in T^\infty$ as follows:
for a state $s$, $s\xfire{\sigma}$ iff $s\xfire{\sigma'}$ for each finite prefix $\sigma'$ of~$\sigma$.

For a state $s\in S$, $\rset{s}=\{s'\in S\mid\exists\sigma\in T^*\colon s\xfire{\sigma}s'\}$
denotes the set of states reachable from $s$. 

Two \lts{} with the same label set, $(S,\to,T,s_0)$ and $(S',\to',T,s_0')$,
will be called \emph{isomorphic} if there is a bijection $\beta\colon S\to S'$
such that $s_0'=\beta(s_0)$ and
$(r,t,s)\in\to$ if{}f $(\beta(r),t,\beta(s))\in\to'$.

For a finite sequence $\sigma\in T^*$ of labels, the \emph{Parikh vector}
$\Parikh(\sigma)$ is a $T$-vector (i.e., a~vector of natural
numbers with index set $T$), where $\Parikh(\sigma)(t)$ denotes
the number of occurrences of $t$ in $\sigma$.
This may be extended to infinite sequences if we allow vectors of natural or infinite numbers:
if $\sigma\in T^\infty$, $\Parikh(\sigma)(t)$ is $\infty$ if $t$ occurs infinitely often in $\sigma$;
otherwise, it is the number of occurrences of $t$ in $\sigma$.
Two sequences $\sigma,\tau\in T^*$ are called \emph{Parikh-equivalent}
if $\Parikh(\sigma)=\Parikh(\tau)$.
\ENDDEF{lts.def}

\DEF{lts-prop.def}{Basic properties of labelled transition systems}
A labelled transition system $(S,\to,T,s_0)$ is called
\emph{finite} if $S$ and $T$ (hence also $\to$) are finite sets;
\emph{totally reachable} if $S=\rset{s_0}$;
\emph{deterministic} if for any states $s,s',s''\in\rset{s_0}$
and sequences $\sigma,\tau\in T^*$ with $\Parikh(\sigma)=\Parikh(\tau)$:
$(s\xfire{\sigma} s'\land s\xfire{\tau}s'')\impl s'=s''$ as well as
 $(s'\xfire{\sigma}s\land s''\xfire{\tau}s)\impl s'=s''$
(i.e., from any one state, Parikh-equivalent sequences may not lead to two different successor states,
nor come from two different predecessor states);\footnote{Often,
determinism is defined only with respect to successor states.}
\emph{persistent} \cite{DBLP:journals/jacm/LandweberR78} if
for all reachable states $s,s',s''\in\rset{s_0}$, 
and labels $t,u\in T$ with $t\neq u$,
if $s\xfire{t} s'$ and $s\xfire{u}s''$,
then there is some reachable state $r\in\rset{s_0}$ such that both
$s'\xfire{u}r$ and $s''\xfire{t}r$
(i.e., once two different labels are both enabled at some state,
none of them can disable the other,
and executing both, in any order, leads to the same state).
A state $s\in\rset{s_0}$ is called a \emph{deadlock} if $\neg(\exists t\in T\colon s\xfire{t})$.
\ENDDEF{lts-prop.def}

\section{Petri nets and their reachability graphs}
\label{pn.sct}

We define Petri nets as place-transition nets, as in \cite{DBLP:books/sp/Reisig85a}.
The reachability graph of a Petri net turns out to be a labelled transition system
whose states are the reachable markings of the net.
Conversely, a labelled transition system may, or may not, be
isomorphic to the reachability graph of some Petri net.
In that sense, labelled transition systems are more general.
In the correspondence between Petri net reachability graphs
and labelled transition systems, the set of labels of an \lts{} and
the set of transitions of a Petri net correspond uniquely to each other.
This is why we use the same letter, $T$, for both.

\DEF{pn.def}{Petri nets}
A (finite, initially marked, arc-weighted)
Petri net is a quadruple
$N=(P,T,F,M_0)$ such that
$P$ is a finite set of \emph{places},
$T$ is a finite set of \emph{transitions}, with $P\cap T=\es$,
$F$ is a \emph{flow} function $F\colon((P\times T)\cup(T\times P))\to\nsymbol$,
and $M_0$ is the \emph{initial marking},
where a \emph{marking} is a mapping $M\colon P\to\nsymbol$
(indicating a number of \emph{tokens} in each place).\footnote{If
the initial marking plays no role, we shall sometimes allow $N$ to denote
only an unmarked Petri net $(P,T,F)$.}
A~transition $t\in T$ is \emph{enabled by} a marking $M$,
denoted by $M\xfire{t}$, if $\forall p\in P\colon M(p)\geq F(p,t)$.
If $t$ is enabled at $M$, then $t$ can \emph{occur} (or \emph{fire})
in $M$, leading to the marking $M'$ defined by $M'(p)=M(p)-F(p,t)+F(t,p)$ (denoted by $M\xfire{t}M'$).
$\rset{M}$ is the set of markings reachable from $M$.
The \emph{reachability graph of $N$}, $RG(N)$, is the
labelled transition system with the set of vertices $\rset{M_0}$,
the set of edges $\{(M,t,M')\mid M,M'\in\rset{M_0}\land M\xfire{t}M'\}$,
and initial state $M_0$.
\ENDDEF{pn.def}

\DEF{pn-classes.def}{Basic structural properties of Petri nets}
If we omit the initial marking, we can write $N=(P,T,F)$.
For a place $p$ and a transition $t$ of a Petri net $N=(P,T,F)$,
let ${}^\dt t=\{p\in P\mid F(p,t)>0\}$ be the \emph{preset of $t$} containing \emph{pre-places},
$t^\dt=\{p\in P\mid F(t,p)>0\}$ its \emph{postset} containing \emph{post-places},
${}^\dt p=\{t\in T\mid F(t,p)>0\}$ the \emph{preset of $p$}, and $p^\dt=\{t\in T\mid F(p,t)>0\}$ its 
\emph{postset}.
$N$ is called
\emph{plain} if $cod(F)\subseteq\{0,1\}$ (i.e., there are no multiple arcs);\footnote{In
this paper, only plain nets will be considered (except in Example \ref{ts1-ts2.exa} and in Section \ref{spe-fpe-ec.sct}).}
\emph{pure} or \emph{side-condition free} if $p^\dt\cap{}^\dt p=\es$ for all places $p\in P$;
\emph{choice-free} (CF) \cite{DBLP:journals/ipl/Crespi-ReghizziM75}
if $|\post{p}|\leq 1$ for each place $p\in P$;
\emph{free-choice} (FC) \cite{hack72,de95}
if $N$ is plain and for all $t,t'\in T$,
$(\pre{t}\cap\pre{t'}\neq\emptyset)$ entails $(\pre{t}=\pre{t'})$;
\emph{equal conflict} (EC)
\cite{DBLP:conf/apn/TeruelS93}
if for all $t,t'\in T$,
$(\pre{t}\cap\pre{t'}\neq\emptyset)$ entails $F(.,t)=F(.,t')$, 
where for $x\in T$, $F(.,x)$ is the $P$-vector $F(.,x)\colon P\to\nsymbol$ with $F(.,x)(p)=F(p,x)$;
\emph{dissymmetric choice} (DC)
if $N$ is plain and for all $t,t'\in T$,
$(\pre{t}\cap\pre{t'}\neq\emptyset)$ entails $(\pre{t}\subseteq\pre{t'})\lor(\pre{t'}\subseteq\pre{t})$;
and \emph{asymmetric choice} (AC) \cite{hack72,DBLP:books/sp/BestD24}
if $N$ is plain and for all $p,p'\in P$,
$(\post{p}\cap\post{p'}\neq\emptyset)$ entails $(\post{p}\subseteq\post{p'})\lor(\post{p'}\subseteq\post{p})$.
The \emph{reverse dual} ${\mathcal RD}(N)$ of $N$ is obtained by exchanging the roles of places and transitions
and reversing all the arcs: ${\mathcal RD}(N)=(T,P,F')$, where $F'(x,y)=F(y,x)$
for all $x,y\in(P\times T)\cup(T\times P)$.
\ENDDEF{pn-classes.def}

\DEF{pnprop.def}{Basic behavioural properties of Petri nets}
An initially marked Petri net $N=(P,T,F,M_0)$ is
\emph{$k$-bounded}, for some $k\in\nsymbol$, if
$\forall M\in\rset{M_0}\colon\forall p\in P\colon M(p)\leq k$
(i.e., the number of tokens on any place never exceeds $k$);
\emph{safe} if it is $1$-bounded;
\emph{bounded} if $\exists k\in\nsymbol\colon N\text{ is $k$-bounded}$;
\emph{pps} if it is plain, pure, and safe;
\emph{totally reachable / deterministic / persistent} if so is its reachability graph.
A finite firing sequence $M_0\xfire{\sigma}M$ is called \emph{deadlocking}
(or \emph{maximal}, or \emph{non-extendable}) if $M$ is a deadlock in the reachability graph.
\ENDDEF{pnprop.def}

We shall use letters $a,b,c,\ldots\in T$
(but also $t,u\in T$) for the labels of a~transition system
(or for the corresponding transitions of a Petri net);
$\alpha,\beta,\sigma,\tau\in T^*$ for sequences of transitions;
$r,s$ for the states of an \lts;
$p,q$ for the places of a net;
and $M,J,K,L$ for the markings of a net;
those denotations may also be ``decorated'',
such as in $a_1$, $\sigma'$, $\widetilde{p}$, and $\widehat{M}$.

The class of pps Petri nets
is closely related to \emph{elementary nets} \cite{DBLP:conf/ac/RozenbergE96}.
Elementary Petri nets have a strengthened firing rule: namely, $t$ can occur if all its pre-places have exactly
one token \emph{and} all its post-places have exactly zero tokens.
Every elementary net with this strengthened firing rule can be turned into an equivalent
(in terms of reachability graph isomorphism)
pps net
with the usual firing rule, by adding appropriate complement places.\footnote{I.e., 
for each place $p$ there is a complement place $\tilde{p}$ such that for each reachable
marking $M$, $M(p)+M(\tilde{p})=1$.}
Conversely, for pps nets, the two firing rules coincide.

The following two results belong to basic Petri net knowledge.

\PROP{pn1.prop}{Finite reachability graphs}
A Petri net $N$ is bounded if{}f $RG(N)$ is finite.
\ENXPROPP{pn1.prop}

This result depends on the finiteness of $N$.
If we allow $N$ to be infinite, we may get bounded
nets with infinite reachability graphs.

\Needspace{5\baselineskip}
\PROP{pn2.prop}{Total reachability and determinism of Petri nets}
Any Petri net $N$ is totally reachable and deterministic.
In particular:
If $N=(P,T,F,M_0)$ is a net and if $M_0\xfire{\sigma}M_1$ and $M_0\xfire{\tau}M_2$
are two firing sequences with $\sigma,\tau\in T^*$ and $\Parikh(\sigma)=\Parikh(\tau)$,
then $M_1=M_2$.
\ENXPROPP{pn2.prop}

\DEF{solv.def}{Solvability}
A Petri net $N$ \emph{solves} a transition system $\TS$ if $RG(N)$ and $\TS$ are isomorphic.
\ENDDEF{solv.def}

\REM{synth.rem}{Reachability graphs are special \lts}
By Definitions \ref{pn.def} and \ref{solv.def}, the class of Petri net reachability graphs determines
a class of labelled transition systems,
and by Proposition \ref{pn2.prop}, this class is a proper subclass,
since the properties of total reachability and determinism are not shared by
arbitrary labelled transition systems.
There are other such properties \cite{DBLP:journals/scp/BestD18},
and region theory
\cite{DBLP:conf/ac/BadouelD96,
DBLP:journals/acta/DeselR96, DBLP:journals/acta/EhrenfeuchtR89a}
provides full algorithmic and graph-theoretical characterisations of
the class of Petri net solvable \lts{} (both characterisations are in \cite{DBLP:books/sp/BestD24}).
\ENDREM{synth.rem}

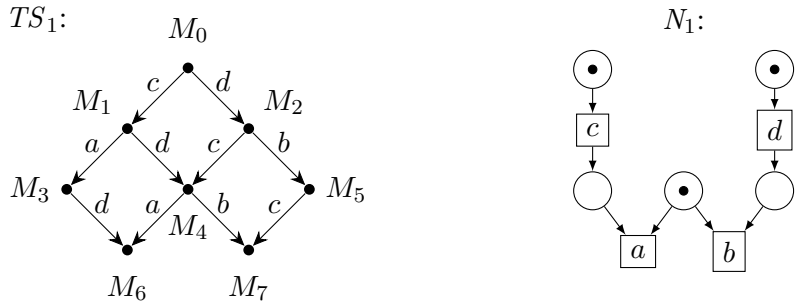
\begin{figure}[htb]
\begin{center}
\begin{tikzpicture}[scale=0.8]
\refstepcounter{exampleTScounter}\label{basic-exa.ts}
\node[]()at(-0.5,3.8)[]{$\TS_\theexampleTScounter$:};
\node[circle,fill=black!100,inner sep=0.05cm](0)at(2,3)[label=above:]{};
 \node[above of=0,node distance=0.5cm]{$M_0$};
\node[circle,fill=black!100,inner sep=0.05cm](1)at(1,2)[label=above left:$M_1$]{};
\node[circle,fill=black!100,inner sep=0.05cm](2)at(3,2)[label=above right:$M_2$]{};
\node[circle,fill=black!100,inner sep=0.05cm](3)at(0,1)[label=left:$M_3$]{};
\node[circle,fill=black!100,inner sep=0.05cm](4)at(2,1)[label=below:]{};
 \node[below of=4,node distance=0.5cm]{$M_4$};
\node[circle,fill=black!100,inner sep=0.05cm](5)at(4,1)[label=right:$M_5$]{};
\node[circle,fill=black!100,inner sep=0.05cm](6)at(1,0)[label=below:]{};
 \node[below of=6,node distance=0.5cm]{$M_6$};
\node[circle,fill=black!100,inner sep=0.05cm](7)at(3,0)[label=below:]{};
 \node[below of=7,node distance=0.5cm]{$M_7$};
\draw[-{Stealth[length=2mm,width=1.7mm]},bend right=0](0)edge node[above,inner sep=0.15cm,pos=0.6]{$c$}(1);
\draw[-{Stealth[length=2mm,width=1.7mm]},bend right=0](0)edge node[above,inner sep=0.15cm,pos=0.6]{$d$}(2);
\draw[-{Stealth[length=2mm,width=1.7mm]},bend right=0](1)edge node[above,inner sep=0.15cm,pos=0.6]{$a$}(3);
\draw[-{Stealth[length=2mm,width=1.7mm]},bend right=0](1)edge node[above,inner sep=0.15cm,pos=0.6]{$d$}(4);
\draw[-{Stealth[length=2mm,width=1.7mm]},bend right=0](2)edge node[above,inner sep=0.15cm,pos=0.6]{$c$}(4);
\draw[-{Stealth[length=2mm,width=1.7mm]},bend right=0](2)edge node[above,inner sep=0.15cm,pos=0.6]{$b$}(5);
\draw[-{Stealth[length=2mm,width=1.7mm]},bend right=0](3)edge node[above,inner sep=0.15cm,pos=0.6]{$d$}(6);
\draw[-{Stealth[length=2mm,width=1.7mm]},bend right=0](4)edge node[above,inner sep=0.15cm,pos=0.6]{$a$}(6);
\draw[-{Stealth[length=2mm,width=1.7mm]},bend right=0](4)edge node[above,inner sep=0.15cm,pos=0.6]{$b$}(7);
\draw[-{Stealth[length=2mm,width=1.7mm]},bend right=0](5)edge node[above,inner sep=0.15cm,pos=0.6]{$c$}(7);
\end{tikzpicture}\hspace*{2.5cm}
\raisebox{0.5cm}{\begin{tikzpicture}[scale=0.8]
\refstepcounter{examplePNcounter}\label{basic-exa.pn}
\node[]()at(1.5,3.8)[]{$\PN_\theexamplePNcounter$:};
\node[circle,draw,minimum size=0.5cm](p0)at(0,3)[]{};
 \filldraw[black](0,3)circle(2pt);
\node[circle,draw,minimum size=0.5cm](p1)at(3,3)[]{};
 \filldraw[black](3,3)circle(2pt);
\node[circle,draw,minimum size=0.5cm](p2)at(0,1)[]{};
\node[circle,draw,minimum size=0.5cm](p3)at(1.5,1)[]{};
 \filldraw[black](1.5,1)circle(2pt);
\node[circle,draw,minimum size=0.5cm](p4)at(3,1)[label=above:]{};
\node[draw,minimum size=0.4cm](a)at(0.75,0){$a$}; 
\node[draw,minimum size=0.4cm](b)at(2.25,0){$b$};
\node[draw,minimum size=0.4cm](c)at(0,2){$c$};
\node[draw,minimum size=0.4cm](d)at(3,2){$d$};
\draw[-latex](p0)--(c);
\draw[-latex](p1)--(d);
\draw[-latex](c)--(p2);
\draw[-latex](d)--(p4);
\draw[-latex](p2)--(a);
\draw[-latex](p3)--(a);
\draw[-latex](p3)--(b);
\draw[-latex](p4)--(b);
\end{tikzpicture}}
\end{center}
\caption{An \lts{} $\TSref{basic-exa.ts}$
and a Petri net solution $\PNref{basic-exa.pn}$.
The net $\PNref{basic-exa.pn}$ is plain; pure; safe; and asymmetric choice;
but not dissymmetric choice
because $(\pre{a}\cap\pre{b}\neq\es)$, $\neg(\pre{a}\subseteq\pre{b})$, and $\neg(\pre{b}\subseteq\pre{a})$;
and, {\it a fortiori}, not free-choice.
It has two deadlocks, $M_6$ and $M_7$.}
\label{basic-exa.fig}
\end{figure}

\EXA{basic-exa.rem}{Figure \ref{basic-exa.fig}:
an \lts{} $\TSref{basic-exa.ts}$ (l.h.s.) and a solution $\PNref{basic-exa.pn}$ (r.h.s.)}
A transition system is represented as a directed graph whose
nodes are states and whose edges are labelled with labels from the set $T$.
$\TSref{basic-exa.fig}$
has eight states, ten edges, and an initial state $M_0$.
A Petri net is represented by circles for places;
tokens inside places to represent markings;
squares for transitions; and directed arrows, inscribed by their weights, for arcs.
For simplicity, arcs with weight zero are omitted altogether,
and arcs with weight one are drawn without any explicit inscription.
$\PNref{basic-exa.fig}$ has five places, four transitions,
eight arcs with weight $1$, and an initial marking $M_0$ comprising three tokens.
The states in $\TSref{basic-exa.ts}$ are already named in such a way that
they correspond uniquely to the reachable markings of $\PNref{basic-exa.pn}$,
with $M_0$ being the initial state.
Formally,
\[\TSref{basic-exa.ts}\;=\;\Bigl(\{M_0,\ldots,M_7\},\{(M_0,c,M_1),\ldots,(M_5,c,M_7)\},\{a,b,c,d\},M_0\Bigr)
\]
but such a description is somewhat unwieldy and we shall not also specify it for the net
(the places would have to be named first anyway).
\ENDEXA{basic-exa.rem}

\Needspace{5\baselineskip}
\REM{acdc.rem}{Free-choice, dissymmetric, and asymmetric Petri nets}
In Definition \ref{pn-classes.def},
the free-choice property has been defined as
\begin{equation}\label{fc1.eq}
\forall t,t'\in T\colon\;\;(\pre{t}\cap\pre{t'}\neq\emptyset)\;\;\impl\;\;(\pre{t}=\pre{t'})
\end{equation}
which is equivalent to its reverse dual
\begin{equation}\label{fc2.eq}
\forall p,p'\in P\colon\;\;(\post{p}\cap\post{p'}\neq\emptyset)\;\;\impl\;\;(\post{p}=\post{p'})
\end{equation}
The dissymmetric choice property arises from (\ref{fc1.eq})
by weakening the right-hand side to $(\pre{t}\subseteq\pre{t'})\lor(\pre{t'}\subseteq\pre{t})$.
The asymmetric choice property, as in \cite{de95,DBLP:books/sp/BestD24} and other works,
arises from (\ref{fc2.eq}) by weakening the right-hand side to
$(\post{p}\subseteq\post{p'})\lor(\post{p'}\subseteq\post{p})$.

\begin{figure}[htb]
\begin{center}
\begin{tikzpicture}[scale=0.8]
\node[circle,very thick,draw,minimum size=0.5cm](p0)at(0,1.3)[]{};
\node[circle,very thick,draw,minimum size=0.5cm](p1)at(2,1.3)[]{};
\node[circle,very thick,draw,minimum size=0.5cm](p2)at(4,1.3){};
\node[draw,very thick,minimum size=0.4cm](a)at(1,0)[label=left:$t$]{$$}; 
\node[draw,very thick,minimum size=0.4cm](b)at(3,0)[label=right:$t'$]{$$};
\draw[-latex,very thick,bend left=10](p0)edge node[above,inner sep=0.15cm,pos=0.6]{$$}(a);
\draw[-latex,very thick,bend right=33](p1)edge node[above,inner sep=0.15cm,pos=0.6]{$$}(a);
\draw[-latex,very thick,bend left=33](p1)edge node[above,inner sep=0.15cm,pos=0.6]{$$}(b);
\draw[-latex,very thick,bend right=10](p2)edge node[above,inner sep=0.15cm,pos=0.6]{$$}(b);
  \draw [-latex](p0)edge node[below]{$$} coordinate[pos=0.85] (p0b)(b);
  \draw[red,very thick,shift={(p0b)}](-0.1,-0.1)--(0.1,+0.1); \draw[red,very thick,shift={(p0b)}](-0.1,+0.1)--(+0.1,-0.1);
  \draw [-latex](p2)edge node[below]{$$} coordinate[pos=0.85] (p2a)(a);
  \draw[red,very thick,shift={(p2a)}](-0.1,-0.1)--(0.1,+0.1); \draw[red,very thick,shift={(p2a)}](-0.1,+0.1)--(+0.1,-0.1);
\end{tikzpicture}\hspace*{2cm}
\begin{tikzpicture}[scale=0.8]
\node[circle,very thick,draw,minimum size=0.5cm](p0)at(1,1.3)[label={[label distance=-0.08cm]175:$p$}]{};
\node[circle,very thick,draw,minimum size=0.5cm](p1)at(3,1.3)[label={[label distance=-0.08cm]5:$p'$}]{};
\node[draw,very thick,minimum size=0.4cm](a)at(0,0){$$};
\node[draw,very thick,minimum size=0.4cm](b)at(2,0){$$}; 
\node[draw,very thick,minimum size=0.4cm](c)at(4,0){$$};
\draw[-latex,very thick,bend right=33](p0)edge node[above,inner sep=0.15cm,pos=0.6]{$$}(a);
\draw[-latex,very thick,bend left=25](p0)edge node[above,inner sep=0.15cm,pos=0.6]{$$}(b);
\draw[-latex,very thick,bend right=25](p1)edge node[above,inner sep=0.15cm,pos=0.6]{$$}(b);
\draw[-latex,very thick,bend left=33](p1)edge node[above,inner sep=0.15cm,pos=0.6]{$$}(c);
  \draw [-latex,bend right=5](p1)edge node[below]{$$} coordinate[pos=0.8] (p1a)(a);
  \draw[red,very thick,shift={(p1a)}](-0.1,-0.1)--(0.1,+0.1); \draw[red,very thick,shift={(p1a)}](-0.1,+0.1)--(+0.1,-0.1);
  \draw [-latex,bend left=5](p0)edge node[below]{$$} coordinate[pos=0.8] (p0c)(c);
  \draw[red,very thick,shift={(p0c)}](-0.1,-0.1)--(0.1,+0.1); \draw[red,very thick,shift={(p0c)}](-0.1,+0.1)--(+0.1,-0.1);
\end{tikzpicture}
\end{center}
\caption{DC excludes the Petri net pattern shown on the left-hand side
whereas AC excludes the pattern shown on the right-hand side.
An arc with a cross on top means ``no such arc''.}
\label{acdc.fig}
\end{figure}
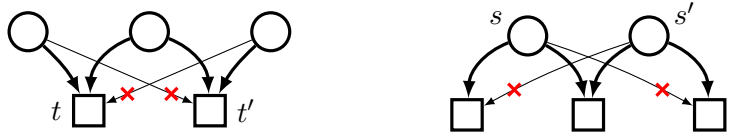

The AC and DC properties are reverse duals of each other,
but they are not equivalent.
The excluded structures are shown in Figure~\ref{acdc.fig}.
We refer the reader to Appendix~\ref{history.app} for a historical synopsis of these concepts.
\ENDREM{acdc.rem}

\section{Patterns and embeddings}
\label{pt.sct}

In the following, we shall capture interesting substructures of an LTS by means of patterns.
In general, patterns specify that some features are mandatory and other ones are excluded.
In the present paper, we shall restrict our attention to rather simple patterns, 
where the mandatory features are the presence of some labelled arcs
(prescribing enabling situations) and the excluded ones
are the absence of some labelled arcs (requiring non-enabling situations).

\Needspace{5\baselineskip}
\DEF{pattern.def}{Pattern of an \lts}
A \emph{pattern} is a quadruple $\PT=(S,\to,T,D)$
where $S$ is a set of states, $T$ a set of labels, $\to\;\subseteq(S\times T\times S)$ (represented as usual),
and $D\subseteq(S\times T)$ (represented by labelled arcs without specified endpoint and a red cross).
\ENDDEF{pattern.def}

The fact that a pattern is present in a transition system will be captured by an embedding notion.

\DEF{embed.def}{Embedding of a pattern in an \lts}
Let $\PT=(S_1,\to_1,T_1,D_1)$ be a pattern and $\TS=(S,\to,T,s_0)$ an \lts{}.
Then $\PT$ is \emph{embedded in $\TS$} if there is a function
$f=f_1\cup f_2$, with $f_1\colon (S_1\to S)$ and $f_2\colon(T_1\to T)$, such that
\begin{itemize}
\item[(i)]
$\forall(s,a,s')\in\to_1\colon(f(s),f(a),f(s'))\in\to$ (mandatory arcs), and
\item[(ii)]
$\forall(s,a)\in D_1\colon\neg(f(s)\xfire{f(a)})$ (excluded arcs).
\end{itemize}
Moreover, we shall assume that $f_2$ is injective, meaning that it does not ``fuse'' labels:
$\forall t,t'\in T_1:t\neq t'\impl f_2(t)\neq f_2(t')$.
\ENDDEF{embed.def}

\begin{figure}[htb]
\begin{center}
\raisebox{0.15cm}{\begin{tikzpicture}[scale=0.8]
\node[]()at(-0.8,1.4)[]{$\PT_{\mathit{nonpers}}$:};
\node[circle,fill=black!100,inner sep=0.05cm](4)at(2,1)[label=above:$1$]{};
 \node[below of=4,node distance=0.5cm]{$$};
\node[circle,fill=black!100,inner sep=0.05cm](6)at(1,0)[label=below:$2$]{};
\node[circle,fill=black!100,inner sep=0.05cm](7)at(3,0)[label=below:$3$]{};
 \node[below of=6,node distance=0.5cm]{$$};
  \node[left=0.8cm of 6](6a){};\draw [-{Stealth[length=2mm,width=1.7mm]}](6)edge node[below]{$b$} coordinate (m6)(6a);
  \draw[red,thick,shift={(m6)}](-0.1,-0.1)--(0.1,+0.1); \draw[red,thick,shift={(m6)}](-0.1,+0.1)--(+0.1,-0.1);
\draw[-{Stealth[length=2mm,width=1.7mm]},very thick,bend right=0](4)edge node[above,inner sep=0.15cm,pos=0.6]{$a$}(6);
\draw[-{Stealth[length=2mm,width=1.7mm]},very thick,bend right=0](4)edge node[above,inner sep=0.15cm,pos=0.6]{$b$}(7);
\end{tikzpicture}}
\end{center}
\caption{The pattern $\PT_{\mathit{nonpers}}=(\{1,2,3\},\{(1,a,2),(1,b,3)\},\{a,b\},\{(2,b)\})$
highlights a violated persistence at state $1$,
due to the disabling of $b$ by executing $a$.
The set $\{(2,b)\}$ specifies the excluded arrow.}
 \label{PT1}
\end{figure}
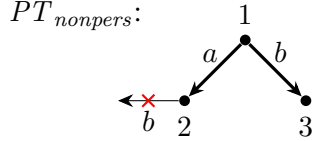

An \lts{} pattern is illustrated in Figure~\ref{PT1}.
This pattern is designed
 to characterise a non-persistent part of an \lts{}
(both $a$ and $b$ are enabled at $1$, but after executing $a$, $b$ is disabled at $2$).

\begin{figure}[htb]
\begin{center}
\begin{tikzpicture}[scale=0.8]
\refstepcounter{exampleTScounter}\label{confuse-exa.ts}
\node[]()at(-1.5,3.6)[]{$\TS_\theexampleTScounter$:};
\node[circle,fill=black!100,inner sep=0.05cm](0)at(0,3)[label=above:$s_0$]{};
\node[circle,fill=black!100,inner sep=0.05cm](1)at(2,3)[label=above:$s_1$]{};
\node[circle,fill=black!100,inner sep=0.05cm](2)at(4,3)[label=above:$s_2$]{};
\draw[-{Stealth[length=2mm,width=1.7mm]},bend right=20](0)edge node[below,inner sep=0.15cm,pos=0.5]{$y$}(1);
\draw[-{Stealth[length=2mm,width=1.7mm]},bend right=-20](0)edge node[above,inner sep=0.15cm,pos=0.5]{$x$}(1);
\draw[-{Stealth[length=2mm,width=1.7mm]},bend right=0](1)edge node[above,inner sep=0.15cm,pos=0.5]{$x$}(2); 
\end{tikzpicture}\hspace*{1.5cm}
\raisebox{0.5cm}{\begin{tikzpicture}[scale=0.8]
\refstepcounter{examplePNcounter}\label{confuse-exa.pn}
\node[]()at(-0.1,1.5)[]{$\PN_\theexamplePNcounter$:};
\node[circle,draw,minimum size=0.5cm](p0)at(3,1)[]{};
 \filldraw[black](3,0.89)circle(2pt);\filldraw[black](3,1.11)circle(2pt);
\node[draw,minimum size=0.4cm](x)at(5,1){$x$}; 
\node[draw,minimum size=0.4cm](y)at(1,1){$y$};
\draw[-latex](p0)--(x);
\draw[-latex,bend right=20](p0)edge node[above,inner sep=0.15cm,pos=0.5]{$2$}(y);
\draw[-latex,bend right=20](y)edge node[above,inner sep=0.15cm,pos=0.5]{$$}(p0);
\end{tikzpicture}}
\end{center}
\caption{An \lts{} $\TSref{confuse-exa.ts}$ containing the pattern $\PT_{\mathit{nonpers}}$,
fusing states $2$ and $3$ into $s_1$.
The impure, arc-weighted net $\PNref{confuse-exa.pn}$ solves $\TSref{confuse-exa.ts}$,
and the embedding of $\PT_{\mathit{nonpers}}$ into $\TSref{confuse-exa.ts}$
witnesses the non-persistence of~$\PNref{confuse-exa.pn}$.}
\label{confuse-exa.fig}
\end{figure}
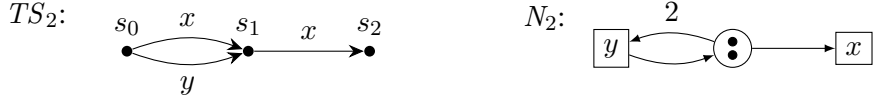

\EXA{ts1-ts2.exa}{Embedding $\PT_{\mathit{nonpers}}$ into $\TSref{basic-exa.fig}$ and into $\TSref{confuse-exa.ts}$
(Figures \ref{basic-exa.fig} and \ref{confuse-exa.fig})}
A possible embedding of the pattern $\PT_{\mathit{nonpers}}$ into the LTS $\TSref{basic-exa.fig}$
is given by the function $f$ such that $f(1)=M_4$, $f(2)=M_6$, $f(3)=M_7$, $f(a)=a$ and $f(b)=b$.
Another possible embedding is $f(1)=M_4$, $f(2)=M_7$, $f(3)=M_6$, $f(a)=b$ and $f(b)=a$.

Both embeddings are injective on the states as well as on the transitions,
but Figure~\ref{confuse-exa.fig} exemplifies a situation 
where some states are fused:
$f(1)=s_0$, $f(2)=s_1=f(3)$, $f(a)=x$, and $f(b)=y$.
This still captures a non-persistent situation
($s_0$ enables $x$ and $y$ but executing $x$ disables $y\neq x$).

We require injectivity on labels, since if
$f(a)=f(b)$ for $a\neq b$, 
we lose the fact that, in the definition of persistence, the two considered labels are different
(for example, in Figure~\ref{confuse-exa.fig}, if $f(a)=x=f(b)$ in $\PT_{\mathit{nonpers}}$, we could drop $y$ and we 
would simply have two $x$'s in a row, which would have nothing to do with a situation of non-persistence).

By Lemma \ref{nonpers.lem} below, the
embedding of $\PT_{\mathit{nonpers}}$ indicates that $\TSref{confuse-exa.ts}$ cannot be solved by a persistent Petri net.
As it happens, it cannot be realised by any pure or plain Petri net, as considered in this paper,
but it could occur in the reachability graph of a non-pure, non-plain Petri net such as $\PNref{confuse-exa.pn}$.
\ENDEXA{ts1-ts2.exa}

\Needspace{5\baselineskip}
\LEM{nonpers.lem}{$\PT_{\mathit{nonpers}}$ can recognise non-persistent nets}
Suppose that $N=(P,T,F,M_0)$ is a Petri net and $\TS=(S,\to,T,s_0)$
is (isomorphic to) its reachability graph.
Suppose that $\PT_{\mathit{nonpers}}$ is embedded in $\TS$.
Then $N$ is not persistent.
\ENDLEM

\BEW
Let $f$ be the embedding function from $\PT_{\mathit{nonpers}}$ to $\TS$.

The marking (corresponding to) $f(1)$ is reachable in $N$ since
$\PT_{\mathit{nonpers}}$ is embedded in an \lts{} that is isomorphic to
$RG(N)$, which is totally reachable.
It enables both $f(a)$ and $f(b)$, by the definition of an embedding,
and $f(a)\neq f(b)$ by the injectivity of $f$ on $\{a,b\}$. 
The non-enabling $(2,b)$ implies that $f(b)$ is not enabled at $f(2)$.
Together with $f(1)\xfire{f(a)}f(2)$ and $f(1)\xfire{f(b)}f(3)$ (since $f$ is an embedding),
this implies, according to Definitions \ref{lts.def} and \ref{pnprop.def},
that $N$ is not persistent at the marking~$f(1)$.

Hence $N$ is not persistent, since a non-persistent marking is reachable.
\ENDBEW{nonpers.lem}

\section{Persistent sequences}
\label{perm1.sct}

The transition system $\TSref{basic-exa.fig}$ shown in Figure \ref{basic-exa.fig} is not persistent,
nor is, therefore, its generating Petri net $\PNref{basic-exa.fig}$.
The offending state is $M_4$ where both $a$ and $b$ are enabled
in such a way that executing $a$ disables $b$ (and vice versa).
This is witnessed by the pattern $\PT_{\mathit{nonpers}}$ which is embedded twice
in $\TSref{basic-exa.fig}$.
$M_4$ might be called a ``proper choice state'',
as opposed to the ``fake choice states'' $M_0$, $M_1$ and~$M_2$.
Those latter states denote a choice in the order of subsequent execution of two concurrent,
independent, transitions,
whereas $M_4$ does not allow the concurrent execution of $a$ and $b$.
One tends to say that, from $M_0$,
executing $cd$ and executing $dc$ really denote the same
(concurrent) computation, differing only in the order of the occurrence of
$c$ and $d$. Formally, $M_0\xfire{cd}$ and $M_0\xfire{dc}$ are only
``a permutation of two transitions apart'' from each other.
The \lts{} structure they span is colloquially known as a ``diamond''  with ``corners'' $M_1,M_2$.

\Needspace{0.5\baselineskip}
More generally, we shall call two sequences, say $M_0\xfire{dca}$ and $M_0\xfire{cad}$,
to be in relation $\equiv$ if they are a number of transition permutations apart; e.g.,
\[M_0\xfire{dca}\;\;\;\;\equiv\;\;\;\;M_0\xfire{cda}\;\;\;\;\equiv\;\;\;\;M_0\xfire{cad}
\]
with single permutations in between, and two permutations between $M_0\xfire{dca}$ and $M_0\xfire{cad}\,$.

But note that there is a difference between these sequences in terms of persistence:
it happens that one of them ($M_0\xfire{dca}\,$) 
goes through an arc which switches an enabled transition into disabled
while this does not happen for the other sequence ($M_0\xfire{cad}\,$)$\,$:
\begin{equation}\label{equiv-perm.eq}
M_0\xfire{d}M_2\xfire{c}M_4\xfire{a}M_6\;\;\;\;\equiv\;\;\;\;M_0\xfire{c}M_1\xfire{a}M_3\xfire{d}M_6
\;\;\;\;\;\;\;\;\text{(see Figure \ref{basic-exa.fig})}
\end{equation}
The critical arc is $M_4\xfire{a}M_6$ for which $M_4\xfire{b}$, but not $M_6\xfire{b}$ (and $b\neq a$).
The first sequence is then called non-persistent while the other one is called persistent.

This paper is concerned with circumstances in which non-persistent firing sequences
have equivalent persistent permutations, as in (\ref{equiv-perm.eq})
where the non-persistent sequence $M_0\xfire{dca}$ (touching the choice state $M_4$)
can be permuted equivalently to $M_0\xfire{cad}$ (not touching any choice state).

In this section, we define the notion of a persistent firing sequence.
In the next section, we then go on to define what it means for a sequence to be permutable.

In the following, let $N=(P,T,F,M_0)$ be a Petri net with some (initial) marking $M_0$.

\DEF{pers-seq.def}{Persistence of firing sequences \cite{och,folco-2014}}
A finite or infinite firing sequence $M_0\fire{a_1}M_1\fire{a_2}\ldots$ is called
\emph{persistent} if for every used index $i>0$ and transition $t\neq a_i$,
if $M_{i-1}\fire{t}$ then also $M_i\fire{t}\;$.
\ENDDEF{pers-seq.def}

\REM{pers-local.rem}{Locality of persistent sequences} 
The persistence of a sequence $M_0\xfire{a_1}M_1\xfire{a_2}\ldots$
means that no single step $M_{i-1}\xfire{a_i}M_i$ may switch
some transition $t\neq a_i$ from being enabled at $M_{i-1}$ to being disabled at $M_i$.
This is similar to the persistence of a net, but restricted to a single firing sequence.

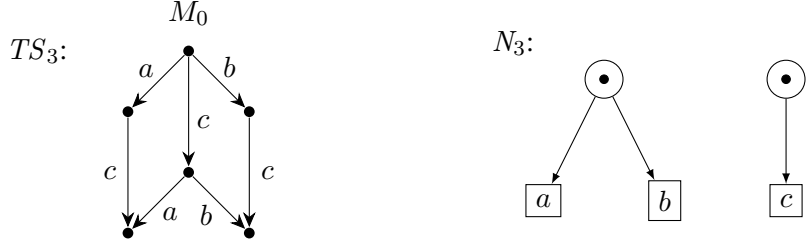
\begin{figure}[htb] 
\begin{center}
\begin{tikzpicture}[scale=0.8]
\refstepcounter{exampleTScounter}\label{pers-local.ts}
\node[]()at(-0.5,3)[]{$\TS_\theexampleTScounter$:};
\node[circle,fill=black!100,inner sep=0.05cm](0)at(2,3)[label=above:]{};
 \node[above of=0,node distance=0.5cm]{$M_0$};
\node[circle,fill=black!100,inner sep=0.05cm](1)at(1,2)[label=above left:$$]{};
\node[circle,fill=black!100,inner sep=0.05cm](2)at(3,2)[label=above right:$$]{};
\node[circle,fill=black!100,inner sep=0.05cm](3)at(2,1)[label=below:]{};
\node[circle,fill=black!100,inner sep=0.05cm](4)at(1,0)[label=below:]{};
\node[circle,fill=black!100,inner sep=0.05cm](5)at(3,0)[label=below:]{};
\draw[-{Stealth[length=2mm,width=1.7mm]},bend right=0](0)edge node[above left,inner sep=0.05cm,pos=0.5]{$a$}(1);
\draw[-{Stealth[length=2mm,width=1.7mm]},bend right=0](0)edge node[above right,inner sep=0.05cm,pos=0.5]{$b$}(2);
\draw[-{Stealth[length=2mm,width=1.7mm]},bend right=0](0)edge node[right,inner sep=0.1cm,pos=0.6]{$c$}(3);
\draw[-{Stealth[length=2mm,width=1.7mm]},bend right=0](1)edge node[left,inner sep=0.15cm,pos=0.5]{$c$}(4);
\draw[-{Stealth[length=2mm,width=1.7mm]},bend right=0](2)edge node[right,inner sep=0.15cm,pos=0.5]{$c$}(5);
\draw[-{Stealth[length=2mm,width=1.7mm]},bend right=0](3)edge node[below right,inner sep=0.05cm,pos=0.5]{$a$}(4);
\draw[-{Stealth[length=2mm,width=1.7mm]},bend right=0](3)edge node[below left,inner sep=0.05cm,pos=0.45]{$b$}(5);
\end{tikzpicture}\hspace*{2.5cm}
\raisebox{0.5cm}{\begin{tikzpicture}[scale=0.8]
\refstepcounter{examplePNcounter}\label{pers-local.pn}
\node[]()at(-0.5,2.6)[]{$\PN_\theexamplePNcounter$:};
\node[circle,draw,minimum size=0.5cm](p0)at(1,2)[]{};
 \filldraw[black](1,2)circle(2pt);
\node[circle,draw,minimum size=0.4cm](p0a)at(-0.2,1.2)[]{};
 \filldraw[black](-0.2,1.2)circle(2pt);
\node[circle,draw,minimum size=0.4cm](p0b)at(2.2,1.2)[]{};
 \filldraw[black](2.2,1.2)circle(2pt);
\node[circle,draw,minimum size=0.5cm](p1)at(4,2)[]{};
 \filldraw[black](4,2)circle(2pt);
\node[draw,minimum size=0.4cm](a)at(0,0){$a$}; 
\node[draw,minimum size=0.4cm](b)at(2,0){$b$};
\node[draw,minimum size=0.4cm](c)at(4,0){$c$};
\draw[-latex](p0)--(a);
\draw[-latex](p0a)--(a);
\draw[-latex](p0)--(b);
\draw[-latex](p0b)--(b);
\draw[-latex](p1)--(c);
\end{tikzpicture}}
\end{center}
\caption{An \lts{} $\TSref{pers-local.ts}$
and a Petri net solution $\PNref{pers-local.pn}$.
The sequence $M_0\xfire{c}$ is  persistent
while the sequences $M_0\xfire{a}$ and $M_0\xfire{b}$ are not.}
\label{pers-local.fig}
\end{figure}

It may be the case that a sequence ``touches'' a non-persistent state
while being persistent itself.
In Figure \ref{pers-local.fig}, the sequence $M_0\xfire{c}$ 
is persistent and starts with $c$, but the initial marking $M_0$ is also a non-persistent state involving
$a$ and $b$, rather than $c$.
The sequences $M_0\xfire{ac}$ and $M_0\xfire{bc}$ are not persistent.
Here we have two full diamonds, involving $a,c$ and $b,c$, and two 1/2 diamonds,  involving $a,b$.
In general, there may also be 3/4 diamonds; an example will be given later (Figure \ref{dia-lem.fig}).
A persistent sequence may thus in some circumstances go through one or more non-persistent states,
but if a sequence does not go through any non-persistent state, it is certainly persistent.
\ENDREM{pers-local.rem}

The following corollary is a straightforward consequence of Definition \ref{pers-seq.def}: 

\KOR{persist.cor}{Persistence factorisation}
A firing sequence $M\xfire{\alpha}M'\xfire{\beta}$ is persistent iff so are $M\xfire{\alpha}M'$ and $M'\xfire{\beta}\;$. \\
A sequence $M_0\xfire{a_1}M_1\xfire{a_2}\ldots$ is persistent iff the singleton sequences $M_{i-1}\xfire{a_i}M_i$
are persistent for every used index $i>0$.
\ENDKOR{persist.cor}

The relationship between the persistence of $N$,
the persistence of its firing sequences,
and choice-freeness,
is clarified by the next propositions.

\PROP{pers.prop}{Persistent nets and persistent firing sequences}
$N=(P,T,F,M_0)$ is persistent if and only if all of its
firing sequences $M_0\xfire{\sigma}$ are persistent.
\ENDPROP

To shorten the proof, let us define the predicate $\mathit{nonpers}(M,a,M',t)$ to be true
if $M$ is a reachable marking, $a$ and $t$ are two different transitions, $M\xfire{a}M'$,
$M\xfire{t}$ and $\neg(M'\xfire{t})$; thus $\mathit{nonpers}$ describes a situation
of non-persistence in the reachability graph of a net.

\BEW
By contraposition in both directions.

($\Rightarrow$):
Suppose that $M_0\xfire{a_1}M_1\xfire{a_2}M_2\xfire{a_3}\ldots$ is a (finite or infinite)
non-persistent firing sequence of $(N,M_0)$.
By Definition \ref{pers-seq.def}, there is some index $i\geq1$ and a transition $t\neq a_i$ such that
$\mathit{nonpers}(M_{i-1},a_i,M_i,t)$.
Since $M_{i-1}$ is reachable, $(N,M_0)$ is non-persistent, by Definition \ref{lts-prop.def}.

($\Leftarrow$):
Suppose that $(N,M_0)$ is non-persistent.
By Definition \ref{lts-prop.def},
for some reachable markings $M,M'$ and transitions $a,t$,
the predicate $\mathit{nonpers}(M,a,M',t)$ is true.
Since $M$ is reachable, $M_0\xfire{\alpha}M$ for some $\alpha\in T^*$.
Then, by Definition~\ref{pers-seq.def}, $M_0\xfire{\alpha\,a}M'$ is a finite non-persistent sequence.
\ENDBEW{pers.prop}

\PROP{CF.prop}{Choice-free nets are persistent}
Any choice-free Petri net is persistent, whatever its initial marking.
\ENDPROP

\BEW
Let $t,u$ with $t\neq u$ be two distinct transitions and let $M,M',M''$ be reachable markings
such that $M\xfire{t}M'$ and $M\xfire{u}M''$.
By $t\neq u$ and choice-freeness, $\pre{t}\cap\pre{u}=\es$.
Thus firing $t$ does not disable $u$ and vice versa, so that both
$M\xfire{tu}$ and $M\xfire{ut}$ are firable and lead to the same state, by determinism.
\ENDBEW{CF.prop}

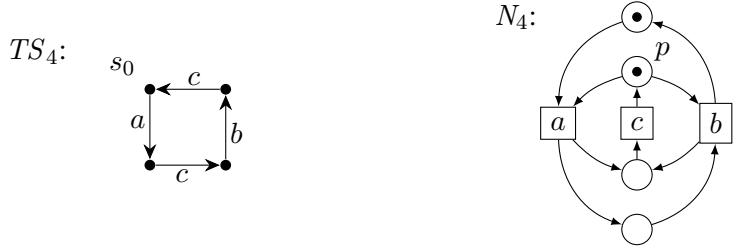
\begin{figure}[htb]
\begin{center}
\begin{tikzpicture}[scale=1]
\refstepcounter{exampleTScounter}\label{acbc.ts}
\node[]()at(-1.5,1.5)[]{$\TS_\theexampleTScounter$:};
\node[circle,fill=black!100,inner sep=0.05cm](0)at(0,1)[label=above left:$s_0$]{};
\node[circle,fill=black!100,inner sep=0.05cm](1)at(0,0)[label=below left:$$]{};
\node[circle,fill=black!100,inner sep=0.05cm](2)at(1,0)[label=below right:$$]{};
\node[circle,fill=black!100,inner sep=0.05cm](3)at(1,1)[label=above right:$$]{};
\draw[-{Stealth[length=2mm,width=1.7mm]},bend right=0](0)edge node[left,inner sep=0.5mm,pos=0.4]{$a$}(1);
\draw[-{Stealth[length=2mm,width=1.7mm]},bend right=0](1)edge node[below,inner sep=0.5mm,pos=0.4]{$c$}(2);
\draw[-{Stealth[length=2mm,width=1.7mm]},bend right=0](2)edge node[right,inner sep=0.5mm,pos=0.4]{$b$}(3);
\draw[-{Stealth[length=2mm,width=1.7mm]},bend right=0](3)edge node[above,inner sep=0.5mm,pos=0.4]{$c$}(0);
\end{tikzpicture}\hspace*{3cm}
\raisebox{-1cm}{\begin{tikzpicture}[scale=0.8] 
\refstepcounter{examplePNcounter}\label{acbc.pn}
\node[]()at(-1,3)[]{$\PN_\theexamplePNcounter$:};
\node[circle,draw,minimum size=0.4cm](pab)at(1,-0.5)[label=below:$$]{};
\node[circle,draw,minimum size=0.4cm](pabc)at(1,0.4)[label=below:$$]{};
\node[circle,draw,minimum size=0.4cm](pcab)at(1,2.1)[label=below:$$][label={[label distance=-0.1cm]5:$p$}]{};
\filldraw[black](1,2.1)circle(2pt);
\node[circle,draw,minimum size=0.4cm](pba)at(1,3)[label=below:$$]{};
\filldraw[black](1,3)circle(2pt);
\node[draw,minimum size=0.3cm](a)at(-0.3,1.25){$a$};
\node[draw,minimum size=0.3cm](c)at(1,1.25){$c$};
\node[draw,minimum size=0.3cm](b)at(2.3,1.25){$b$};
\draw[-latex,bend right=35](a)edge node[above,pos=0.5]{$$}(pab);
\draw[-latex,bend right=35](pab)edge node[above,pos=0.5]{$$}(b);
\draw[-latex,bend right=35](b)edge node[above,pos=0.5]{$$}(pba);
\draw[-latex,bend right=35](pba)edge node[above,pos=0.5]{$$}(a);
\draw[-latex,bend right=15](a)edge node[above,pos=0.5]{$$}(pabc);
\draw[-latex,bend left=15](b)edge node[above,pos=0.5]{$$}(pabc);
\draw[-latex,bend right=15](pcab)edge node[above,pos=0.5]{$$}(a);
\draw[-latex,bend left=15](pcab)edge node[above,pos=0.5]{$$}(b);
\draw[-latex,bend right=0](pabc)edge node[above,pos=0.5]{$$}(c);
\draw[-latex,bend right=0](c)edge node[above,pos=0.5]{$$}(pcab);
\end{tikzpicture}}
\end{center}
\caption{A transition system $\TSref{acbc.ts}$ and a Petri net $\PNref{acbc.pn}$ solving it.
$\PNref{acbc.pn}$ is
deadlock-free,
plain, pure, safe, and persistent, but not choice-free
since $a$ and $b$ share a pre-place $p$.
It is asymmetric choice, but neither dissymmetric choice nor free-choice.} 
\label{acbc.fig}
\end{figure}

Proposition \ref{CF.prop} cannot be reversed in general. Figure \ref{acbc.fig} shows
a persistent transition system $\TSref{acbc.ts}$ and a persistent pps Petri net solution $\PNref{acbc.pn}$ 
which is not choice-free.

\section{Permutation equivalence}
\label{perm2.sct}

The formal definition of permutation equivalence, $\equiv$, has three parts:
(i): $\equiv_0$, referring to a single permutation;
(ii): $\equiv_0^*$, referring to a finite concatenation of single permutations, including the identity as a special case,
i.e., referring  to the reflexive and transitive closure of $\equiv_0$;
and (iii): $\equiv$, referring, in addition, to infinite concatenations of single permutations.

\Needspace{5\baselineskip}
\DEF{exch.def}{Permutation equivalence of firing sequences \cite{DBLP:journals/tcs/BestD87}}
Whenever $M_0\xfire{\alpha t_it_{i+1}\beta}$ and $M_0\xfire{\alpha t_{i+1}t_i\beta}$
for $\alpha\in T^*$ and $\beta\in T^*\cup  T^\infty$,
then
\[M_0\xfire{\alpha t_it_{i+1}\beta}\;\;\equiv_0\;\;M_0\xfire{\alpha t_{i+1}t_i\beta}
\]
by definition; moreover, $\equiv_0^*$ is the reflexive and transitive closure of $\equiv_0$,
and $\equiv\;=\;\equiv_0^*$ for finite firing sequences, by definition.
If $\sigma_1,\sigma_2\in T^\infty$ are infinite firable sequences,
then $M_0\xfire{\sigma_1}\;\equiv\;M_0\xfire{\sigma_2}$ if
\begin{itemize}
\item
either there are ``finitely many permutations'', i.e.: $M_0\xfire{\sigma_1}\;\;\equiv_0^*\;M_0\xfire{\sigma_2}$
\item
or there are ``infinitely many permutations'', defined as follows: \\
for every $n\geq0$, there are $M_0\xfire{\sigma_1'}$ and $M_0\xfire{\sigma_2'}$
such that
\begin{itemize}
\item[] 
$\!\!\!\!\!\!M_0\xfire{\sigma_1}\;\;\equiv_0^*\;M_0\xfire{\sigma_1'}$
and $\sigma_1'$, $\sigma_2$ have the same prefix of length $n$
\item[]
$\!\!\!\!\!\!M_0\xfire{\sigma_2}\;\;\equiv_0^*\;M_0\xfire{\sigma_2'}$
and $\sigma_2'$, $\sigma_1$ have the same prefix of length $n$.
\BX{\ref{exch.def}}
\end{itemize}
\end{itemize}
\ENXDEF

\begin{figure}[htp]
\begin{center}
\begin{tikzpicture}[scale=0.8]
\node[circle,draw,minimum size=0.5cm](0)at(2,1)[label=above:$$]{};
\filldraw[black](2,1)circle(2pt);
\node[draw,minimum size=0.4cm](a)at(1,1)[label=above:]{$a$};
\node[draw,minimum size=0.4cm](b)at(3,1)[label=above:]{$b$};
\draw([yshift=0.09cm]0.west)[]edge[-latex,bend right=10]
node[ellipse,below,inner sep=2pt,pos=0.5]{}([yshift=0.09cm]a.east);
\draw([yshift=-0.09cm]a.east)[]edge[-latex,bend right=10]
node[ellipse,below,inner sep=2pt,pos=0.5]{}([yshift=-0.09cm]0.west);
\draw([yshift=0.09cm]0.east)[]edge[-latex,bend left=10]
node[ellipse,below,inner sep=2pt,pos=0.5]{}([yshift=0.09cm]b.west);
\draw([yshift=-0.09cm]b.west)[]edge[-latex,bend left=10]
node[ellipse,below,inner sep=2pt,pos=0.5]{}([yshift=-0.09cm]0.east);
\end{tikzpicture}\hspace*{2cm}
\begin{tikzpicture}[scale=0.8]
\node[circle,draw,minimum size=0.5cm](0)at(2,1)[label=above:$$]{};
\filldraw[black](2,1)circle(2pt);
\node[circle,draw,minimum size=0.5cm](1)at(3,1)[label=above:$$]{};
\filldraw[black](3,1)circle(2pt);
\node[draw,minimum size=0.4cm](a)at(1,1)[label=above:]{$a$};
\node[draw,minimum size=0.4cm](b)at(4,1)[label=above:]{$b$};
\draw([yshift=0.09cm]0.west)[]edge[-latex,bend right=10]
node[ellipse,below,inner sep=2pt,pos=0.5]{}([yshift=0.09cm]a.east);
\draw([yshift=-0.09cm]a.east)[]edge[-latex,bend right=10]
node[ellipse,below,inner sep=2pt,pos=0.5]{}([yshift=-0.09cm]0.west);
\draw([yshift=0.09cm]1.east)[]edge[-latex,bend left=10]
node[ellipse,below,inner sep=2pt,pos=0.5]{}([yshift=0.09cm]b.west);
\draw([yshift=-0.09cm]b.west)[]edge[-latex,bend left=10]
node[ellipse,below,inner sep=2pt,pos=0.5]{}([yshift=-0.09cm]1.east);
\end{tikzpicture}\hspace*{2cm}
\raisebox{-0.7cm}{\begin{tikzpicture}[scale=1]
\node[circle,fill=black!100,inner sep=0.05cm](0)at(0,0)[label=above:$M_0$]{};
\draw[-{Stealth[length=2mm,width=1.7mm]}](0)to[out=140,in=230,looseness=30]node[left]{$a$}(0);
\draw[-{Stealth[length=2mm,width=1.7mm]}](0)to[out=40,in=310,looseness=30]node[right]{$b$}(0);
\end{tikzpicture}}
\end{center}
\caption{Two nets generating isomorphic reachability graphs (namely, the one shown on the right-hand side).
In both cases, the sequences $M_0\xfire{ababa\ldots}$ and $M_0\xfire{babab\ldots}$
are $\equiv$-equivalent but not $\equiv_0^*$-equivalent.}
\label{abab.fig}
\end{figure}
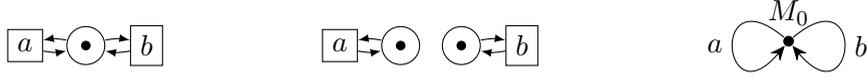

The last part of Definition \ref{exch.def} is relevant in cases such as shown in Figure \ref{abab.fig}
where infinite firing sequences are in relation $\equiv$ but not in relation $\equiv_0^*$.

\PROP{perm-parikh.kor}{Permutation equivalence implies Parikh-equivalence}
If $(M_0\xfire{\sigma_1}\;\equiv\;M_0\xfire{\sigma_2})$,
then $( \Parikh(\sigma_1)=\Parikh(\sigma_2))$.
\ENDPROP

\BEW

Clearly, $(M_0\fire{\sigma_1}\;\equiv_0M_0\fire{\sigma_2})\;\impl\;(\Parikh(\sigma_1)=\Parikh(\sigma_2))$,
hence also $(M_0\fire{\sigma_1}\;\equiv_0^*M_0\fire{\sigma_2})\;\impl\;( \Parikh(\sigma_1)=\Parikh(\sigma_2))$,
and the property is true for finite firing sequences.

If $\sigma_1,\sigma_2\in T^\infty$, $M_0\fire{\sigma_1}\;\equiv M_0\fire{\sigma_2}$ and $t\in T$, then
\begin{itemize}
\item 
 if $t$ occurs infinitely often in $\sigma_1$, 
let $n\in\nsymbol$ and $n_1$ be the position of the $n^{th}$ occurrence of $t$ in~$\sigma_1$.
By definition, there is $\sigma'_2$ such that $M_0\fire{\sigma_2}\;\equiv_0^*M_0\fire{\sigma_2'}$
and $\sigma_1$, $\sigma_2'$ agree on their prefixes of length $n_1$, meaning that $\Parikh(\sigma_2)(t)\geq n$,
hence that $\Parikh(\sigma_2)(t)=\infty=\Parikh(\sigma_1)(t)$ 
(and similarly if $t$ occurs infinitely often in $\sigma_2$);
we thus only have to consider the case where $t$ occurs finitely often, both in $\sigma_1$ and $\sigma_2$;
\item
 if $\Parikh(\sigma_1)(t)= n\in\nsymbol$ and $n_1$ is the position of the $n^{th}$ occurrence of $t$ in $\sigma_1$,
by definition there is $\sigma'_2$ such that $M_0\fire{\sigma_2}\;\equiv_0^*M_0\fire{\sigma_2'}$
and $\sigma_1$, $\sigma_2'$ agree on their prefixes of length $n_1$, meaning that $\Parikh(\sigma_2)(t)\geq n=\Parikh(\sigma_1)(t)$:
similarly, $\Parikh(\sigma_1)(t)\geq \Parikh(\sigma_2)(t)$, so that $\Parikh(\sigma_1)(t)= \Parikh(\sigma_2)(t)$.
\BX{\ref{perm-parikh.kor}}
\end{itemize}
\ENXBEW

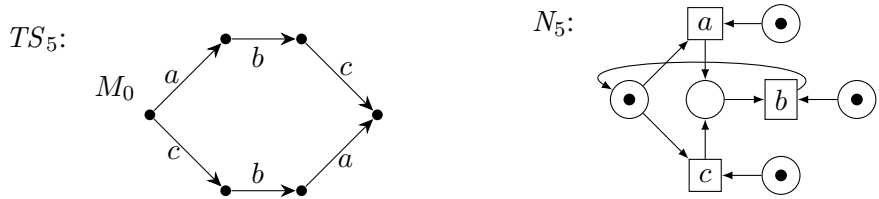
\begin{figure}[htb]
\begin{center}
\begin{tikzpicture}[scale=0.9]
\refstepcounter{exampleTScounter}\label{spar.ts}
\node[]()at(-1.5,1)[]{$\TS_\theexampleTScounter$:};
\node[circle,fill=black!100,inner sep=0.05cm](0)at(0,0)[label=above left:$M_0$]{};
\node[circle,fill=black!100,inner sep=0.05cm](1)at(1,1)[label=below left:$$]{};
\node[circle,fill=black!100,inner sep=0.05cm](2)at(2,1)[label=below right:$$]{};
\node[circle,fill=black!100,inner sep=0.05cm](3)at(3,0)[label=above right:$$]{};
\node[circle,fill=black!100,inner sep=0.05cm](4)at(1,-1)[label=below left:$$]{};
\node[circle,fill=black!100,inner sep=0.05cm](5)at(2,-1)[label=below right:$$]{};
\draw[-{Stealth[length=2mm,width=1.7mm]},bend right=0](0)edge node[left,inner sep=1mm,pos=0.5]{$a$}(1);
\draw[-{Stealth[length=2mm,width=1.7mm]},bend right=0](1)edge node[below,inner sep=1mm,pos=0.4]{$b$}(2);
\draw[-{Stealth[length=2mm,width=1.7mm]},bend right=0](2)edge node[above,inner sep=1mm,pos=0.6]{$c$}(3);
\draw[-{Stealth[length=2mm,width=1.7mm]},bend right=0](0)edge node[left,inner sep=1mm,pos=0.5]{$c$}(4);
\draw[-{Stealth[length=2mm,width=1.7mm]},bend right=0](4)edge node[above,inner sep=1mm,pos=0.4]{$b$}(5);
\draw[-{Stealth[length=2mm,width=1.7mm]},bend right=0](5)edge node[below,inner sep=1mm,pos=0.6]{$a$}(3);
\end{tikzpicture}\hspace*{1.5cm}
\begin{tikzpicture}[scale=0.9]
\refstepcounter{examplePNcounter}\label{spar.pn}
\node[]()at(-1,1)[]{$\PN_\theexamplePNcounter$:};
\node[circle,draw,minimum size=0.5cm](0)at(0,0)[label=below:$$]{};\filldraw[black](0,0)circle(2pt);
\node[circle,draw,minimum size=0.5cm](1)at(1,0)[label=below:$$]{};
\node[circle,draw,minimum size=0.5cm](2)at(3,0)[label=below:$$]{};\filldraw[black](3,0)circle(2pt);
\node[circle,draw,minimum size=0.5cm](3)at(2,1)[label=below:$$]{};\filldraw[black](2,1)circle(2pt);
\node[circle,draw,minimum size=0.5cm](4)at(2,-1)[label=below:$$]{};\filldraw[black](2,-1)circle(2pt);
\node[draw,minimum size=0.4cm](a)at(1,1){$a$};
\node[draw,minimum size=0.4cm](c)at(1,-1){$c$};
\node[draw,minimum size=0.4cm](b)at(2,0){$b$};
\draw[-latex,bend right=0](0)edge node[above,pos=0.5]{$$}(a);
\draw[-latex,bend right=0](0)edge node[above,pos=0.5]{$$}(c);
\draw[-latex,bend right=0](a)edge node[above,pos=0.5]{$$}(1);
\draw[-latex,bend right=0](c)edge node[above,pos=0.5]{$$}(1);
\draw[-latex,bend right=0](1)edge node[above,pos=0.5]{$$}(b);
\draw[-latex,bend left=0](2)edge node[above,pos=0.5]{$$}(b);
\draw[-latex,bend right=0](3)edge node[above,pos=0.5]{$$}(a);
\draw[-latex,bend left=0](4)edge node[above,pos=0.5]{$$}(c);
\draw[-latex,bend right=150](b)edge node[above,pos=0.5]{$$}(0);
\end{tikzpicture}
\end{center}
\caption{A pps net $\PNref{spar.pn}$ on the right and its reachability graph $\TSref{spar.ts}$ on the left:
we have $\Parikh(abc)=(1,1,1)=\Parikh(cba)$ but not $M_0\xfire{abc}\;\equiv M_0\xfire{cba}$,
since neither $M_0\xfire{bac}$ (so, the first sequence does not allow
$a$ and $b$ to be permuted) nor $M_0\xfire{acb}\;$ (so, a permutation of $b$ and $c$
is also not permitted in the first sequence).
Thus, the first sequence may not be permuted into something else, and the same is true for the second sequence.}
\label{spar.fig}
\end{figure}

But having the same Parikh vector does not imply permutation equivalence,
even if the two sequences are firable from the same marking,
as illustrated by Figure~\ref{spar.fig}.

\section{\text{Fairness, SPE, S$\widetilde{\text{P}}$E, FPE, and Ochma\'nski's conjecture}}
\label{och.sct}

It is easier to detect whether some \emph{finite}
sequence can be permuted into an equivalent persistent finite sequence
than to check whether an \emph{infinite} sequence can be permuted into a persistent infinite equivalent.
Therefore, it is interesting to know whether
the permutability of non-persistent into persistent sequences
can be inherited by infinite sequences from finite ones.
Example \ref{unfair.exa} shows that even in the restricted context of pps Petri nets,
this is not always the case.

\begin{figure}[htb]
\begin{center}
\begin{tikzpicture}[scale=0.8]
\refstepcounter{examplePNcounter}\label{unfair.pn}
\node[]()at(-1,2.3)[]{$\PN_\theexamplePNcounter$:};
\node[circle,draw,minimum size=0.5cm](p0)at(7,1)[]{};
 \filldraw[black](7,1)circle(2pt);
\node[circle,draw,minimum size=0.5cm](p1)at(0.5,2)[]{};
\node[circle,draw,minimum size=0.5cm](p2)at(0.5,0)[]{};
 \filldraw[black](0.5,0)circle(2pt);
\node[circle,draw,minimum size=0.5cm](p3)at(2,2)[]{};
 \filldraw[black](2,2)circle(2pt);
\node[circle,draw,minimum size=0.5cm](p4)at(2,0)[]{};
\node[circle,draw,minimum size=0.5cm](p5)at(5,1)[]{};
\node[draw,minimum size=0.4cm](a)at(1.25,1){$a$};
\node[draw,minimum size=0.4cm](b)at(4,1){$b$};
\node[draw,minimum size=0.4cm](c)at(2.5,1){$c$};
\node[draw,minimum size=0.4cm](x)at(0,1){$x$};
\node[draw,minimum size=0.4cm](y)at(6,1){$y$};
\draw[-latex](x)--(p1);
\draw[-latex](p1)--(a);
\draw[-latex](a)--(p2);
\draw[-latex](p2)--(x);
\draw[-latex](a)--(p4);
\draw[-latex](p4)--(c);
\draw[-latex](c)--(p3);
\draw[-latex](p3)--(a);
\draw[-latex](p3)--(b);
\draw[-latex](p0)--(y);
\draw[-latex](y)--(p5);
\draw[-latex](p5)--(b);
\end{tikzpicture}\hspace*{1cm}
\raisebox{-0.65cm}{\begin{tikzpicture}[scale=0.9]
\refstepcounter{exampleTScounter}\label{unfair.ts}
\node[]()at(-2,3.2)[]{$\TS_\theexampleTScounter$:};
\node[circle,fill=black!100,inner sep=0.05cm](M0)at(0,3.2)[label=left:$M_0$]{};
\node[circle,fill=black!100,inner sep=0.05cm](M1)at(4,3.2)[label=below left:$$]{};
\node[circle,fill=black!100,inner sep=0.05cm](M2)at(2,3.2)[label=below left:$$]{};
\node[circle,fill=black!100,inner sep=0.05cm](M3)at(3.15,1.3)[label=above:$$]{};
\node[circle,fill=black!100,inner sep=0.05cm](M4)at(2,2)[label=above:$$]{};
\node[circle,fill=black!100,inner sep=0.05cm](M5)at(0,2)[label=above:$$]{};
\node[circle,fill=black!100,inner sep=0.05cm](M6)at(4,2)[label=above:$$]{};
\node[circle,fill=black!100,inner sep=0.05cm](M7)at(1.15,1.3)[label=above:$$]{};
\draw[-{Stealth[length=2mm,width=1.7mm]},bend right=30](M1)edge node[below,inner sep=0.6mm,pos=0.5]{$c$}(M0);
\draw[-{Stealth[length=2mm,width=1.7mm]},bend left=0](M1)edge node[right,inner sep=1mm,pos=0.5]{$y$}(M6);
\draw[-{Stealth[length=2mm,width=1.7mm]},bend right=0](M0)edge node[below,inner sep=1mm,pos=0.5]{$x$}(M2);
\draw[-{Stealth[length=2mm,width=1.7mm]},bend right=0](M2)edge node[below,inner sep=1mm,pos=0.5]{$a$}(M1);
\draw[-{Stealth[length=2mm,width=1.7mm]},bend right=0](M2)edge node[right,inner sep=1mm,pos=0.5]{$y$}(M4);
\draw[-{Stealth[length=2mm,width=1.7mm]},bend left=0](M0)edge node[right,inner sep=1mm,pos=0.5]{$y$}(M5);
\draw[-{Stealth[length=2mm,width=1.7mm]}](M6)to[out=-120,in=300,looseness=1.3]node[above]{$c$}(M5);
\draw[-{Stealth[length=2mm,width=1.7mm]},bend right=0](M5)edge node[above,inner sep=1mm,pos=0.5]{$x$}(M4);
\draw[-{Stealth[length=2mm,width=1.7mm]},bend right=0](M4)edge node[above,inner sep=1mm,pos=0.5]{$a$}(M6);
\draw[-{Stealth[length=2mm,width=1.7mm]},bend right=0](M4)edge node[above right,inner sep=0.4mm,pos=0.7]{$b$}(M3);
\draw[-{Stealth[length=2mm,width=1.7mm]},bend right=0](M5)edge node[above right,inner sep=0.4mm,pos=0.7]{$b$}(M7);
\draw[-{Stealth[length=2mm,width=1.7mm]},bend right=0](M7)edge node[above,inner sep=1mm,pos=0.4]{$x$}(M3);
\end{tikzpicture}}
\end{center}
\caption{The pps net $\PNref{unfair.pn}$ -- with reachability graph $RG(\PNref{unfair.pn}){=}\TSref{unfair.ts}$ --
satisfies SPE; essentially, the conflict between $a$ and $b$ can be delayed 
by executing $y$ as late as necessary.
It also has an infinite, but unfair, firing sequence
$M_0\xfire{y(xac)^\infty}$ which cannot be permuted into a persistent one;
essentially, $y$ cannot be delayed forever
without violating permutation equivalence.}
\label{unfair.fig}
\end{figure}
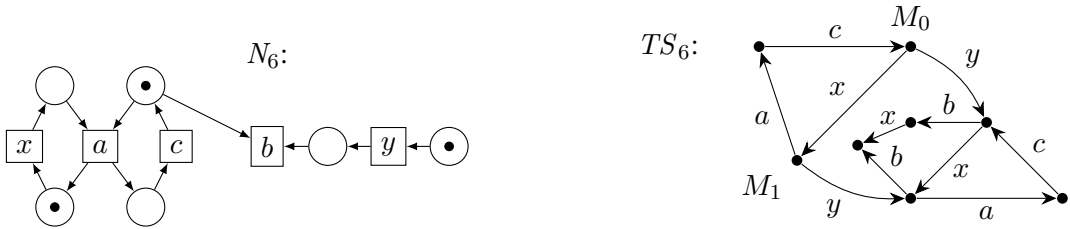

\EXA{unfair.exa}{Figure \ref{unfair.fig} (taken from \cite{folco-2014})}
In $\PNref{unfair.pn}$, every finite firing sequence can be permuted in such a way
as to avoid enabling $a$ and $b$ (the only two transitions which are in a conflict
due to their shared pre-place) simultaneously;
for instance, the sequence $y(xac)^n$ may be permuted into $(xac)^ny$.
However, there is an infinite non-persistent firing sequence
\begin{equation}\label{unfair.eq}
M_0\;\xfire{\;y\;(\;x\;a\;c\;)^\infty}
\end{equation}
which has no equivalent persistent permutation.

Observe, however, that the critical transition $b$ which is the cause
of this non-persistence (since it is enabled everywhere in (\ref{unfair.eq})
except after $yx(acx)^ia$ for $i=0,1,\ldots$) does not occur in (\ref{unfair.eq}).
In normal parlance, (\ref{unfair.eq}) is \emph{unfair} towards $b$.
It may therefore be asked whether from the permutability
of finite sequences, at least the permutability of \emph{fair} infinite sequences follows.
This question is at the core of Ochma\'nski's hypothesis (as explained later in this section).
\ENDEXA{unfair.exa}

Example \ref{unfair.exa} motivates why fairness is an essential
ingredient of Ochma\'nski's conjecture.
The fairness notion he employs is very much standard
and is called \emph{strong fairness},
or just \emph{fairness}, in the literature.
Strong fairness applies to infinite firing sequences.
For finite firing sequences, Ochma\'nski adopts the idea
that only \emph{deadlocking} sequences should be called fair.
There are other possibilities, and we shall discuss them
in Section \ref{fair.sct} below.\footnote{The results, proofs and (counter-) examples in
the main part of this paper (Sections~\ref{main.sct} and \ref{counter.sct})
do not depend on this choice.}

Let $\exists_i^{\infty}$ denote ``there are infinitely many $i$ with $\ldots$'' and
$\forall_i^{\infty}$ denote ``for all but finitely many $i$, $\ldots$''.

\DEF{fair.def}{(Strong) Fairness}
A finite firing sequence $M_0\xfire{\sigma}M$ is fair if $M$ is a deadlock.

An infinite firing sequence
$M_0\xfire{t_1}M_1\xfire{t_2}M_2\xfire{t_3}M_3\xfire{t_4}\ldots$
is \emph{fair with respect to $t\in T$} if
\[(\exists_i^{\infty}\colon(t=t_i))\;\;\vee\;\;(\forall_i^{\infty}\colon\neg(M_i\xfire{t}))
\]
and it is \emph{fair} if it is fair with respect to every transition $t\in T$.
\ENDDEF{fair.def}

\DEF{spe-fpe.def}{Persistent permutation equivalents \cite{och} and Parikh equivalents}
The marked net $(N,M_0)$ is called SPE (for ``short persistent equivalent'') if
every finite firing sequence starting from $M_0$
has a persistent permutation equivalent;
S$\widetilde{\text{P}}$E (for ``short Parikh equivalent'') if every finite firing sequence starting from $M_0$
has a persistent Parikh equivalent;
and FPE (for ``fair persistent equivalent'') if every fair firing sequence starting from $M_0$
has a persistent permutation equivalent.
\ENDDEF{spe-fpe.def} 

Proposition~\ref{perm-parikh.kor} implies that SPE is stronger than S$\widetilde{\text{P}}$E.

\EXA{spe-fpe-1.exa}{$(\PNref{basic-exa.fig},M_0)$ in Figure \ref{basic-exa.fig}}
This net is pps and satisfies both SPE and S$\widetilde{\text{P}}$E.
For example, the non-persistent sequence $M_0\xfire{cda}$ has the persistent (and Parikh) equivalent $M_0\xfire{cad}$.
It also satisfies FPE, since there are no infinite firing sequences,
so that all fair sequences are deadlocking and hence (of course) finite,
and because it satisfies SPE.
As a consequence of Proposition~\ref{pers.prop},
all persistent nets (for instance, $\PNref{acbc.pn}$ in Figure \ref{acbc.fig}
and the two nets in Figure \ref{abab.fig})
satisfy both SPE (and S$\widetilde{\text{P}}$E) and FPE, using identity permutations.
\ENDEXA{spe-fpe-1.exa}

\CONJ{och.conj}{(extended to pps Petri nets, based on Ochma\'nski \cite{och,folco-2014})} 

Let $(N,M_0)$ be a pps Petri net which satisfies SPE. Then it also satisfies FPE.
\ENDCONJ{och.conj}

In \cite{och,folco-2014}, the conjecture has been formulated in terms of elementary nets,
rather than in terms of pps Petri nets.
If it turns
out to be true for pps Petri nets, then it is also true for elementary nets (with an immediate proof).
By SPE, every fair finite sequence has a persistent equivalent,
simply by its finiteness.
The key problem is to investigate whether SPE $\impl$ FPE for \emph{infinite} firing sequences.

\begin{figure}[htb]
\begin{center}
\begin{tikzpicture}[scale=0.8]
\node[]()at(-1,3.5)[]{$\TSref{basic-exa.ts}':$};
\node[circle,fill=black!100,inner sep=0.05cm](0)at(2,3)[label=above:]{};
 \node[above of=0,node distance=0.4cm]{$M_0$};
\node[circle,fill=black!100,inner sep=0.05cm](1)at(1,2)[label=above left:$M_1$]{};
\node[circle,fill=black!100,inner sep=0.05cm](2)at(3,2)[label=above right:$M_2$]{};
\node[circle,fill=black!100,inner sep=0.05cm](3)at(0,1)[label=left:$M_3$]{};
\node[circle,fill=black!100,inner sep=0.05cm](4)at(2,1)[label=below:]{};
 \node[below of=4,node distance=0.5cm]{$M_4$};
\node[circle,fill=black!100,inner sep=0.05cm](5)at(4,1)[label=right:$M_5$]{};
\node[circle,fill=black!100,inner sep=0.05cm](6)at(1,0)[label=below:]{};
 \node[below of=6,node distance=0.5cm]{$M_6$};
\node[circle,fill=black!100,inner sep=0.05cm](7)at(3,0)[label=below:]{};
 \node[below of=7,node distance=0.5cm]{$M_7$};
\draw[-{Stealth[length=2mm,width=1.7mm]},bend right=0](0)edge node[above,inner sep=0.15cm,pos=0.6]{$c$}(1);
\draw[-{Stealth[length=2mm,width=1.7mm]},bend right=0](0)edge node[above,inner sep=0.15cm,pos=0.6]{$d$}(2);
\draw[-{Stealth[length=2mm,width=1.7mm]},bend right=0](1)edge node[above,inner sep=0.15cm,pos=0.6]{$a$}(3);
\draw[-{Stealth[length=2mm,width=1.7mm]},bend right=0](1)edge node[above,inner sep=0.15cm,pos=0.6]{$d$}(4);
\draw[-{Stealth[length=2mm,width=1.7mm]},bend right=0](2)edge node[above,inner sep=0.15cm,pos=0.6]{$c$}(4);
\draw[-{Stealth[length=2mm,width=1.7mm]},bend right=0](2)edge node[above,inner sep=0.15cm,pos=0.6]{$b$}(5);
\draw[-{Stealth[length=2mm,width=1.7mm]},bend right=0](3)edge node[above,inner sep=0.15cm,pos=0.6]{$d$}(6);
\draw[-{Stealth[length=2mm,width=1.7mm]},bend right=0](4)edge node[above,inner sep=0.15cm,pos=0.6]{$a$}(6);
\draw[-{Stealth[length=2mm,width=1.7mm]},bend right=0](4)edge node[above,inner sep=0.15cm,pos=0.6]{$b$}(7);
\draw[-{Stealth[length=2mm,width=1.7mm]},bend right=0](5)edge node[above,inner sep=0.15cm,pos=0.6]{$c$}(7);
\draw[-{Stealth[length=2mm,width=1.7mm]}](6)to[out=180,in=170,looseness=3]node[left]{$e$}(0);
\draw[-{Stealth[length=2mm,width=1.7mm]}](7)to[out=0,in=10,looseness=3]node[right]{$\mathit{f}$}(0);
\end{tikzpicture}\hspace*{-0.2cm}
\raisebox{-0.0cm}{\begin{tikzpicture}[scale=0.8]
\node[]()at(-1.8,4)[]{$\PNref{basic-exa.pn}':$};
\node[circle,draw,minimum size=0.5cm](p0)at(0,3)[]{};
 \filldraw[black](0,3)circle(2pt);
\node[circle,draw,minimum size=0.5cm](p1)at(3,3)[]{};
 \filldraw[black](3,3)circle(2pt);
\node[circle,draw,minimum size=0.5cm](p2)at(0,1)[]{};
\node[circle,draw,minimum size=0.5cm](p3)at(1.5,1)[]{};
 \filldraw[black](1.5,1)circle(2pt);
\node[circle,draw,minimum size=0.5cm](p5)at(-1.0,1)[]{};
\node[circle,draw,minimum size=0.5cm](p6)at(4,1)[]{};
\node[circle,draw,minimum size=0.5cm](p4)at(3,1)[label=above:]{};
\node[draw,minimum size=0.4cm](a)at(0,0){$a$};
\node[draw,minimum size=0.4cm](b)at(3,0){$b$};
\node[draw,minimum size=0.4cm](c)at(0,2){$c$};
\node[draw,minimum size=0.4cm](d)at(3,2){$d$};
\node[draw,minimum size=0.4cm](e)at(0.7,4.2){$e$};
\node[draw,minimum size=0.4cm](f)at(2.3,4.2){$\mathit{f}$};
\draw[-latex](p0)--(c);\draw[-latex](p1)--(d);
\draw[-latex](c)--(p2);\draw[-latex](d)--(p4);
\draw[-latex](p2)--(a);\draw[-latex](p3)--(a);
\draw[-latex](p3)--(b);\draw[-latex](p4)--(b);
\draw[-latex,bend left=25](a)edge(p5);\draw[-latex,bend right=25](b)edge(p6);
\draw[-latex](e)--(p3);\draw[-latex](f)--(p3);
\draw[-latex](e)--(p0);\draw[-latex](e)--(p1);
\draw[-latex](f)--(p0);\draw[-latex](f)--(p1);
\draw[-latex,bend left=45](p5)edge(e);
\draw[-latex,bend right=45](p6)edge(f);
\draw[-latex](p4) -- (e);
\draw[-latex](p2) -- (f);
\end{tikzpicture}}
\end{center}
\caption{A variant of Figure~\ref{basic-exa.fig} generating infinite sequences.
The additional transitions $e$ and $\mathit{f}$ lead back, respectively, from the
final (deadlock) states $M_6$ and $M_7$ to the initial state $M_0$ of $\TSref{basic-exa.ts}'$.
The augmented net $\PNref{basic-exa.pn}'$ is still plain, pure, and safe,
and it is a solution of $\TSref{basic-exa.ts}'$.}
\label{variant.fig}
\end{figure}
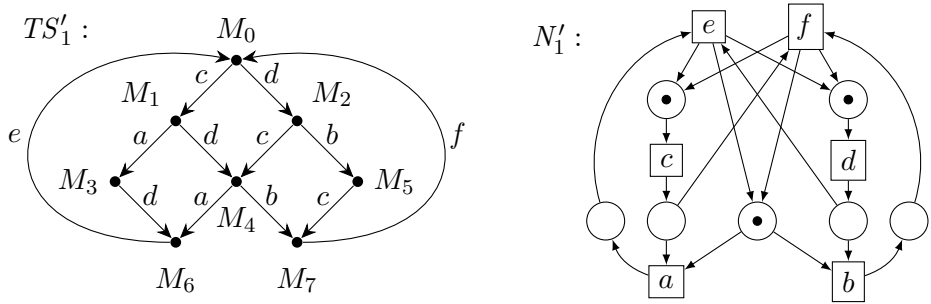

\EXA{spe-fpe-2.exa}{Figures \ref{acbc.fig} and \ref{variant.fig}}
Figure \ref{acbc.fig} illustrates a case where FPE is trivially satisfied:
there is a single infinite firing sequence $(acbc)^\infty$ which is both fair and persistent; 
hence FPE and Ochma\'nski's conjecture are satisfied.

As a more involved example,
we may consider the system shown in Figure~\ref{variant.fig}, which is a variant of
the one shown in Figure~\ref{basic-exa.fig}, adding loops from the two deadlock states back to the initial state.
Its infinite firing sequences are of the form\footnote{Where
$|$ means a choice between several patterns,
and $(...)^\infty$ denotes an infinite repetition of finite patterns.}
$(cade|cdae|dcae|dbc\mathit{f}|dcb\mathit{f}|cdb\mathit{f})^\infty$.
They are fair unless they contain infinitely many $cdae$ or $dcae$ 
but only finitely many $dbc\mathit{f}$, $dcb\mathit{f}$ and $cdb\mathit{f}$,
since then transition $b$ is enabled infinitely often but finitely often executed,
or if they contain infinitely many $dcb\mathit{f}$ or $cdb\mathit{f}$ but only finitely many $cade$, $cdae$ and $dcae$,
since then transition $a$ is enabled infinitely often but finitely often executed.
They are non-persistent if they contain at least one $cdae$, $dcae$, $dcb\mathit{f}$ or $cdb\mathit{f}$, 
since no path traversing $M_4$ is persistent (hence the net itself is non-persistent).
However, all of these infinite firing sequences (even the non-fair ones in this case, 
contrary to what happened in Example~\ref{unfair.exa})
are equivalent to a persistent sequence in $(cade|dbc\mathit{f})^\infty$;
indeed, any $cdae$ may be transformed into $cade$ with one permutation;
any $dcae$ may be transformed into $cade$ with two permutations;
any $dcb\mathit{f}$ may be transformed into $dbc\mathit{f}$ with one permutation; and
any $cdb\mathit{f}$ may be transformed into $dbc\mathit{f}$ with two permutations.
Hence, $\PNref{basic-exa.pn}'$ (which inherits the pps property from $\PNref{basic-exa.pn}$) satisfies FPE,
and thus Ochma\'nski's conjecture (while being outside the classes for which we shall prove the latter in the next section).
\ENDEXA{spe-fpe-2.exa}

\REM{hered.rem}{Non-inheritance of the SPE property} 
If a net $N$ is fixed, a marking $M$ of it will be said to satisfy SPE if so does $(N,M$). 
It may be observed that if $M$ satisfies SPE, it may happen that this is not true for some of its successor markings.
For example, in Figure~\ref{basic-exa.fig}, $M_0$ satisfies SPE for $\PNref{basic-exa.pn}$; in particular,
the sequence $M_0\fire{c}M_1\fire{db}$ has a persistent equivalent,
namely $M_0\fire{dbc}$; but SPE is not valid for $M_1$ since 
 the sequence $M_1\fire{db}$ does not have a persistent equivalent.
The same phenomenon occurs in Figure \ref{unfair.fig} where, after
$M_0\fire{x}M_1$, the sequence $M_1\fire{yb}$ is enabled
but does not have a persistent equivalent.
This  lack of SPE inheritance complicates the analysis of the relationship
between SPE and FPE in the general case. 
\ENDREM{hered.rem}

\Needspace{7\baselineskip} 
\section{Two Petri net classes satisfying SPE $\impl$ FPE}
\label{main.sct}

We shall now identify two Petri net classes for which the implication
\[\text{SPE$\impl$FPE}\;\;\;\;\text{or rather, the even stronger implication}\;\;\;\;\text{S$\widetilde{\text{P}}$E$\impl$FPE}
\]
is valid. The nontrivial parts of these proofs concern the implications
\[\text{SPE$\impl$persistence}\;\;\;\;\text{or, more strongly,}\;\;\;\;\text{S$\widetilde{\text{P}}$E$\impl$persistence}
\]
because persistence$\impl$FPE is trivially true by identity permutations, as has been pointed out above.
The two classes are: (not necessarily bounded)
equal-conflict nets which generalise free-choice nets (Section \ref{spe-fpe-ec.sct}), 
and pure (not necessarily bounded) dissymmetric choice nets
(Section \ref{spe-fpe-dc.sct}).

\subsection{Equal conflict Petri nets satisfying S$\widetilde{\text{P}}$E are persistent}
\label{spe-fpe-ec.sct}

The result we prove in this section generalises a result from \cite{DBLP:conf/apn/BestD25}
in two ways: from safe to bounded or unbounded, and from FC to EC (Definition \ref{pn-classes.def}).
We shall prove that the implication S$\widetilde{\text{P}}$E $\impl$ persistent
holds within the full class of equal conflict, and hence also in the class of free-choice,
Petri nets, whether they are bounded or not.
A characteristic property of free-choice nets is that if two transitions $t$ and $t'$ are \emph{conflicting}, i.e., if they
share a pre-place
($\pre{t}\cap\pre{t'}\neq\es$), then
any marking $M$ enables $t$ if and only if it enables $t'$.
EC nets $N=(P,T,F)$ are defined by the property (see Definition~\ref{pn-classes.def})
\begin{equation}\label{ec.eq}
\forall t,t'\in T\colon\pre{t}\cap\pre{t'}\neq\es\impl F(.,t)=F(.,t')
\end{equation}
This ensures that
the pre-place interfaces
of two conflicting transitions $t,t'$ are the same,
and thus, again, at any marking $M$, $M\xfire{t}$ iff $M\xfire{t'}$.
In the plain case, EC reduces to FC.

\SATZ{ec.th}{S$\widetilde{\text{P}}$E $\impl$ persistent for EC nets}
Let $(N,M_0)$ be an equal conflict Petri net and
suppose that $(N,M_0)$ satisfies S$\widetilde{\text{P}}$E. \\
Then $(N,M_0)$ is persistent.
\ENDSATZ

\BEW
By contradiction.
Assume that $(N,M_0)$ is EC and satisfies S$\widetilde{\text{P}}$E.
Further, assume that $(N,M_0)$ is non-persistent.
We shall derive a contradiction, showing that the last assumption is wrong.

\begin{figure}[htb]
\begin{center}
\begin{tikzpicture}[scale=1]
\node[circle,fill=black!100,inner sep=0.05cm](M0)at(0,2)[label=left:$M_0$]{};
\node[circle,fill=black!100,inner sep=0.05cm](Md)at(2,0)[label=below:$M'$]{};
\node[circle,fill=black!100,inner sep=0.05cm](Kd)at(3.6,0)[label=below:$K'$]{};
  \node (Kda) at ($(Kd)+(29:12mm)$) {};
  \draw[-{Stealth[length=2mm,width=1.7mm]}](Kd)edge node[above left,inner sep=0.3mm,pos=0.5]{$a_2$} coordinate[pos=0.6] (mKd)(Kda);
\node[circle,fill=black!100,inner sep=0.05cm](Ld)at(3,0.5)[label=right:]{};
\node[above=0.1cm of Ld](Lda){$L'$};
\node[circle,fill=black!100,inner sep=0.05cm](Mdd)at(6,0)[label=below:$M''$]{};
\node[circle,fill=black!100,inner sep=0.05cm](Kdd)at(7,0.5)[label=below:]{};
\node[above=0.1cm of Kdd](Kdda){$K''$};
\node[circle,fill=black!100,inner sep=0.05cm](Ldd)at(7.6,0)[label=below:$L''$]{};
  \node (Ldda) at ($(Ldd)+(29:12mm)$) {};
  \draw[-{Stealth[length=2mm,width=1.7mm]}](Ldd)edge node[above left,inner sep=0.3mm,pos=0.5]{$a_1$} coordinate[pos=0.6] (mLdd)(Ldda);
\node[circle,fill=black!100,inner sep=0.05cm](M)at(10,2)[label=above:$M$]{};
\node[circle,fill=black!100,inner sep=0.05cm](K)at(11,1)[label=left:$K$]{};
  \node[above right=of K](Ka){};\draw[-{Stealth[length=2mm,width=1.7mm]},bend left=15](K)edge node[below right]{$a_2$} coordinate[pos=0.7] (mK)(Ka);
  \draw[red,very thick,shift={(mK)}](0,-0.15)--(0,+0.15);\draw[red,very thick,shift={(mK)}](-0.15,+0)--(+0.15,0);
  \node[below right=of K](Kb){};\draw[-{Stealth[length=2mm,width=1.7mm]}, bend right=15](K)edge node[above right]{$a_1$} coordinate[pos=0.7] (mKb)(Kb);
  \draw[red,very thick,shift={(mKb)}](0,-0.15)--(0,+0.15);\draw[red,very thick,shift={(mKb)}](-0.15,+0)--(+0.15,0);
\node[circle,fill=black!100,inner sep=0.05cm](L)at(11,3)[label=above right:$L$]{};
\node[circle,fill=black!100,inner sep=0.05cm](Mhat)at(10,0)[label=below:$\widehat{M}$]{};
\draw[-{Stealth[length=2mm,width=1.7mm]},bend left=10](M0)edge node[above,inner sep=1mm,pos=0.5]{$\delta$}(M);
\draw[-{Stealth[length=2mm,width=1.7mm]},bend left=15](M)edge node[above right,inner sep=0.4mm,pos=0.55]{$a_1$}(K);
\draw[-{Stealth[length=2mm,width=1.7mm]},bend right=15](M)edge node[below right,inner sep=0.4mm,pos=0.55]{$a_2$}(L);
\draw[-{Stealth[length=2mm,width=1.7mm]},bend right=20](M0)edge node[below,inner sep=1mm,pos=0.5]{$\beta_1$}(Md);
\draw[-{Stealth[length=2mm,width=1.7mm]},bend left=0](Md)edge node[below,inner sep=1mm,pos=0.5]{$a_1$}(Kd);
\draw[-{Stealth[length=2mm,width=1.7mm]},bend left=0](Md)edge node[above left,inner sep=0.5mm,pos=0.5]{$a_2$}(Ld);
\draw[-{Stealth[length=2mm,width=1.7mm]},bend right=10](Kd)edge node[below,inner sep=1mm,pos=0.5]{$\beta_2$}(Mdd);
\draw[-{Stealth[length=2mm,width=1.7mm]},bend right=0](Mdd)edge node[above left,inner sep=1mm,pos=0.5]{$a_1$}(Kdd);
\draw[-{Stealth[length=2mm,width=1.7mm]},bend left=0](Mdd)edge node[below,inner sep=0.5mm,pos=0.5]{$a_2$}(Ldd);
\draw[-{Stealth[length=2mm,width=1.7mm]},bend right=10](Ldd)edge node[below,inner sep=1mm,pos=0.5]{$\beta_3$}(Mhat);
\draw[-{Stealth[length=2mm,width=1.7mm]},bend right=20](Mhat)edge node[below right,inner sep=1mm,pos=0.5]{$b$}(K);
\draw[decorate,decoration={brace,amplitude=10pt,mirror,raise=1cm},xshift=0cm,yshift=0cm]
(7.6,0)--(11,0)node [black,midway,xshift=0.0cm,yshift=-1.6cm]{$\beta''$};
\draw[decorate,decoration={brace,amplitude=10pt,mirror,raise=1cm},xshift=0cm,yshift=0cm]
(3.6,-1)--(11,-1)node [black,midway,xshift=0.0cm,yshift=-1.6cm]{$\beta'$};
\draw[decorate,decoration={brace,amplitude=10pt,mirror,raise=1cm},xshift=0cm,yshift=0cm]
(0,-2)--(11,-2)node [black,midway,xshift=0.0cm,yshift=-1.6cm]{$\beta$};
\end{tikzpicture}
\end{center}
\caption{The markings $M$, $K$, $L$ stem from the non-persistence of $(N,M_0)$.
A persistent sequence leading from $M_0$ through $\beta_1$,
$M'$, $a_1$, $K'$, $\beta_2$, $M''$, $a_2$, $L''$, $\beta_3$, $\widehat{M}$, all the way to $K$,
which is Parikh-equivalent to the sequence $M_0\xfire{\delta\,a_1}K$,
can be constructed by means of the properties S$\widetilde{\text{P}}$E and EC.
However, this leads to a contradiction, as shown in the proof.} 
\label{main-fc.fig}
\end{figure}
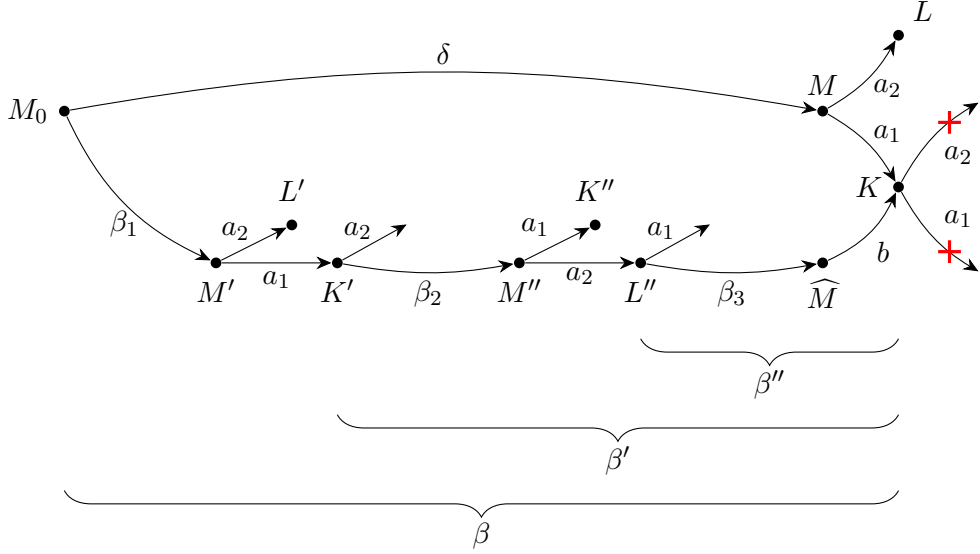

By non-persistence, there is a reachable marking $M$ violating the persistence property.
That is, we have
$M_0\xfire{\delta}M$ and $M\xfire{a_1}K$ and $M\xfire{a_2}L$ and $\neg(K\xfire{a_2})$
for some sequence $\delta\in T^*$ and some transitions
$a_1,a_2$ with $a_1\neq a_2$.
(See the upper part of Figure \ref{main-fc.fig}.)

By $M\xfire{a_1}K$, $M\xfire{a_2}$, and $\neg(K\xfire{a_2})$, $\pre{a_1}\cap\pre{a_2}\neq\emptyset$.
From (\ref{ec.eq}), $F(.,a_1)=F(.,a_2)$, which implies that $a_1$ is enabled (at any marking)
if and only if $a_2$ is enabled (at this marking).
In particular, since $\pre{a_1}\cap\pre{a_2}\neq\emptyset$ and $a_2$ is disabled at $K$,
we can deduce that $\neg(K\xfire{a_1})$ (right-hand side of Figure \ref{main-fc.fig}).

By S$\widetilde{\text{P}}$E, we find a persistent firing sequence $M_0\xfire{\beta}K$
which is Parikh-equivalent to the non-persistent firing sequence $M_0\xfire{\delta\,a_1}K$.
We shall show that $\beta$ can be split
in the way shown in the lower part of Figure \ref{main-fc.fig}.

By $\Parikh(\delta a_1)=\Parikh(\beta)$, $a_1$ occurs in $\beta$.
Consider the \emph{last} occurrence of $a_1$ in $\beta$ and
let the marking just before that occurrence be called $M'$.
Then $M_0\xfire{\beta}K$ is split as $M_0\xfire{\beta_1}M'\xfire{a_1}K'\xfire{\beta'}K$
where $\beta'$ is the (persistent, by Corollary \ref{persist.cor}) tail of $\beta$ starting at $K'$.
By the EC property, $M'\xfire{a_2}L'$, for some marking $L'$,
and by the persistence of $\beta'$, in combination with $a_1\neq a_2$,
also $K'\xfire{a_2}$.

But $\neg(K\xfire{a_2})$, so that
$a_2$ must occur somewhere in $\beta'$
(otherwise it cannot be turned from enabled at $K'$ to disabled at $K$,
since $\beta'$ is persistent).
In particular, $\beta'\neq\leer$ and $a_1$ cannot be the last transition $b$ in $\beta$
(since otherwise, by determinism, $\widehat{M}=M$ and $\beta'$ would not be persistent).
Say that $M''$ is the marking just before such an occurrence of $a_2$
(it does not matter whether this occurrence is the last or any other
occurrence of $a_2$ in $\beta'$).
So we now have $M''\xfire{a_2}L''$ in $\beta'$,
and by the equal conflict  property, $M''\xfire{a_1}K''$ for some marking $K''$.
We cannot have the degenerate case that $a_2$
is the last transition $b$ in $\beta$, because then we would have $M''=\widehat{M}$ and $L''=K$,
and $\widehat{M}\xfire{b}K$ would not be persistent because it would
turn $a_1$ from enabled to disabled (contradicting the choice of~$\beta$).
Instead, $L''\neq K$ and $L''\xfire{a_1}$.
Thus, $K'\xfire{\beta'}K$ is split as
$K'\xfire{\beta_2}M''\xfire{a_2}L''\xfire{\beta_3}\widehat{M}\xfire{b}K$,
with a persistent tail $\beta''=\beta_3\,b$ of $\beta'$ which starts at $L''$.

So, $L''$ enables $a_1$ but,
as we have seen above, $K$ does not enable $a_1$.
This implies, using the same argument as before, in combination
with the persistence of $\beta''$,
that $a_1$ must occur in $\beta''$,
contradicting the choice of the other $a_1$ (firable at $M'$)
as the last such occurrence in $\beta$.

This contradiction shows that the assumption that $(N,M_0)$ is non-persistent was wrong
and that $(N,M_0)$ is instead persistent, as was claimed.
\ENDBEW{ec.th}

\REM{ec.rem}{On the proof of Theorem \ref{ec.th}}
The proof uses the EC property for $a_1$ and $a_2$ with $\pre{a_1}\cap\pre{a_2}\neq\es$
in both directions ($a_1$ enabled $\impl$ $a_2$~enabled as well as $a_2$ enabled $\impl$ $a_1$ enabled)
in order to construct an infinite sequence of alternations of $a_1$'s and $a_2$'s in $\beta$,
provided that one of them occurs at all.
But one of them must occur by the Parikh equivalence
of $M_0\xfire{\beta}$ and $M_0\xfire{\delta\,a_1}$; hence the contradiction.
If the premise is changed from EC to DC, the proof breaks down,
because DC only allows us to reason in one of the two directions.
\ENDREM{ec.rem}

\Needspace{5\baselineskip}
\KOR{ec.cor}{EC or FC nets satisfying SPE or S$\widetilde{\text{P}}$E are persistent and satisfy FPE}
Let $N$ be an EC or FC Petri net satisfying SPE or S$\widetilde{\text{P}}$E.
Then $N$ is persistent and satisfies FPE.
\ENXKOR

\BEW
This follows directly from the previous theorem, from the fact that EC is stronger than FC,
from the fact that SPE is stronger that S$\widetilde{\text{P}}$E (Proposition~\ref{perm-parikh.kor}),
and from Proposition \ref{pers.prop} (implying that in a persistent net,
every firing sequence has a trivial persistent equivalent, be it finite or infinite, fair or unfair).
\ENDBEW{ec.cor}

\Needspace{7\baselineskip} 
\subsection{Pure DC Petri nets satisfying S$\widetilde{\text{P}}$E are persistent}
\label{spe-fpe-dc.sct}

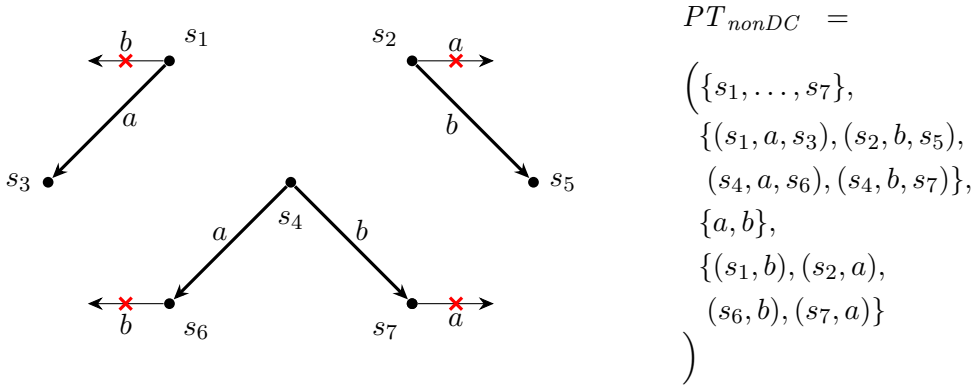
\begin{figure}[htb]
\begin{center}
\begin{tikzpicture}[scale=0.8]
\node[circle,fill=black!100,inner sep=0.05cm](1)at(2,4)[label=above right:$s_1$]{}; 
  \node[left=of 1](1a){};\draw [-{Stealth[length=2mm,width=1.7mm]}](1)edge node[above]{$b$} coordinate (m1)(1a);
  \draw[red,very thick,shift={(m1)}](-0.1,-0.1)--(0.1,+0.1); \draw[red,very thick,shift={(m1)}](-0.1,+0.1)--(+0.1,-0.1);
\node[circle,fill=black!100,inner sep=0.05cm](2)at(6,4)[label=above left:$s_2$]{}; 
  \node[right=of 2](2a){};\draw [-{Stealth[length=2mm,width=1.7mm]}](2)edge node[above]{$a$} coordinate (m2)(2a);
  \draw[red,very thick,shift={(m2)}](-0.1,-0.1)--(0.1,+0.1); \draw[red,very thick,shift={(m2)}](-0.1,+0.1)--(+0.1,-0.1);
\node[circle,fill=black!100,inner sep=0.05cm](3)at(0,2)[label=left:$s_3$]{}; 
\node[circle,fill=black!100,inner sep=0.05cm](4)at(4,2)[label=below:]{};
 \node[below of=4,node distance=0.5cm]{$s_4$}; 
\node[circle,fill=black!100,inner sep=0.05cm](5)at(8,2)[label=right:$s_5$]{}; 
\node[circle,fill=black!100,inner sep=0.05cm](6)at(2,0)[label=below:]{};
 \node[below right of=6,node distance=0.5cm]{$s_6$}; 
\node[circle,fill=black!100,inner sep=0.05cm](7)at(6,0)[label=below:]{};
 \node[left=of 6](6a){};\draw [-{Stealth[length=2mm,width=1.7mm]}](6)edge node[below]{$b$} coordinate (m6)(6a);
  \draw[red,very thick,shift={(m6)}](-0.1,-0.1)--(0.1,+0.1); \draw[red,very thick,shift={(m6)}](-0.1,+0.1)--(+0.1,-0.1);
\node[circle,fill=black!100,inner sep=0.05cm](7)at(6,0)[label=below:]{};
 \node[below left of=7,node distance=0.5cm]{$s_7$}; 
  \node[right=of 7](7a){};\draw [-{Stealth[length=2mm,width=1.7mm]}](7)edge node[below]{$a$} coordinate (m7)(7a);
  \draw[red,very thick,shift={(m7)}](-0.1,-0.1)--(0.1,+0.1); \draw[red,very thick,shift={(m7)}](-0.1,+0.1)--(+0.1,-0.1);
\draw[-{Stealth[length=2mm,width=1.7mm]},very thick,bend right=0](1)edge node[below,inner sep=0.15cm,pos=0.3]{$a$}(3);
\draw[-{Stealth[length=2mm,width=1.7mm]},very thick,bend right=0](2)edge node[below,inner sep=0.15cm,pos=0.3]{$b$}(5);
\draw[-{Stealth[length=2mm,width=1.7mm]},very thick,bend right=0](4)edge node[above,inner sep=0.15cm,pos=0.6]{$a$}(6);
\draw[-{Stealth[length=2mm,width=1.7mm]},very thick,bend right=0](4)edge node[above,inner sep=0.15cm,pos=0.6]{$b$}(7);
\end{tikzpicture}\hspace*{1cm}
\raisebox{2cm}{$\begin{array}{l}
\PT_{\mathit{nonDC}}\;\;= \\[0.3cm]
\Bigl(\{s_1,\ldots,s_7\}, \\
\;\;\{(s_1,a,s_3),(s_2,b,s_5), \\
\;\;\;(s_4,a,s_6),(s_4,b,s_7)\}, \\
\;\;\{a,b\}, \\
\;\;\{(s_1,b),(s_2,a), \\
\;\;\;(s_6,b),(s_7,a)\} \\
\Bigr)
\end{array}
$}
\end{center}
\caption{The pattern $\PT_{\mathit{nonDC}}$ indicating that the DC property is violated.}
\label{non-DC-lts.fig}
\end{figure}

Now we shall prove that if a pure, dissymmmetric choice, Petri net
satisfies S$\widetilde{\text{P}}$E, then it is necessarily persistent (and satisfies SPE).
Remark \ref{ec.rem} indicates that a separate argument is needed for that result.
We first argue that no \lts{} that contains the pattern $\PT_{\mathit{nonDC}}$
shown in Figure \ref{non-DC-lts.fig} can be realised by a DC net.

\Needspace{5\baselineskip}
\LEM{nondc.lem}{$\PT_{\mathit{nonDC}}$ can recognise non-DC nets}
Suppose that $N=(P,T,F,M_0)$ is a Petri net and $\TS=(S,\to,T,s_0)$
is (isomorphic to) its reachability graph.
Suppose that $\PT_{\mathit{nonDC}}$ is embedded in $\TS$.
Then $N$ is not dissymmmetric choice.
\ENDLEM

\BEW
Let $f$ be the embedding function from $\PT_{\mathit{nonDC}}$ to $\TS$.
Then:
\begin{itemize}
\item
The marking (corresponding to) $f(s_4)$ enables both $f(a)$ and $f(b)$.
The two non-enablings $(s_6,b)$ and $(s_7,a)$ imply that
$f(a)$ and $f(b)$ have at least one common
preplace, that is, that $\pre{f(a)}\cap\pre{f(b)}\neq\es$ in $N$.
\item
The marking corresponding to $f(s_1)$ enables $f(a)$ but, by $(s_1,b)\in D$
and the definition of embedding, not $f(b)$.
This entails $\neg(\pre{f(b)}\subseteq\pre{f(a)})$ in $N$.
\item
Similarly, the marking corresponding to $f(s_2)$ enables $f(b)$ but, by $(s_2,a)\in D$, not $f(a)$,
which implies $\neg(\pre{f(a)}\subseteq\pre{f(b)})$ in $N$.
\end{itemize}
Taken together, this means that $N$ is not DC.
\ENDBEW{nondc.lem}

The proof of this lemma is somewhat analogous to the proof of
Lemma \ref{nonpers.lem} in Section~\ref{pt.sct},
except that here, there has been no need to use the injectivity of $f$ on transitions
(the injectivity is implied by the pattern itself, because if $f(a)=f(b)$, from $f(s_1)$, 
we have both $f(a)$ allowed and not allowed).

Now we turn our attention to the claim that if a pure, dissymmmetric choice, Petri net
satisfies S$\widetilde{\text{P}}$E, then it is necessarily persistent.
This claim is proved in the following logically equivalent way:
Let a given net be pure and assume it satisfies S$\widetilde{\text{P}}$E and is non-persistent.
Then it cannot be dissymmmetric choice.
The proof is done in three steps as follows:
\begin{itemize}[beginpenalty=10000]
\item[1.]
{\bf There are no 3/4-diamonds:}
By pureness, any choice state in the reachability graph of a plain net
either starts a half-diamond (proper choice) or a full diamond (proper concurrency).
There are no 3/4-diamonds.
\item[2.]
{\bf Parikh-equivalent sequences can be unified:}
Any two persistent sequences from $M_0$ having the same
Parikh vectors (thus coming together at some $M$ by determinism)
can be permuted such that they go through some state just
two labels before $M$ (unless they are very short, and provided all
shorter sequences are also persistent).
\item[3.]
{\bf A non-DC pattern can be found:}
Non-persistence of $N$ means that some non-persistent state $M$ can be reached,
which we can assume to be reachable by some shortest possible sequence~$\gamma$
(so that all sequences of lengths $<|\gamma|$ are persistent).
 By non-persistence and Step 1., there is a half-diamond at $M$, which is a symmetric situation.
S$\widetilde{\text{P}}$E implies that the two ends of the half-diamond can be reached
by persistent sequences $\alpha,\beta$ from $M_0$ which are (almost) Parikh-equivalent
to $\gamma$ (except for the two legs of the half-diamond).
Using Step 2. twice, we can reconstruct the pattern shown in Figure \ref{non-DC-lts.fig}
and apply Lemma \ref{nondc.lem}.
\end{itemize}

\Needspace{5\baselineskip}
\subsection*{There are no 3/4-diamonds}
\label{pure-diamond.sct}

\LEM{pure-diamond.lem}{3/4-diamonds can be completed to 4/4-diamonds}
Let $N=(P,T,F)$ be a pure, plain net.
Let $M$ be a marking of $N$ and let $x,y\in T$. \\
Suppose that $M\xfire{y}M''\xfire{x}\widehat{M}$ is a firing sequence and that $M\xfire{x}M'$. \\
Then $M\xfire{x}M'\xfire{y}\widehat{M}$ is also a firing sequence.
\ENDLEM

\begin{figure}[htb]
\begin{center}
\begin{tikzpicture}[scale=1]
\refstepcounter{exampleTScounter}\label{dia-lem.ts}
\node[]()at(2.5,3.6)[]{$\TS_\theexampleTScounter$:};
\node[circle,fill=black!100,inner sep=0.05cm](M)at(2.5,2)[label=left:$M$]{};
\node[circle,fill=black!100,inner sep=0.05cm](M1)at(4,1)[label=below:$M'$]{};
\node[circle,fill=black!100,inner sep=0.05cm](M2)at(4,3)[label=above:$M''$]{};
\node[circle,fill=black!100,inner sep=0.05cm](Mh)at(5.5,2)[label=right:$\widehat{M}$]{};
\draw[-triangle 45,bend right=0](M)edge node[below left,inner sep=1mm,pos=0.5]{$x$}(M1);
\draw[-triangle 45,bend right=0](M)edge node[above,inner sep=1.3mm,pos=0.4]{$y$}(M2);
\draw[-triangle 45,bend right=0](M2)edge node[above right,inner sep=1mm,pos=0.5]{$x$}(Mh);
\end{tikzpicture}\hspace*{2cm}
\raisebox{0.7cm}{\begin{tikzpicture}[scale=0.8]
\refstepcounter{examplePNcounter}\label{dia-lem.pn}
\node[]()at(-0.8,3.5)[]{$\PN_\theexamplePNcounter$:};
\node[circle,draw,minimum size=0.5cm](p0)at(1,3)[]{};
 \filldraw[black](1,3)circle(2pt);
\node[circle,draw,minimum size=0.5cm](p1)at(3,3)[]{};
 \filldraw[black](3,3)circle(2pt);
\node[draw,minimum size=0.4cm](x)at(0,1){$x$}; 
\node[draw,minimum size=0.4cm](y)at(2,1){$y$};
\draw[-latex](p0)--(x);
\draw[-latex,bend right=20](p0)edge node[above,inner sep=0.15cm,pos=0.5]{$$}(y);
\draw[-latex,bend right=20](y)edge node[above,inner sep=0.15cm,pos=0.5]{$$}(p0);
\draw[-latex](p1)--(y);
\end{tikzpicture}
}
\end{center}
\caption{A 3/4-diamond $\TSref{dia-lem.ts}$. The net $\PNref{dia-lem.pn}$ proves that
the 3/4-diamond has a Petri net solution
which is plain, safe, AC, and DC, but impure.
If $M=M''$ and $M'=\widehat{M}$, then there is even an impure,  plain and safe FC solution,
while we still have a (degenerate) 3/4 diamond.}
\label{dia-lem.fig}
\end{figure}
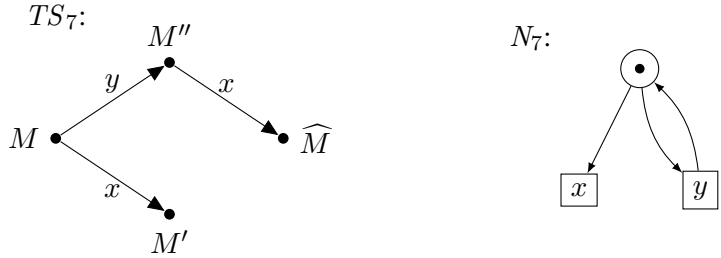

\BEW
We show that $M\xfire{xy}$ is a firing sequence:

By $M\xfire{yx}$, pureness and plainness, the places $p\in\pre{x}\cap\pre{y}$ satisfy $M(p)\geq2$,
and by $M\xfire{x}$,
the places $q\in(\pre{x}\minus\pre{y})\cup(\pre{y}\minus\pre{x})$ satisfy $M(q)\geq1$.
Hence, after firing $x$ from $M$ by $M\xfire{x}M'$, the places $p\in\pre{y}$ satisfy $M'(p)\geq1$,
and therefore, $y$ is firable from $M'$.
Of course, $\widehat{M}$ is reached after $M\xfire{xy}$ because $\Parikh(xy)=\Parikh(yx)$.
\ENDBEW{pure-diamond.lem}

Note that if $N$ is not safe, it may be the case that $x=y$,
but the proof is unaffected.
Informally, Lemma \ref{pure-diamond.lem} means that in a pure and plain net, there may
only be 1/2 (half; meaning proper choice) or full (1/1; indicating proper concurrency)
diamonds at choice points of the reachability graph, and no 3/4-diamonds.
The pureness requirement is essential for this property, as exhibited by the net $\PNref{dia-lem.pn}$
on the right-hand side of Figure \ref{dia-lem.fig}.

\Needspace{7\baselineskip}
\subsection*{Unifying Parikh-equivalent firing sequences}
\label{unifying.sct}

\PROP{fs-unify.prop}{Parikh-equivalent sequences can be permuted into each other}
Suppose that $N=(P,T,F,M_0)$ is a pure, plain Petri net with initial marking $M_0$  
and that $\widetilde{M}$ is a marking that can be reached
by two equally long sequences $\alpha=a_1\ldots a_n$ and $\beta=b_1\ldots b_n$
as follows:
\[M_0\xfire{a_1\ldots a_{n-1}}\widetilde{M_1}\xfire{a_n}\widetilde{M}\;\;\text{ and }\;\;
M_0\xfire{b_1\ldots b_{n-1}}\widetilde{M_2}\xfire{b_n}\widetilde{M}
\]
such that $a_n\neq b_n$, i.e., the last two transitions are different,
and $\Parikh(\alpha)=\Parikh(\beta)$.
Further, assume that all firing sequences $M_0\xfire{\tau}$ of length
$|\tau|\leq n-1$, in particular $M_0\xfire{a_1\ldots a_{n-1}}$ and $M_0\xfire{b_1\ldots b_{n-1}}$,
are persistent.
Then there is a marking $J$ which can be reached by a persistent sequence
$M_0\xfire{\sigma}J$ satisfying $J\xfire{a_n}\widetilde{M_2}$, $J\xfire{b_n}\widetilde{M_1}$,
and $\Parikh(\sigma a_nb_n)\;(=\Parikh(\sigma b_na_n))=\Parikh(\alpha)\;(=\Parikh(\beta))$.
\ENDPROP

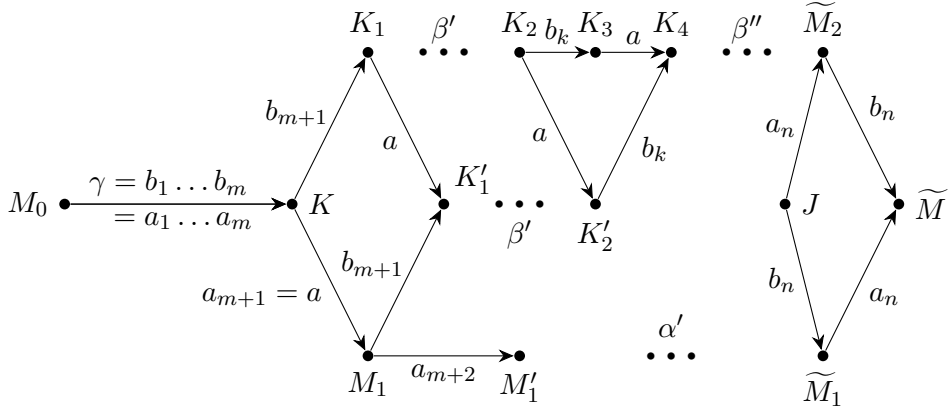
\begin{figure}[htb] 
\begin{center}
\begin{tikzpicture}[scale=1]
\node[circle,fill=black!100,inner sep=0.05cm](M0)at(-3,6)[label=left:$M_0$]{};
\node[circle,fill=black!100,inner sep=0.05cm](M)at(0,6)[label=right:$K$]{};
\node[circle,fill=black!100,inner sep=0.05cm](M2)at(1,8)[label=above:$K_1$]{};
\node[circle,fill=black!100,inner sep=0.05cm](K1d)at(2,6)[label={[label distance=-0.08cm]85:$K_1'$}]{};
\node[circle,fill=black!100,inner sep=0.05cm](M1d)at(3,4)[label=below:$M_1'$]{};
\node[circle,fill=black!100,inner sep=0.05cm](M1)at(1,4)[label=below:$M_1$]{};
\node[circle,fill=black!100,inner sep=0.05cm](K2)at(3,8)[label=above:$K_2$]{};
\node[circle,fill=black!100,inner sep=0.05cm](K2d)at(4,6)[label=below:$K_2'$]{};
\node[circle,fill=black!100,inner sep=0.05cm](K3)at(4,8)[label=above:$K_3$]{};
\node[circle,fill=black!100,inner sep=0.05cm](K4)at(5,8)[label=above:$K_4$]{};
\node[circle,fill=black!100,inner sep=0.05cm](Mt1)at(7,4)[label=below:$\widetilde{M}_1$]{};
\node[circle,fill=black!100,inner sep=0.05cm](Mt2)at(7,8)[label=above:$\widetilde{M}_2$]{};
\node[circle,fill=black!100,inner sep=0.05cm](Mt)at(8,6)[label=right:$\widetilde{M}$]{};
\node[circle,fill=black!100,inner sep=0.05cm](J)at(6.5,6)[label=right:$J$]{};
\draw[-{Stealth[length=2mm,width=1.7mm]},bend right=0](M0)edge node[above,inner sep=1mm,pos=0.45]{$\gamma=b_1\ldots b_m$}(M);
\draw[-{Stealth[length=2mm,width=1.7mm]},bend right=0](M0)edge node[below,inner sep=1mm,pos=0.45]{$\;\;\;\,=a_1\ldots a_m$}(M);
\draw[-{Stealth[length=2mm,width=1.7mm]},bend right=0](M)edge node[below left,inner sep=1mm,pos=0.5]{$a_{m+1}=a$}(M1);
\draw[-{Stealth[length=2mm,width=1.7mm]},bend right=0](M1)edge node[below,inner sep=1mm,pos=0.5]{$a_{m+2}$}(M1d);
\node[](lowerdots)at(5,4)[]{{\tiny $\dt\;\dt\;\dt$}};
\node[](loweralpha1)at(5,4.4)[]{$\alpha'$};
\draw[-{Stealth[length=2mm,width=1.7mm]},bend right=0](Mt1)edge node[below right,inner sep=1mm,pos=0.5]{$a_n$}(Mt);
\draw[-{Stealth[length=2mm,width=1.7mm]},bend right=0](M)edge node[above left,inner sep=0.2mm,pos=0.5]{$b_{m+1}$}(M2);
\draw[-{Stealth[length=2mm,width=1.7mm]},bend right=0](M1)edge node[above left,inner sep=0.2mm,pos=0.5]{$b_{m+1}$}(K1d);
\node[](upperdots1)at(2,8)[]{{\tiny $\dt\;\dt\;\dt$}};
\node[](middledots1)at(3,6)[]{{\tiny $\dt\;\dt\;\dt$}};
\node[](upperbeta1)at(2,8.3)[]{$\beta'$};
\node[](middlebeta1)at(3,5.6)[]{$\beta'$};
\draw[-{Stealth[length=2mm,width=1.7mm]},bend right=0](K2)edge node[above,inner sep=1mm,pos=0.5]{$b_k$}(K3);
\draw[-{Stealth[length=2mm,width=1.7mm]},bend right=0](K3)edge node[above,inner sep=1mm,pos=0.5]{$a$}(K4);
\node[](upperdots2)at(6,8)[]{{\tiny $\dt\;\dt\;\dt$}};
\node[](upperbeta2)at(6,8.3)[]{$\beta''$};
\draw[-{Stealth[length=2mm,width=1.7mm]},bend right=0](Mt2)edge node[above right,inner sep=1mm,pos=0.5]{$b_n$}(Mt);
\draw[-{Stealth[length=2mm,width=1.7mm]},bend right=0](M2)edge node[below left,inner sep=1mm,pos=0.5]{\black{$a$}}(K1d);
\draw[-{Stealth[length=2mm,width=1.7mm]},bend right=0](K2)edge node[below left,inner sep=0.8mm,pos=0.45]{\black{$a$}}(K2d);
\draw[-{Stealth[length=2mm,width=1.7mm]},bend right=0](K2d)edge node[below right,inner sep=1mm,pos=0.5]{\black{$b_k$}}(K4);
\draw[-{Stealth[length=2mm,width=1.7mm]},bend right=0](J)edge node[left,inner sep=1mm,pos=0.5]{$a_n$}(Mt2);
\draw[-{Stealth[length=2mm,width=1.7mm]},bend right=0](J)edge node[left,inner sep=1mm,pos=0.5]{$b_n$}(Mt1);
\end{tikzpicture}
\end{center}
\caption{Sketch of $\gamma$ (left), $\alpha=a_1\ldots a_n$ (bottom), and $\beta=b_1\ldots b_n$ (top)
with $\Parikh(\alpha)=\Parikh(\beta)$.
Node $J$ results from the construction in the proposition.
Eventually, $J$ can be reached from $M_0$ by a (persistent) sequence $\sigma$
which satisfies $\Parikh(\sigma a_nb_n)=\Parikh(\alpha)=\Parikh(\beta)$.} 
\label{dia-prop.fig}
\end{figure}

\emph{Remark}: 
The proposition states that there is a full 1/1-diamond with starting
point $J$, legs $a_n,b_n$, and end point~$\widetilde{M}$
(see the right-hand side of Figure \ref{dia-prop.fig}).

\BEW
Suppose that all premises are satisfied for $N$, $\widetilde{M}$,
$\alpha=a_1\ldots a_n$, and $\beta=b_1\ldots b_n$.
We shall prove the existence of $J$.

Let $\gamma=a_1\ldots a_m=b_1\ldots b_m$ be the longest
common prefix of $\alpha$ and~$\beta$
(see the left-hand side of Figure \ref{dia-prop.fig}).
The length of $\gamma$ can be zero (if $a_1\neq b_1$), and it can
be at most $n-2$, since lengths $n-1$ and $n$ are prohibited by
$\Parikh(\alpha)=\Parikh(\beta)$ in combination with $a_n\neq b_n$.
Thus, $0\leq m\leq n-2$, and $a_{m+1}\neq b_{m+1}$ since $\gamma$
is defined as the longest common prefix.
The proof proceeds by downward induction on $n-2-m\geq0$
(i.e., by upward induction on $m$).

{\bf Base case:} $n-2-m=0$, that is, $m=n-2$.

Then the marking just before $\widetilde{M_1}$ in $\alpha$ must,
by definition of $\gamma$, be the same as the marking just before $\widetilde{M_2}$ in $\beta$.
Call it $J$. By $\Parikh(\alpha)=\Parikh(\beta)$, it follows that
$a_{n-1}=b_n$ and $b_{n-1}=a_n$, yielding the desired diamond.

{\bf Inductive step:} 
Let us assume that the claim has been proved when two Parikh-equivalent firing sequences 
of length $n$ have a common prefix of length smaller 
than $m$;
we now prove it for $\gamma$ of length $m$ (with $0\leq m<n-2$).

Figure \ref{dia-prop.fig} depicts the general setup.
From $M_0$ up to $K$, $\alpha$ and $\beta$ agree with each other,
and $\gamma$ is their common prefix.
Because of $a_{m+1}\neq b_{m+1}$, they deviate from point $K$ onwards.
The sequence $a_{m+1}\ldots a_{n-1}$ leads from $K$ to $\widetilde{M_1}$
via $a_{m+1}$, $M_1$, $a_{m+2}$, $M_1'$, and $\alpha'$ in the lower part of the figure.
The sequence $b_{m+1}\ldots b_{n-1}$ leads from $K$ to $\widetilde{M_2}$
via $b_{m+1}$, $K_1$, $\beta'$, $K_2$, $b_k$, $K_3$, $b_{k+1}$, $K_4$, and $\beta''$ in the upper part of the figure.
Moreover, both sequences are persistent because they are tails
of the persistent sequences $a_1\ldots a_{n-1}$ and $b_1\ldots b_{n-1}$, respectively.
(Compare Corollary~\ref{persist.cor}.)

Abbreviate $a=a_{m+1}$.
Because $a\neq b_{m+1}$ and $\Parikh(a_{m+1}\ldots a_n)=\Parikh(b_{m+1}\ldots b_n)$,
$a$ also occurs amongst the $b_{m+2},\ldots,b_n$;
we shall assume that the first such occurrence is
just after $b_k$ ($m+2\leq k\leq n-1$), i.e., $a=b_{k+1}$.
Because $a\neq b_{m+1}$ and because $b_{m+1}\ldots b_{n-1}$ is persistent,
$K_1$ enables $a$.
In fact, by the same reason
and because $b_{m+1}\ldots b_k$ do not contain~$a$, all markings up to
and including $K_2$ (and $K_3$, of course) enable~$a$.
At $K_2$, there is thus a 3/4-diamond formed by $K_2,K_3,K_2',K_4$
and legs $a,b_k$, which can be ``closed'' by a $b_k$-arrow from $K_2'$ to $K_4$
according to Lemma \ref{pure-diamond.lem}.

Now replace $\beta$ by the sequence $\widetilde{\beta}$
which leads from $M_0$ to $\widetilde{M}$
via $\gamma$, $a_{m+1}$, $M_1$, $b_{m+1}$, $K_1'$,
$\beta'$, $K_2'$, $b_k$, $K_4$, $\beta''$, $\widetilde{M_2}$, and $b_n$.
The pair $\alpha, \widetilde{\beta}$ has the same salient properties
as the pair $\alpha,\beta$ before, but it has a longer
common initial prefix.

Hence the induction hypothesis can be applied. Eventually, the base case
takes effect, proving the existence of a marking $J$, as was claimed.
\ENDBEW{fs-unify.prop}

\REM{p1.rem}{On Proposition \ref{fs-unify.prop} and its proof}
Various special cases may arise. For example, the new
initial prefix may be much longer than the previous one (not just
by one transition). Essentially, we have permuted $a$ backwards in front of $b_{m+1}\ldots b_k$.
Also, the markings $\widetilde{M_1}$ and $\widetilde{M_2}$
may coincide if $a_n$ and $b_n$ are duplicated transitions, 
i.e., transitions having the same preplace- and postplace-interfaces. 
It may even be the case that the whole diamond collapses to a single point.

Note that Proposition~\ref{fs-unify.prop} provides a special case in which
there is a kind of reverse of Proposition~\ref{perm-parikh.kor},
in the sense that we construct a permutation from two
Parikh-equivalent sequences. Such a construction does
not work in general because the sequences do not need to be persistent in general.
The proof bears some similarity to the proof of Keller's theorem
\cite{DBLP:journals/acta/BestD09,DBLP:conf/sagamore/Keller74}.
\ENDREM{p1.rem}

\Needspace{7\baselineskip}
\subsection*{Constructing a non-DC pattern from a non-persistent state}
\label{nonac-pattern.sct}

\PROP{nonac.prop}{Using a non-persistent state to derive Figure \ref{non-DC-lts.fig}}
Suppose that $N=(P,T,F,M_0)$ is a pure, plain, non-persistent Petri net 
satisfying the S$\widetilde{\text{P}}$E property.
Then in its reachability graph, the pattern $\PT_{\mathit{nonDC}}$
shown in Figure~{\rm \ref{non-DC-lts.fig}} is embedded.
\ENDPROP

\BEW
Suppose that $N$ satisfies all premises.
We shall prove that $\PT_{\mathit{nonDC}}$ (Figure \ref{non-DC-lts.fig}) can be embedded into $RG(N)$.

\begin{figure}[htb]
\begin{center}
\begin{tikzpicture}[scale=1]
\node[circle,fill=black!100,inner sep=0.05cm](M0)at(0,2)[label=left:$M_0$]{};
\node[circle,fill=black!100,inner sep=0.05cm](J2)at(7,3)[label=below left:$J_2$]{};
  \node[above=0.9cm of J2](J2a){};
  \draw[-{Stealth[length=2mm,width=1.7mm]}](J2)edge node[left,inner sep=1.5mm,pos=0.3]{$a$} coordinate (j2)(J2a);
  \draw[red,thick,shift={(j2)}](-0.1,-0.1)--(0.1,+0.1);
  \draw[red,thick,shift={(j2)}](-0.1,+0.1)--(+0.1,-0.1);
\node[circle,fill=black!100,inner sep=0.05cm](J1)at(7,1)[label=above left:$J_1$]{};
  \node[below=0.9cm of J1](J1a){};
  \draw[-{Stealth[length=2mm,width=1.7mm]}](J1)edge node[left,inner sep=1.5mm,pos=0.3]{$b$} coordinate (j1)(J1a);
  \draw[red,thick,shift={(j1)}](-0.1,-0.1)--(0.1,+0.1);
  \draw[red,thick,shift={(j1)}](-0.1,+0.1)--(+0.1,-0.1);
\node[circle,fill=black!100,inner sep=0.05cm](K2)at(8,4)[label=right:$K_2$]{};
  \node[above=0.9cm of K2](K2a){};
  \draw[-{Stealth[length=2mm,width=1.7mm]}](K2)edge node[left,inner sep=1.5mm,pos=0.3]{$a$} coordinate (k2)(K2a);
  \draw[red,thick,shift={(k2)}](-0.1,-0.1)--(0.1,+0.1);
  \draw[red,thick,shift={(k2)}](-0.1,+0.1)--(+0.1,-0.1);
\node[circle,fill=black!100,inner sep=0.05cm](M)at(8,2)[label=right:$M$]{};
\node[circle,fill=black!100,inner sep=0.05cm](K1)at(8,0)[label=right:$K_1$]{};
  \node[below=0.9cm of K1](K1a){};
  \draw[-{Stealth[length=2mm,width=1.7mm]}](K1)edge node[right,inner sep=1.5mm,pos=0.3]{$b$} coordinate (k1)(K1a);
  \draw[red,thick,shift={(k1)}](-0.1,-0.1)--(0.1,+0.1);
  \draw[red,thick,shift={(k1)}](-0.1,+0.1)--(+0.1,-0.1);
\node[circle,fill=black!100,inner sep=0.05cm](M2)at(9,3)[label=above right:$M_2$]{};
  \node[below right=0.9cm of M2](M2a){};
  \draw[-{Stealth[length=2mm,width=1.7mm]}](M2)edge node[below left,pos=0.4]{$a$} coordinate (m2)(M2a);
  \draw[red,thick,shift={(m2)}](-0.1,0)--(0.1,0);
  \draw[red,thick,shift={(m2)}](0,+0.1)--(0,-0.1);
\node[circle,fill=black!100,inner sep=0.05cm](M1)at(9,1)[label=below right:$M_1$]{};
  \node[above right=0.9cm of M1](M1a){};
  \draw[-{Stealth[length=2mm,width=1.7mm]}](M1)edge node[above left,pos=0.4]{$b$} coordinate (m1)(M1a);
  \draw[red,thick,shift={(m1)}](-0.1,0)--(0.1,0);
  \draw[red,thick,shift={(m1)}](0,+0.1)--(0,-0.1);
\draw[-{Stealth[length=2mm,width=1.7mm]},bend left=25](M0)edge node[above,inner sep=1mm,pos=0.45]{$\beta'$}(K2);
\draw[-{Stealth[length=2mm,width=1.7mm]},bend left=15](M0)edge node[above,inner sep=1mm,pos=0.45]{$\gamma_2$}(J2);
\draw[-{Stealth[length=2mm,width=1.7mm]},bend left=0](M0)edge node[above,inner sep=1mm,pos=0.45]{$\delta$}(M);
\draw[-{Stealth[length=2mm,width=1.7mm]},bend right=15](M0)edge node[below,inner sep=1mm,pos=0.45]{$\gamma_1$}(J1);
\draw[-{Stealth[length=2mm,width=1.7mm]},bend right=25](M0)edge node[below,inner sep=1mm,pos=0.45]{$\alpha'$}(K1);
\draw[-{Stealth[length=2mm,width=1.7mm]},bend left=0](J2)edge node[below right,inner sep=0.5mm,pos=0.40]{$b$}(K2);
\draw[-{Stealth[length=2mm,width=1.7mm]},bend left=0](M)edge node[above left,inner sep=0.5mm,pos=0.40]{$b$}(M2);
\draw[-{Stealth[length=2mm,width=1.7mm]},bend left=0](J2)edge node[above right,inner sep=0.5mm,pos=0.40]{$y$}(M);
\draw[-{Stealth[length=2mm,width=1.7mm]},bend left=0](K2)edge node[below left,inner sep=0.5mm,pos=0.40]{$y$}(M2);
\draw[-{Stealth[length=2mm,width=1.7mm]},bend left=0](J1)edge node[below right,inner sep=0.5mm,pos=0.40]{$x$}(M);
\draw[-{Stealth[length=2mm,width=1.7mm]},bend left=0](K1)edge node[above left,inner sep=0.5mm,pos=0.40]{$x$}(M1);
\draw[-{Stealth[length=2mm,width=1.7mm]},bend left=0](M)edge node[below left,inner sep=0.5mm,pos=0.40]{$a$}(M1);
\draw[-{Stealth[length=2mm,width=1.7mm]},bend left=0](J1)edge node[above right,inner sep=0.5mm,pos=0.40]{$a$}(K1);
\end{tikzpicture}
\end{center}
\caption{There is a proper choice (1/2-diamond with legs $a,b$, $a\neq b$,
and $\neg(M_2\xfire{a})$, $\neg(M_1\xfire{b})$) at state~$M$.
By S$\widetilde{\text{P}}$E, we find a persistent sequence $M_0\xfire{\alpha'x}M_1$
and another persistent sequence $M_0\xfire{\beta'y}M_2$.
Besides $a\neq b$, we also have $a\neq x$ and $b\neq y$.
The non-enablings at $K_1,K_2,J_1,J_2$ are described in the proof.} 
\label{nonac.fig}
\end{figure}
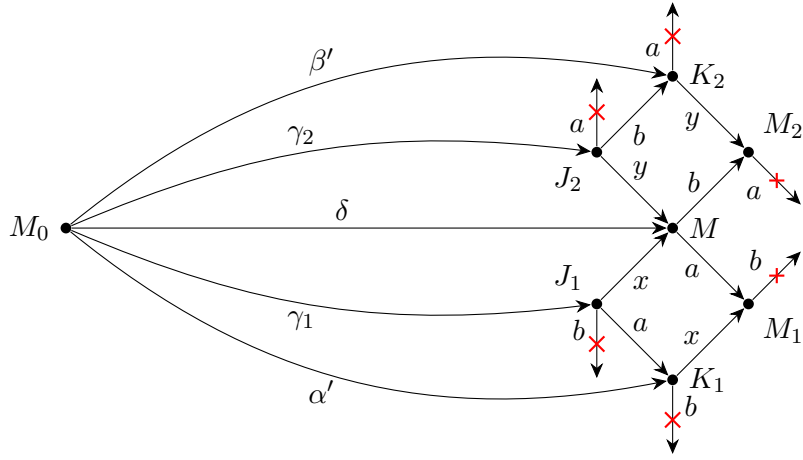

Since $N$ is non-persistent, there is some non-persistent reachable state $M$.
Pick one of them that is nearest to $M_0$, that is, we have $M_0\xfire{\delta}M$
(central horizontal line in Figure \ref{nonac.fig}),
and whenever $M_0\xfire{\tau}M'$ for another non-persistent state $M'$, then $|\tau|\geq|\delta|$.
In particular, no non-persistent state is reachable by a sequence that is shorter than $\delta$,
and $\delta$ is itself persistent.
Also, let $M_1$ and $M_2$ with $M\xfire{a}M_1$ and $M\xfire{b}M_2$
be the corners of the 1/2-diamond starting at $M$ (right-hand side of Figure \ref{nonac.fig}).
Of course, $a\neq b$, since otherwise, there is no proper choice at $M$,
and $M_1\neq M\neq M_2$ since $M$ enables $a$ and $b$
while $M_1$ does not enable $b$ and $M_2$ does not enable $a$.

By the S$\widetilde{\text{P}}$E property, there is some persistent sequence,
Parikh-equivalent to $M_0\xfire{\delta a}M_1$, by which $M_1$ can be reached from $M_0$.
This sequence cannot lead through $M$; 
indeed, if that sequence has the form $M_0\xfire{\delta'}M\xfire{x\delta''}M_1$, 
since $\delta'$ may not be shorter than $\delta$ and $\Parikh(\delta a)=\Parikh(\delta' x\delta'')$,
we must have $\delta''=\leer$; if $x\neq b$ then $b$ is switched
from enabled to non-enabled when going from $M$ to $M_1$,
and if $x=b$ then $M_1=M_2$ and $a$ is switched from enabled to non-enabled when going from $M$ to $M_1$;
in either case, that sequence is not persistent.

Hence there is a marking $K_1\neq M$ such that this persistent sequence has the form
$M_0\xfire{\alpha'}K_1\xfire{x}M_1$, for some transition $x$,
as shown in the lower part of Figure \ref{nonac.fig}.
By $K_1\neq M$ and determinism,
also $a\neq x$.
Similarly, there is a reachable marking $K_2$ and a persistent sequence
$M_0\xfire{\beta'}K_2\xfire{y}M_2$ for some transition $y$, as shown in the upper part of the figure.
By $K_2\neq M$, $b\neq y$.

We can now invoke Proposition \ref{fs-unify.prop} two times.
For the lower part of the figure, set $\alpha=\alpha'x$ and $\beta=\delta a$
(or the other way round).
This results in state $J_1$, reachable by $\gamma_1$,
and the backward diamond completions $J_1\xfire{a}K_1$ and $J_1\xfire{x}M$.
For the upper part of the figure, set $\alpha=\delta b$ and $\beta=\beta'y$
(or the other way round).
This results in state $J_2$ and the backward diamond completions $J_2\xfire{b}K_2$ and $J_2\xfire{y}M$.

We shall now check the failed enablings induced by the
failed enablings of $b$ at $M_1$ and $a$ at $M_2$.
$K_1$ does not enable $b$ because the sequence $\alpha'x$ is persistent
since it stems from the S$\widetilde{\text{P}}$E property.
This implies that $J_1$ also does not enable $b$ since otherwise,
we would have a non-persistent state ($J_1$) that is properly nearer to $M_0$ than $M$,
contradicting the choice of $M$.
Similarly, in the upper part of the figure, we have $\neg(K_2\xfire{a})$ and $\neg(J_2\xfire{a})$.

The pattern $\PT_{\mathit{nonDC}}$
is embedded by a function $f=f_1\cup f_2$ (referring to Definition \ref{embed.def}) as follows:
\begin{equation}\label{embed.eq}
\begin{array}{l}
f_1(s_1)=J_1,\;f_1(s_2)=J_2,\;f_1(s_3)=K_1,\;f_1(s_4)=M, \\
f_1(s_5)=K_2,\;f_1(s_6)=M_1,\;f_1(s_7)=M_2 \\[0.2cm]
\text{and }\;\;f_2\;=\;\text{identity on } \{a,b\}
\end{array}
\end{equation}
This completes the proof.
\ENDBEW{nonac.prop}

\SATZ{main.th}{\scalefont{0.9}{Pure, plain, non-persistent nets
satisfying S$\widetilde{\text{P}}$E are not DC}}
Let $N=(P,T,F,M_0)$ be a pure, plain and non-persistent Petri net satisfying
the S$\widetilde{\text{P}}$E property. Then $N$ is not a dissymmmetric choice net.
\ENDSATZ

\BEW
Referring to Figure \ref{nonac.fig},
the pattern $\PT_{\mathit{nonDC}}$, defined in  Figure~\ref{non-DC-lts.fig}, 
can be embedded in the reachability graph of $N$ as defined above, in (\ref{embed.eq}).
We can now simply invoke Lemma \ref{nondc.lem}.
This completes the proof.
\ENDBEW{main.th}

Actually, the embedding (\ref{embed.eq}) is almost injective:  
We cannot have $M=M_1$ because the former enables $b$ while the latter does not.
We cannot have $J_1=K_1$ because this would imply that $M=M_1$.
We cannot have $J_1=M$ because the former does not enable $b$
while the latter does.
We cannot have $K_1=M_1$ because that would imply $J_1=M$.
Also, we cannot have $M=K_1$ because the former enables $b$ while the latter does not.
We cannot have $J_1=M_1$ by pureness and plainness because in the pure and plain case,
it is not possible that, from $M$, $a$ occurs twice in a row but (still from $M$)
after only one $b$, $a$ is disabled 
(in $M$, the places in $\pre{a}$ must have at least two tokens, and $b$ may only take one of them).  
(The other part of the figure, referring to the diamond $J_2,K_2,M,M_2$, is symmetric.)
However, we might have $K_1=K_2$.

It is presently unknown whether or not the statement of the main theorem  
withstands dropping pureness from the set of premises.
Lemma \ref{pure-diamond.lem} goes out of the window in that case,
and it is doubtful whether it can be compensated easily.
As in Corollary \ref{ec.cor}, we get:

\Needspace{5\baselineskip}
\KOR{main2.cor}{Pure DC nets satisfying SPE or S$\widetilde{\text{P}}$E are persistent and satisfy FPE}
Any pure DC Petri net satisfying SPE or S$\widetilde{\text{P}}$E is persistent and satisfies FPE.
\ENDKOR{main2.cor}

Figure~\ref{unfair.fig} highlights the importance of the DC property in this result. 
Indeed, the net shown there is pps and satisfies SPE, but it is neither dissymmetric choice 
(as exhibited by transitions $a$ and $b$)
nor persistent (since we have the firing sequences $M_0\xfire{xya}$ and $M_0\xfire{xyb}$, but not $M_0\xfire{xyab}$);
note that SPE is satisfied since, for instance, the sequence $M_0\xfire{xya}$ is non-persistent ($b$ is enabled after $xy$
but not after $xya$), but it is equivalent to $M_0\xfire{xay}$
(where $b$ is no longer enabled before the end). 

\section{A plain and pure counterexample to SPE $\impl$ FPE}
\label{counter.sct}

\begin{figure}[htb]
\begin{center}
\begin{tikzpicture}[scale=0.8]
\refstepcounter{exampleTScounter}\label{spe-non-fpe.ts}
\node[]()at(-1,6.2)[]{$\TS_\theexampleTScounter$:};
\node[circle,fill=black!100,inner sep=0.05cm](0)at(0.1,4.5)[label={[label distance=0.02cm]93.5:$M_0$}]{};
\node[circle,fill=black!100,inner sep=0.08cm](1)at(3,4.5)[label=above:$K$]{};
\node[circle,fill=black!100,inner sep=0.08cm](2)at(5.9,4.5)[label=above:$K'$]{};
\node[circle,fill=black!100,inner sep=0.05cm](3)at(8.4,4.5)[label={[label distance=0.00cm]87.5:$L$}]{};
\node[circle,fill=black!100,inner sep=0.05cm](4)at(4.5,6)[label=above:]{};
\node[circle,fill=black!100,inner sep=0.08cm](5)at(0.1,3)[label={[label distance=0.06cm]265:}]{};
\node[circle,fill=black!100,inner sep=0.08cm](6)at(3,3)[label=below left:$M$]{};
\node[circle,fill=black!100,inner sep=0.08cm](7)at(5.9,3)[label=below:$M'$]{};
\node[circle,fill=black!100,inner sep=0.05cm](8)at(8.4,3)[label={[label distance=0.00cm]-87.5:$L'$}]{};
\node[circle,fill=black!100,inner sep=0.05cm](9)at(4.5,0)[label=above:]{};
\node[circle,fill=black!100,inner sep=0.08cm](10)at(1.5,1.5)[label=above:]{};
\node[circle,fill=black!100,inner sep=0.08cm](11)at(4.3,1.5)[label=below:$M''$]{};
\draw[-{Stealth[length=2mm,width=1.7mm]}](0)to[]node[auto,swap]{$x$}(1);
\draw[-{Stealth[length=2mm,width=1.7mm]},very thick](1)to[]node[auto,swap]{$a_1$}(2);
\draw[-{Stealth[length=2mm,width=1.7mm]}](2)to[]node[auto,swap]{$a_2$}(3);
\draw[-{Stealth[length=2mm,width=1.7mm]}](3)to[bend right=20]node[auto,swap]{$b$}(4);
\draw[-{Stealth[length=2mm,width=1.7mm]}](4)to[bend right=20]node[auto,swap]{$c$}(0);
\draw[-{Stealth[length=2mm,width=1.7mm]}](0)to[]node[auto,swap]{$y$}(5);
\draw[-{Stealth[length=2mm,width=1.7mm]},very thick](1)to[]node[auto,swap]{$y$}(6);
\draw[-{Stealth[length=2mm,width=1.7mm]},very thick](2)to[]node[auto,swap]{$y$}(7);
\draw[-{Stealth[length=2mm,width=1.7mm]}](3)to[]node[auto,swap]{$y$}(8);
\draw[-{Stealth[length=2mm,width=1.7mm]}](4)to[bend left=20]node[auto,pos=0.7]{$y$}(9);
\draw[-{Stealth[length=2mm,width=1.7mm]},very thick](5)to[]node[auto,swap]{$x$}(6);
\draw[-{Stealth[length=2mm,width=1.7mm]},very thick](6)to[]node[auto,swap]{$a_1$}(7);
\draw[-{Stealth[length=2mm,width=1.7mm]}](7)to[]node[auto,swap]{$a_2$}(8);
\draw[-{Stealth[length=2mm,width=1.7mm]}](8)to[bend left=40]node[auto]{$b$}(9);
\draw[-{Stealth[length=2mm,width=1.7mm]}](9)to[bend left=40]node[auto]{$c$}(5);
\draw[-{Stealth[length=2mm,width=1.7mm]},very thick](5)to[bend left=0]node[auto,inner sep=0.3]{$b$}(10);
\draw[-{Stealth[length=2mm,width=1.7mm]},very thick](6)to[bend left=0]node[auto,inner sep=0.3]{$b$}(11);
\draw[-{Stealth[length=2mm,width=1.7mm]},very thick](10)to[]node[auto,swap]{$x$}(11);
\end{tikzpicture}\hspace*{1cm}
\raisebox{1cm}{\begin{tikzpicture}[scale=0.8]]
\refstepcounter{examplePNcounter}\label{spe-non-fpe.pn}
\node[]()at(-0.8,5)[]{$\PN_\theexamplePNcounter$:};
\node[draw,very thick,minimum size=0.4cm](a1)at(0,4){$a_1$};
\node[draw,minimum size=0.4cm](a2)at(0,0){$a_2$};
\node[draw,very thick,minimum size=0.4cm](b)at(2,2){$b$};
\node[draw,minimum size=0.4cm](c)at(4,0){$c$};
\node[draw,minimum size=0.4cm](x)at(4,4){$x$};
\node[draw,minimum size=0.4cm](y)at(2,3){$y$};
\node[circle,very thick,draw,minimum size=0.5cm](p0)at(2,4)[]{};
\node[circle,draw,minimum size=0.5cm](p1)at(0,2)[]{};
\node[circle,draw,minimum size=0.5cm](p2)at(2,0)[]{};
\node[circle,draw,minimum size=0.5cm](p3)at(3,1)[]{};
\node[circle,very thick,draw,minimum size=0.5cm](p4)at(1,3)[label=above:$p$]{};
\node[circle,draw,minimum size=0.5cm](p5)at(4,2)[]{$\dt$};
\node[circle,draw,minimum size=0.5cm](p6)at(3,3)[]{$\dt$};
\node[circle,very thick,draw,minimum size=0.5cm](p7)at(1,1)[label=below:$q$]{$\dt$};
\draw[-latex,very thick](p0)--(a1);\draw[-latex,very thick](p7)--(a1);\draw[-latex](a1)--(p1);
\draw[-latex](p1)--(a2);\draw[-latex](a2)--(p2);\draw[-latex](a2)--(p4);\draw[-latex](a2)--(p7);
\draw[-latex,very thick](p4)--(b);\draw[-latex,very thick](p7)--(b);\draw[-latex](b)--(p3);
\draw[-latex](p2)--(c);\draw[-latex](p3)--(c);\draw[-latex](c)--(p5);\draw[-latex](c)--(p7);
\draw[-latex](x)--(p0);\draw[-latex](p5)--(x);
\draw[-latex](p6)--(y);\draw[-latex](y)--(p4);
\end{tikzpicture}}
\end{center}
\caption{An \lts{}
$\TSref{spe-non-fpe.ts}$ satisfying SPE but not FPE and a generating Petri net $\PNref{spe-non-fpe.pn}$.}
\label{spe-non-fpe.fig}
\end{figure}
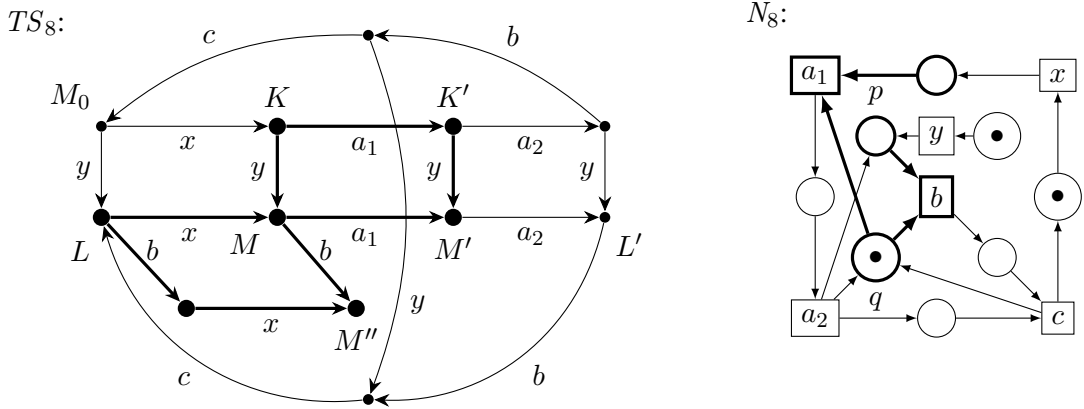

The example shown in Figure \ref{spe-non-fpe.fig},
taken from \cite{DBLP:conf/apn/BestD25,DBLP:journals/topnoc/BarylskaBSS17},
proves that $\neg(\text{SPE$\impl$FPE})$ is possible in the realm of pure, plain, and 2-bounded (rather than safe) nets.
In the labelled transition system $\TSref{spe-non-fpe.ts}$,
any finite firing sequence can be permuted equivalently
in such a way that the last $M'$ in it (if any)
is reached via $y$ rather than via $a_1$,
and $M''$, which can occur at most once, is reached via $x$ rather than via $b$.
Hence, on the one hand, the choice marking $M$ can be avoided and SPE is satisfied.
On the other hand, the infinite firing sequence
\begin{equation}\label{bad-seq.eq}
M_0\;\;\xfire{\;y\;(\;x\;a_1\;a_2\;b\;c\;)^\infty}
\end{equation}
is fair but has no persistent
equivalent, since, for taking account of the fact that the label $y$ must occur at some point or other,
the marking $M$ cannot be avoided, and the firing $M\xfire{a_1}M'$ is non-persistent
(as $M$ enables $b$ but $M'$ does not),
and so is the firing $M\xfire{b}M''$.
Hence FPE is not satisfied.
The intuitive problem is that no permutation of (\ref{bad-seq.eq})
can shift the label $y$ ``to the very end'' without losing it altogether.

The pure and plain
Petri net $\PNref{spe-non-fpe.pn}$ which solves $\TSref{spe-non-fpe.ts}$
is neither dissymmetric choice nor safe (but it is asymmetric choice). 
Indeed, place $q$ and transitions $a_1,b$
violate the DC condition (depicted in bold); and at $L'$, place $p$ carries two tokens.
Note that $q,a_1,b$ is the only non-DC context; that
$p$ is the only unsafe place (with bound~$2$); and that $L'$ is the only non-safe reachable marking.

In fact, (i): no plain, pure, safe net whatsoever can solve $\TSref{spe-non-fpe.ts}$.
Also, (ii): there cannot be any (either safe or unsafe) dissymmetric choice solution.
\begin{itemize}
\item[(i)]
To see that there is no plain, pure, safe net solution, observe that at $L$, we have both $L\xfire{y}$ and $L\xfire{b}$,
but at $K$ (and also at $M_0$), only $yb$, but not $by$, is enabled.
Such a situation cannot occur in a pps net \cite{DBLP:journals/topnoc/BarylskaBSS17}. 
\item[(ii)]
To see that there is no dissymmetric choice net solving $\TSref{spe-non-fpe.ts}$,
observe that the pattern $\PT_{\mathit{nonDC}}$ of Figure \ref{non-DC-lts.fig} 
is embedded in $\TSref{spe-non-fpe.ts}$,
shown in bold, and apply Lemma \ref{nondc.lem}.
\end{itemize}
$\TSref{spe-non-fpe.ts}$ may be compared with $\TSref{unfair.ts}$ shown in Figure \ref{unfair.fig}.
The critical infinite executions (not having persistent permutations) are
$M_0\xfire{y(xac)^\infty}$ in $\TSref{unfair.ts}$, and
$M_0\xfire{y(xa_1a_2bc)^\infty}$ in $\TSref{spe-non-fpe.ts}$.
The difference is that the former is unfair (which means that Ochma\'nski's conjecture is not applicable) 
while the latter is fair but needs unsafeness (which means that Ochma\'nski's conjecture extended to pps nets is not applicable).
In a sense, therefore, Ochma\'nski's conjecture can be
viewed as claiming that the pattern exhibited in $\TSref{unfair.ts}$
cannot be ``refined'' by forcing unfairly treated transitions ($b$, in this case)
into the loop without violating the pps property.
The results presented in this paper are not affected by this problem,
because they apply to different Petri net classes
(restricting structure while relaxing safeness).

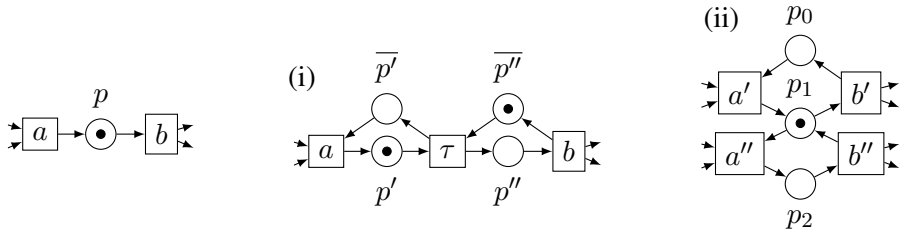
\begin{figure}[htp]
\begin{center}
\raisebox{1cm}{
\begin{tikzpicture}[scale=0.8]
\node[circle,draw,minimum size=0.4cm](p)at(1,1)[label=above:$p$]{};
\filldraw[black](1,1)circle(2pt);
\node[draw,minimum size=0.4cm](a)at(0,1)[label=above:]{$a$};
\node[](d1)at(-0.7,1.2){};\node[](d2)at(-0.7,0.7){};
\node[draw,minimum size=0.4cm](b)at(2,1)[label=above:]{$b$};
\node[](e1)at(2.7,1.2){};\node[](e2)at(2.7,0.7){};
\draw[-latex](a)edge[bend left=0](p);
\draw[-latex](d1)edge[bend left=0](a);\draw[-latex](d2)edge[bend left=0](a);
\draw[-latex](b)edge[bend left=0](e1);\draw[-latex](b)edge[bend left=0](e2);
\draw[-latex](p)edge[bend left=0](b);
\end{tikzpicture}}\hspace*{0.7cm}
\raisebox{0.3cm}{\begin{tikzpicture}[scale=0.8]
\node[]()at(-0.4,2.2)[]{(i)};
\node[circle,draw,minimum size=0.4cm](pd)at(1,1)[label=below:$p'$]{};
\filldraw[black](1,1)circle(2pt);
\node[circle,draw,minimum size=0.4cm](pdd)at(3,1)[label=below:$p''$]{};
\node[circle,draw,minimum size=0.4cm](pdc)at(1,1.7)[label=above:$\overline{p'}$]{};
\node[circle,draw,minimum size=0.4cm](pddc)at(3,1.7)[label=above:$\overline{p''}$]{};
\filldraw[black](3,1.7)circle(2pt);
\node[draw,minimum size=0.4cm](a)at(0,1)[label=above:]{$a$};
\node[](d1)at(-0.7,1.2){};\node[](d2)at(-0.7,0.7){};
\node[draw,minimum size=0.4cm](b)at(4,1)[label=above:]{$b$};
\node[](e1)at(4.7,1.2){};\node[](e2)at(4.7,0.7){};
\node[draw,minimum size=0.4cm](tau)at(2,1)[label=above:]{$\tau$};
\draw[-latex](a)edge[bend left=0](pd);
\draw[-latex](pd)edge[bend left=0](tau);
\draw[-latex](tau)edge[bend left=0](pdd);
\draw[-latex](pdd)edge[bend left=0](b);
\draw[-latex](b)edge[bend left=0](pddc);
\draw[-latex](pddc)edge[bend left=0](tau);
\draw[-latex](tau)edge[bend left=0](pdc);
\draw[-latex](pdc)edge[bend left=0](a);
\draw[-latex](d1)edge[bend left=0](a);\draw[-latex](d2)edge[bend left=0](a);
\draw[-latex](b)edge[bend left=0](e1);\draw[-latex](b)edge[bend left=0](e2);
\end{tikzpicture}}\hspace*{0.7cm}
\begin{tikzpicture}[scale=0.8]
\node[]()at(-0.3,1.7)[]{(ii)};
\node[circle,draw,minimum size=0.4cm](p0)at(1,1.2)[label=above:$p_0$]{};
\node[circle,draw,minimum size=0.4cm](p1)at(1,0)[label=above:$p_1$]{};
\filldraw[black](1,0)circle(2pt);
\node[circle,draw,minimum size=0.4cm](p2)at(1,-1)[label=below:$p_2$]{};
\node[draw,minimum size=0.4cm](a1)at(0,0.5)[label=above:]{$a'$};
\node[](d11)at(-0.8,0.7){};\node[](d12)at(-0.8,0.2){};
\node[draw,minimum size=0.4cm](b1)at(2,0.5)[label=above:]{$b'$};
\node[](e11)at(2.8,0.7){};\node[](e12)at(2.8,0.2){};
\node[draw,minimum size=0.4cm](a2)at(0,-0.5)[label=above:]{$a''$};
\node[](d21)at(-0.8,-0.3){};\node[](d22)at(-0.8,-0.8){};
\node[draw,minimum size=0.4cm](b2)at(2,-0.5)[label=above:]{$b''$};
\node[](e21)at(2.8,-0.3){};\node[](e22)at(2.8,-0.8){};
\draw[-latex](d11)edge[bend left=0](a1);\draw[-latex](d12)edge[bend left=0](a1);
\draw[-latex](b1)edge[bend left=0](e11);\draw[-latex](b1)edge[bend left=0](e12);
\draw[-latex](d21)edge[bend left=0](a2);\draw[-latex](d22)edge[bend left=0](a2);
\draw[-latex](b2)edge[bend left=0](e21);\draw[-latex](b2)edge[bend left=0](e22);
\draw[-latex](a1)edge[bend left=0](p1);
\draw[-latex](p1)edge[bend left=0](b1);
\draw[-latex](a2)edge[bend left=0](p2);
\draw[-latex](p2)edge[bend left=0](b2);
\draw[-latex](b1)edge[bend left=0](p0);
\draw[-latex](p0)edge[bend left=0](a1);
\draw[-latex](b2)edge[bend left=0](p1);
\draw[-latex](p1)edge[bend left=0](a2);
\end{tikzpicture}
\end{center}
\caption{Sketch of two constructions turning a 2-bounded net into a safe net:
(i) pipelining using complement places, and (ii) simulation.
In (i), the marking on $p$ corresponds to the sum of tokens on $p',p''$,
while $\tau$ is a silent (unobservable) transition.
In (ii), $p_i$ corresponds to $i$ tokens on $p$.
Both constructions achieve strong similarity, but they do not
preserve the concurrency between $y$ and $a_2$
starting at state $K'$ in Figure \ref{spe-non-fpe.fig}.}
\label{2to1.fig}
\end{figure}

When trying to disprove -- or prove -- Ochma\'nski's conjecture for
pps nets, two ideas may come to mind; we show that both of them are insufficient:
First, finding a counterexample by turning the bound~2 in Figure~\ref{spe-non-fpe.fig}
into bound 1 by known constructions
(for example, pipelining the tokens on $p$, or simulating $p$ by a set of safe places,
as in Figure \ref{2to1.fig}) fails.
Such constructions necessarily introduce causal
dependencies between (instances of) $y$ and (instances of) $a_2$,
by which the essence of the counterexample is destroyed.
Secondly, one might try to reduce the hoped-for proof to a known result
such as Theorem \ref{ec.th} or Theorem \ref{main.th}
by applying the well-known transformation of a Petri net into a
free-choice Petri net (replacing a place-transition arc by a
sequence of arcs place-transition-place-transition).
But this construction destroys the SPE property;
thus, this idea is unsuitable as well.

\section{Fairness, justice, progress, APE, and JPE}
\label{fair.sct}

In Section \ref{och.sct}, fairness was defined as in \cite{folco-2014}.
For infinite sequences, this corresponds to the standard notion of strong fairness as
known from the literature (see \cite{DBLP:journals/csur/GlabbeekH19}, amongst many others).
For finite sequences, strong fairness has been defined as maximality.
There is at least one alternative definition which will be discussed next.

\subsection{Fairness of finite sequences}

Finite firing sequences play a minor role with regard to fairness.
We might either just call \emph{every} finite sequence fair by default,
or we might pick a subset of finite sequences which are
called fair while others are not.
Indeed, the literature essentially suggests a choice between two alternative possibilities:
\begin{itemize}
\item
The \emph{finite-is-fair principle}
simply considers all finite sequences to be fair.
Intuitively, this principle asserts that unfairness occurs only when some
activity is neglected infinitely often.
This regime has been adopted, for example, in
\cite{DBLP:journals/dc/AptFK88,DBLP:books/sp/BestD24,DBLP:conf/icalp/LehmannPS81,
DBLP:conf/concur/Reisig96}.
\item
The \emph{maximal-finite-is-fair principle}
requires that a finite firing sequence must be maximal
in order to be fair.
Intuitively, this principle asserts that unfairness also occurs
if in some execution, some possible (unimpeded) activity is delayed forever,
for no external reason.
The maximal-finite-is-fair regime
is used (and justified, in various contexts) in
\cite{DBLP:journals/csur/GlabbeekH19,DBLP:journals/ipl/KindlerA99,
folco-2014,DBLP:journals/ipl/RomijnV96}.
\end{itemize}
Although the results in the previous part of this paper do not depend on this choice,
let us consider what happens if we adopt the former regime,
i.e., if the first sentence in Definition \ref{fair.def} is replaced by: 
\begin{quote} 
A finite firing sequence $M_0\xfire{\sigma}M$ is fair, by definition.
\end{quote}
Let the resulting permutation equivalence notions be called SPE$^f$ and FPE$^f$, respectively.
Of course, SPE$^f$ equals SPE, since fairness plays no role there.
But FPE$^f$ differs from FPE; it is actually stronger.
Still, Ochma\'nski's conjecture remains unaffected, since
under the premise SPE (or SPE$^f$), FPE is the same as FPE$^f$.
However, for its \emph{converse}, we have the following:

\LEM{och-conv.lem}{Converses of the implication SPE $\impl$ FPE}
\begin{itemize}
\item[{\rm (i)}]
$\;\;\;\;\mathrm{FPE}^f\;\;\impl\;\;\mathrm{SPE}^f\;\;\;\;$ (for any Petri net)
\item[{\rm (ii)}]
$\neg\;(\;\mathrm{FPE}\;\;\impl\;\;\mathrm{SPE}\;)$
\end{itemize}
\ENDLEM

\BEW
(i): By the finite-is-fair regime, all finite sequences are fair.
Hence FPE$^f$ reduces to SPE, thus to SPE$^f$, for finite sequences.

\begin{figure}[htb]
\begin{center}
\begin{tikzpicture}[scale=0.8]
\node[circle,draw,minimum size=0.5cm](p1)at(0,3)[]{};
 \filldraw[black](0,3)circle(2pt);
\node[circle,draw,minimum size=0.5cm](p2)at(4,3)[]{};
 \filldraw[black](4,3)circle(2pt);
\node[circle,draw,minimum size=0.5cm](p3)at(8,3)[]{};
 \filldraw[black](8,3)circle(2pt);
\node[draw,minimum size=0.4cm](t1)at(0,2){$x$};
\node[draw,minimum size=0.4cm](t2)at(4,2){$y$};
\node[draw,minimum size=0.4cm](t3)at(8,2){$z$};
\node[circle,draw,minimum size=0.5cm](p5)at(2,1)[]{};
 \filldraw[black](2,1)circle(2pt);
\node[circle,draw,minimum size=0.5cm](p7)at(6,1)[]{};
 \filldraw[black](6,1)circle(2pt);
\node[circle,draw,minimum size=0.5cm](p4)at(0,1)[]{};
\node[circle,draw,minimum size=0.5cm](p6)at(4,1)[]{};
\node[circle,draw,minimum size=0.5cm](p8)at(8,1)[]{};
\node[draw,minimum size=0.4cm](a)at(1,0){$a$};
\node[draw,minimum size=0.4cm](b)at(3,0){$b$};
\node[draw,minimum size=0.4cm](c)at(5,0){$c$};
\node[draw,minimum size=0.4cm](d)at(7,0){$d$};
\draw[-latex](p1)--(t1);\draw[-latex](p2)--(t2);\draw[-latex](p3)--(t3);
\draw[-latex](t1)--(p4);\draw[-latex](t2)--(p6);\draw[-latex](t3)--(p8);
\draw[-latex](p4)--(a);\draw[-latex](p8)--(d);
\draw[-latex](p5)--(a);\draw[-latex](p5)--(b);
\draw[-latex](p6)--(b);\draw[-latex](p6)--(c);
\draw[-latex](p7)--(c);\draw[-latex](p7)--(d);
\end{tikzpicture}
\end{center}
\caption{A net which satisfies FPE but not SPE.}
\label{och1.fig}
\end{figure}
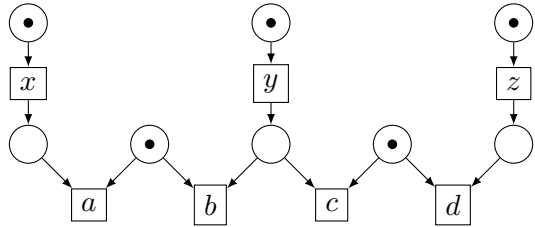

(ii): Figure \ref{och1.fig}, taken from \cite{folco-2014}, shows a pps counterexample.
The net shown there satisfies FPE:
Any maximal firing sequence must perform $x,y,z$ and either $a$ and $c$, or $a$ and $d$, or $b$ and $d$.
The sequence $xyzac$ is not persistent, but the equivalent sequence $xaycz$ is.
Similarly, the sequences $xazdy$ and $zdybx$ are persistent.
But the finite sequences $yb$ and $yc$ have no persistent equivalents, so that SPE is not satisfied.
\ENDBEW{och-conv.lem}

Another distinction between the finite-is-fair and
maximal-finite-is-fair principles is that the latter
-- but not the former -- satisfies a compositionality property
which might be interesting in checking the fairness of sequences:

\LEM{ds.lem}{Compositionality in the maximal-finite-is-fair regime}
Let $N$ be composed of two disjoint nets $N_1$ and $N_2$.
\begin{itemize}
\item[{\rm (i)}]
Under the finite-is-fair regime,
it may be the case that some infinite sequence of $N$ is unfair
while its projections onto $N_1$ and $N_2$ are both fair.
\item[{\rm (ii)}]
Under the maximal-finite-is-fair regime, every (finite or infinite) firing sequence of $N$
is fair if and only if its projections onto $N_1$ and $N_2$ are both fair.
\end{itemize}
\ENDLEM

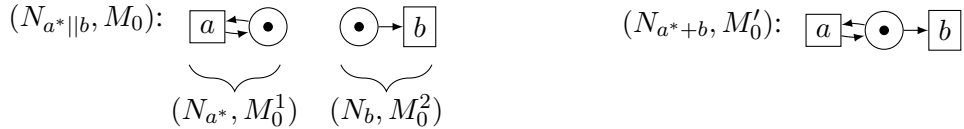
\begin{figure}[htp]
\begin{center}
\begin{tikzpicture}[scale=0.8]
\node[]()at(-1,1.1)[]{$(N_{a^*||b},M_0)$:};
\node[circle,draw,minimum size=0.5cm](ss0)at(2,1)[label=above:$$]{};
\filldraw[black](2,1)circle(2pt);
\node[circle,draw,minimum size=0.5cm](sss0)at(3.5,1)[label=above:$$]{};
\filldraw[black](3.5,1)circle(2pt);
\node[draw,minimum size=0.4cm](tt0)at(1,1)[label=above:]{$a$};
\node[draw,minimum size=0.4cm](tt)at(4.5,1)[label=above:]{$b$};
\draw([yshift=0.09cm]ss0.west)[]edge[-latex,bend right=10]
node[ellipse,below,inner sep=2pt,pos=0.5]{}([yshift=0.09cm]tt0.east);
\draw([yshift=-0.09cm]tt0.east)[]edge[-latex,bend right=10]
node[ellipse,below,inner sep=2pt,pos=0.5]{}([yshift=-0.09cm]ss0.west);
\draw[-latex](sss0)edge[bend left=0](tt);
\draw[decorate,decoration={brace,amplitude=10pt,mirror,raise=0.5cm},xshift=0cm,yshift=0cm]
(0.7,1)--(2.2,1)node [black,midway,xshift=0.0cm,yshift=-1.1cm]{$(N_{a^*},M_0^1)$};
\draw[decorate,decoration={brace,amplitude=10pt,mirror,raise=0.5cm},xshift=0cm,yshift=0cm]
(3.2,1)--(4.7,1)node [black,midway,xshift=0.0cm,yshift=-1.1cm]{$(N_{b},M_0^2)$};
\end{tikzpicture}\hspace*{2cm}
\raisebox{1.1cm}{\begin{tikzpicture}[scale=0.8]
\node[]()at(-1,1.1)[]{$(N_{a^*{+}b},M_0')$:};
\node[circle,draw,minimum size=0.5cm](s0)at(2,1)[label=above:$$]{};
\filldraw[black](2,1)circle(2pt);
\node[draw,minimum size=0.4cm](t0)at(1,1)[label=above:]{$a$};
\node[draw,minimum size=0.4cm](t)at(3,1)[label=above:]{$b$};
\draw([yshift=0.09cm]s0.west)[]edge[-latex,bend right=10]
node[ellipse,below,inner sep=2pt,pos=0.5]{}([yshift=0.09cm]t0.east);
\draw([yshift=-0.09cm]t0.east)[]edge[-latex,bend right=10]
node[ellipse,below,inner sep=2pt,pos=0.5]{}([yshift=-0.09cm]s0.west);
\draw([yshift=0.0cm]s0.east)[]edge[-latex,bend left=0]
node[ellipse,below,inner sep=2pt,pos=0.5]{}([yshift=0.0cm]t.west);
\end{tikzpicture}}
\end{center}
\caption{The net $N_{a^*||b}$ with initial marking $M_0$ is the disjoint sum of two component nets,
$N_{a^*}$ with initial marking $M_0^1$, and $N_{b}$ with initial marking $M_0^2$.
In $N_{a^*||b}$, the sequence $M_0\xfire{a^\infty}$ is unfair and fails to satisfy progress.
In $N_{a^*{+}b}$, with initial marking $M_0'$, the sequence $M_0'\xfire{a^\infty}$
is also (weakly as well as strongly) unfair, but it satisfies progress.}
\label{fair2.fig}
\end{figure}

\BEW
(i):
Consider the net $N_{a^*||b}$ shown on the left-hand side of Figure \ref{fair2.fig}.
The infinite firing sequence $M_0\xfire{a^\infty}$ in $N_{a^*||b}$ is not fair 
since $b$ is constantly neglected.
However, its (empty) projection $M_0^2\xfire{\leer}M_0^2$
on $N_{b}$
is fair by virtue of being finite, and its projection $M_0^1\xfire{a^\infty}$ on $N_{a^*}$
is fair, too, since $a$ occurs infinitely often and $b$ is not part of the scene.
This does not happen with the maximal-finite-is-fair principle,
since $M_0^2$ is not a deadlock in $N_{b}$, 
so that the empty firing sequence is not fair in this case.

(ii):
This follows from the fact that a sequence of $N$ is deadlocking if and only if
its two projections onto $N_1$ and $N_2$ are both deadlocking.
\ENDBEW{ds.lem}

Both \cite{DBLP:journals/csur/GlabbeekH19} and \cite{DBLP:journals/ipl/KindlerA99}
have remarked that the maximal-finite-is-fair principle is a merge of two concepts:
that of strong fairness for infinite sequences and that of
\emph{progress} (as defined in \cite{DBLP:conf/concur/Reisig96}, for instance)
for finite sequences.
We now discuss this in more detail.

\subsection{Continuous and constant neglecting, weak fairness, progress, APE, and JPE}
\label{progress.sct}

In a plain net, a marking $M$ enables two transitions $a$ and $b$ with $a\neq b$ \emph{concurrently}
if $M(p)\geq2$ for every $p\in\pre{a}\cap\pre{b}$ and
$M(p)\geq1$ for every $p\in(\pre{a}\minus\pre{b})\cup(\pre{b}\minus\pre{a})$.
Call a transition $t$ \emph{continuously} (\emph{constantly}) \emph{neglected} along some computation
\[K_0\xfire{a_1}K_1\xfire{a_2}K_2\xfire{a_3}\;\;\ldots
\]
if $t\notin\{a_1,a_2,a_3,\ldots\}$ and
for every $i\geq1$, $K_{i-1}$ enables both $a_i$ and $t$
(respectively: $K_{i-1}$ concurrently enables $a_i$ and $t$).
Thus, in $N_{a^*||b}$ of Figure \ref{fair2.fig}, the sequence $a^\infty$ continuously and constantly
neglects $b$, while in $N_{a^*+b}$, the same sequence continuously neglects $b$
but does not constantly neg\-lect~$b$.
Constant neglecting of a transition in an infinite computation
is akin to the non-maximality property of a finite computation:
some autonomous activity refuses to proceed, even though
no other activity contends with the resources it needs to do so.

An infinite sequence is \emph{weakly fair} (or, in the parlance of \cite{folco-2014}, \emph{just})
if it has no infinite tail in which some transition is continuously neglected,
and it \emph{satisfies progress}
if it has no infinite tail in which some transition is constantly neglected.
To define progress for finite sequences,
there are no two ways of doing so,
since progress corresponds to maximality.
For weak fairness, however, we have the same two possibilities as before: 
all finite sequences, or only the maximal ones, are called weakly fair.

In \cite{folco-2014}, we also find two other properties besides SPE and FPE:
\begin{itemize}
\item
APE requires \emph{all} finite or infinite sequences to have persistent equivalents.
\item
JPE requires all \emph{just} (weakly fair) 
sequences to have persistent equivalents.
\end{itemize}
When these are taken into account, together with
their finite-is-fair variants, we get $\mathrm{APE}^f\Leftrightarrow\mathrm{APE}$
and the following implications: 

\LEM{ajfs.lem}{Relationships between APE, JPE, SPE, and FPE}
\begin{itemize}
\item[{\rm (i)}]
$\;\;\;\;\mathrm{APE}^f\;\;\impl\;\;\mathrm{JPE}^f\;\;\impl\;\;\mathrm{FPE}^f\;\;\impl\;\;\mathrm{SPE}^f\;\;\;\;$ (for any Petri net)
\item[{\rm (ii)}]
$\neg\;(\;\mathrm{FPE}^f\;\;\impl\;\;\mathrm{JPE}^f\;)\;\;\;\;$ and 
$\;\;\;\;\neg\;(\;\mathrm{JPE}^f\;\;\impl\;\;\mathrm{APE}^f\;)$
\item[{\rm (iii)}]
$\;\;\;\;\mathrm{APE}\;\impl\;\mathrm{JPE}\;,\;
\mathrm{APE}\;\impl\;\mathrm{SPE}\;,\;\text{and}\;\;
\mathrm{JPE}\;\impl\;\mathrm{FPE}\;\;$ (for any Petri net)
\item[{\rm (iv)}]
The implications in {\rm (iii)} cannot be reversed;
SPE and JPE are incomparable; and FPE does not imply SPE.
\end{itemize}
\ENDLEM

\BEW
(i): Every fair sequence is just and every just sequence is just a sequence,
together with Lemma~\ref{och-conv.lem}(i).

(ii): Figure \ref{unfair.fig} disproves $\mathrm{FPE}^f\impl\mathrm{JPE}^f$
and Figure \ref{och1.fig}
disproves $\mathrm{JPE}^f\impl\mathrm{APE}^f$.

(iii): Same argument as before.

(iv): Figure \ref{unfair.fig} disproves
$\mathrm{FPE}\impl\mathrm{JPE}$, $\mathrm{SPE}\impl\mathrm{APE}$, and $\mathrm{SPE}\impl\mathrm{JPE}$,
and Figure~\ref{och1.fig} disproves
$\mathrm{JPE}\impl\mathrm{APE}$, $\mathrm{JPE}\impl\mathrm{SPE}$ and $\mathrm{FPE}\impl\mathrm{SPE}$.
\ENDBEW{ajfs.lem}

\begin{figure}[htb]
\begin{center}
\fbox{\begin{tikzpicture}[scale=0.8]
\node[](a)at(0,2)[]{APE$^f$};
\node[](j)at(3,2)[]{JPE$^f$};
\node[](f)at(3,0)[]{FPE$^f$};
\node[](s)at(0,0)[]{SPE$^f$};
\node[black,xshift=0cm,yshift=-0.1cm]()at(1.5,2){$\impl$};
\node[black,rotate=-90,xshift=0cm,yshift=-0.1cm]()at(3,1){$\impl$};
\node[black,rotate=0,xshift=0cm,yshift=-0.2cm]()at(1.5,0){$\impl$?};
\node[black,rotate=180,xshift=0cm,yshift=-0.2cm]()at(1.5,0){$\impl$};
\end{tikzpicture}}\hspace*{2cm}
\raisebox{0.2cm}{\fbox{\begin{tikzpicture}[scale=0.8]
\node[](a)at(0,2)[]{APE};
\node[](j)at(3,2)[]{JPE};
\node[](f)at(3,0)[]{FPE};
\node[](s)at(0,0)[]{SPE};
\node[black,xshift=0cm,yshift=0cm]()at(1.5,2){$\impl$};
\node[black,rotate=-90,xshift=0cm,yshift=0cm]()at(0,1){$\impl$};
\node[black,rotate=-90,xshift=0cm,yshift=0cm]()at(3,1){$\impl$};
\node[black,rotate=0,xshift=0cm,yshift=-0.0cm]()at(1.5,0){$\impl$?};
\end{tikzpicture}}}
\end{center}
\caption{The implications without question mark are true for all nets
(Lemmata \ref{och-conv.lem}(i) and \ref{ajfs.lem}(i,iii)).
The implication(s) provided with a question mark are true for pure dissymmetric choice nets
and for equal conflict nets (Section \ref{main.sct}),
while they are wrong for general, and even for AC, nets (Section \ref{counter.sct})
and are conjectured to be true for pps nets
in general (Section \ref{och.sct}) and for (not necessarily pure) DC nets.
All other possible (bilateral) implications, except those arising from reflexivity and transitivity,
are wrong (Figures \ref{unfair.fig}, \ref{och1.fig}, and Lemmata \ref{och-conv.lem}(ii), \ref{ajfs.lem}(ii,iv)).
Also, APE$\;\Leftrightarrow\;$APE$^f$ and SPE$\;\Leftrightarrow\;$SPE$^f$
and $\text{SPE}\impl(\text{FPE}\Leftrightarrow\text{FPE}^f)$.}
\label{summ.fig}
\end{figure}
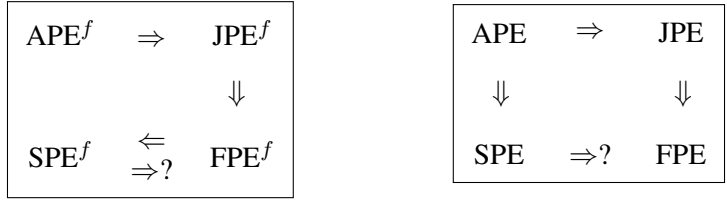

Figure \ref{summ.fig} provides a summary of the relationships proved, disproved, and conjectured so far.

\section{Concluding remarks}
\label{concl.sct}

The core results of this paper are Theorems \ref{ec.th} and \ref{main.th},
together with Corollaries \ref{ec.cor} and \ref{main2.cor}, in Section \ref{main.sct}.
They confirm a generalised version of Ochma\'nski's conjecture to be true
in the class of equal choice Petri nets and in the class of pure dissymmetric Petri nets.
The way to prove this is by first proving that all nets in one of these classes are persistent
if they satisfy the more ``local'' property of S$\widetilde{\text{P}}$E
(all finite firing sequences have persistent Parikh equivalents).
Then the property FPE, demanded by Ochma\'nski's conjecture,
follows immediately by identity permutations.
Both Corollary \ref{ec.cor} and Corollary \ref{main2.cor} are significant extensions of the results
contained in \cite{DBLP:conf/apn/BestD25}.

\Needspace{5\baselineskip} 
In an attempt to elucidate the context of these results,
we have categorised various notions of fairness
and other persistent equivalent notions in Section \ref{fair.sct}.
Also, we have added a brief historical digest of the origins of
the choice-related Petri net classes which have played a role in
our investigations (Appendix \ref{history.app}).
Furthermore, we have examined an example which is pure, plain, AC, and 2-bounded
and which satisfies neither SPE $\impl$ persistent nor SPE $\impl$ FPE,
and we have described some difficulties in turning
this example into a counterexample to Ochma\'nski's conjecture (Section \ref{counter.sct}). 

Several
questions are open.
A first question is whether the premise ``pureness'' can be
dropped from Theorem \ref{main.th} and Corollary \ref{main2.cor} of this paper.
Plainness can of course not be dropped because the
DC property is undefined for non-plain nets. 
However, we might consider extensions of DC (and AC) of the kind $F(.,t)\leq F(.,t')$,
or the converse, in the same way as EC generalises FC.

A second
question is whether or not the following is true:
\begin{equation}\label{och.eq}
\text{SPE$\impl$FPE $\;\;$ for $\;\;$ pure, plain, and safe Petri nets}
\end{equation}
which is a slight strengthening of Ochma\'nski's original conjecture \cite{och,folco-2014}.
From the various results and considerations described in this paper, we can argue that if (\ref{och.eq}) turns
out to be true, then it is a very tight result.
We hope that the proof techniques we have developed
can be re-used in order to contribute to a resolution
of the open questions,
or to generalise the obtained results to non-plain nets.

\Needspace{5\baselineskip}
A third open question concerns the nature of the relationship between
the persistent permutability properties S$\widetilde{\text{P}}$E and SPE.
We have formulated our results, but not (yet) Ochma\'nski's conjecture,
by choosing S$\widetilde{\text{P}}$E, rather than SPE, as a precondition.
Since SPE implies S$\widetilde{\text{P}}$E, all results are still true
if S$\widetilde{\text{P}}$E is replaced by SPE as a premise.
Conversely, it is not, as far as we know,
clear whether S$\widetilde{\text{P}}$E
implies SPE or not (in general, or in some restricted cases,
beyond the pure DC class).
It might be worthwile to explore this question.

\subsection*{Acknowledgments}

This paper is a revised and expanded version of a paper \cite{DBLP:conf/apn/BestD25}
presented at the 46th Petri net conference in Paris in June 2025, 
and of another paper \cite{conf/apn/bestd26}
to be
presented at the 47th Petri net conference in Hamburg in June 2026.
The authors are indebted to the reviewers who read the paper carefully,
detected some mistakes and made helpful suggestions for its presentation.


\bibliographystyle{fundam}
\bibliography{FI-Och-for-arXiv}

\pagebreak[2]

\appendix

\section{Historical remarks on selected choice nets}
\label{history.app}

The class of asymmetric choice Petri nets,
and by (reverse) duality, also the class of dissymmetric choice nets,
has at least two different roots.
Firstly, both classes of nets are pretty immediate
extensions of the class of free-choice nets.
In fact, Michel Hack, the originator of FC net theory,
already saw this \cite{hack72} and proposed
to investigate the class of ``simple nets''
as a promising candidate to which the free-choice net theory could be extended.\footnote{Hack's
original definition of ``simple nets'' requires that
$\forall p,p'\in P,p\neq p'\colon(\post{p}\cap\post{p'}=\es)\lor|\post{p}|\leq1\lor|\post{p'}|\leq1$,
i.e., every transition has at most one shared pre-place.}
As it turns out, simple nets are asymmetric choice,
and conversely, every asymmetric choice net can be simulated
in a strong sense by simple nets \cite{DBLP:books/sp/BestD24}.
Some structural analysis results such as
Fred Commoner's liveness theorem \cite{commoner-1972}
can partly be extended to asymmetric choice nets \cite{DBLP:conf/apn/BarkaouiM92,de95}.
We know of no such research on DC nets, however. We are
not even aware of a name given to that class, which is why we borrowed a suitable moniker
from~\cite{bestdesel-pnnewsletter}.

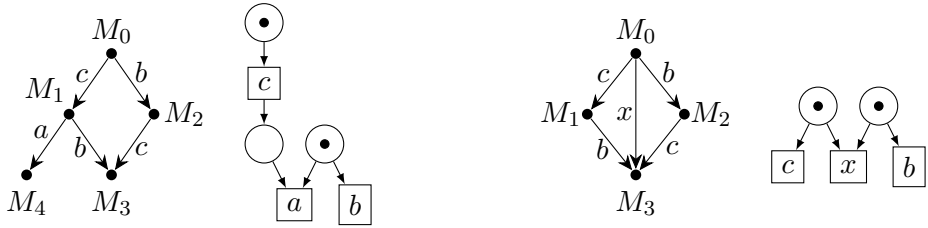
\begin{figure}[htb]
\begin{center}
\begin{tikzpicture}[scale=0.8]
\node[circle,fill=black!100,inner sep=0.05cm](0)at(1.4,2)[label=above:]{};
 \node[above of=0,node distance=0.3cm]{$M_0$};
\node[circle,fill=black!100,inner sep=0.05cm](1)at(0.7,1)[label=above left:$$]{};
 \node[above left of=1,node distance=0.45cm]{$M_1$};
\node[circle,fill=black!100,inner sep=0.05cm](2)at(2.1,1)[label=below right:$$]{};
 \node[right of=2,node distance=0.4cm]{$M_2$};
\node[circle,fill=black!100,inner sep=0.05cm](3)at(1.4,0)[label=below:$M_3$]{};
\node[circle,fill=black!100,inner sep=0.05cm](4)at(0,0)[label=below:$M_4$]{};
\draw[-{Stealth[length=2mm,width=1.7mm]},bend right=0](0)edge node[above left,inner sep=0.02cm,pos=0.5]{$c$}(1);
\draw[-{Stealth[length=2mm,width=1.7mm]},bend right=0](0)edge node[above right,inner sep=0.02cm,pos=0.5]{$b$}(2);
\draw[-{Stealth[length=2mm,width=1.7mm]},bend right=0](1)edge node[below left,inner sep=-0.02cm,pos=0.35]{$b$}(3);
\draw[-{Stealth[length=2mm,width=1.7mm]},bend right=0](2)edge node[below right,inner sep=0.0cm,pos=0.45]{$c$}(3);
\draw[-{Stealth[length=2mm,width=1.7mm]},bend right=0](1)edge node[above left,inner sep=0.02cm,pos=0.4]{$a$}(4);
\end{tikzpicture}\hspace*{0.3cm}
\raisebox{0.0cm}{\begin{tikzpicture}[scale=0.8]
\node[circle,draw,minimum size=0.5cm](p0)at(0.5,3)[]{};
 \filldraw[black](0.5,3)circle(2pt);
\node[circle,draw,minimum size=0.5cm](p1)at(0.5,1)[]{};
\node[circle,draw,minimum size=0.5cm](p2)at(1.5,1)[]{};
 \filldraw[black](1.5,1)circle(2pt);
\node[draw,minimum size=0.4cm](a)at(0.5,2){$c$};
\node[draw,minimum size=0.4cm](b)at(2,0){$b$};
\node[draw,minimum size=0.4cm](x)at(1,0){$a$};
\draw[-latex](p0)--(a);
\draw[-latex](a)--(p1);
\draw[-latex](p1)--(x);
\draw[-latex](p2)--(x);
\draw[-latex](p2)--(b);
\end{tikzpicture}}\hspace*{2cm}
\begin{tikzpicture}[scale=0.8]
\node[circle,fill=black!100,inner sep=0.05cm](0)at(2,2)[label=above:]{};
 \node[above of=0,node distance=0.3cm]{$M_0$};
\node[circle,fill=black!100,inner sep=0.05cm](1)at(1.2,1)[label=above left:$$]{};
 \node[left of=1,node distance=0.35cm]{$M_1$};
\node[circle,fill=black!100,inner sep=0.05cm](2)at(2.8,1)[label=below right:$$]{};
 \node[right of=2,node distance=0.35cm]{$M_2$};
\node[circle,fill=black!100,inner sep=0.05cm](3)at(2,0)[label=below:$M_3$]{};
\draw[-{Stealth[length=2mm,width=1.7mm]},bend right=0](0)edge node[above left,inner sep=0.02cm,pos=0.5]{$c$}(1);
\draw[-{Stealth[length=2mm,width=1.7mm]},bend right=0](0)edge node[above right,inner sep=0.02cm,pos=0.5]{$b$}(2);
\draw[-{Stealth[length=2mm,width=1.7mm]},bend right=0](1)edge node[below left,inner sep=-0.02cm,pos=0.4]{$b$}(3);
\draw[-{Stealth[length=2mm,width=1.7mm]},bend right=0](2)edge node[below right,inner sep=0.02cm,pos=0.4]{$c$}(3);
\draw[-{Stealth[length=2mm,width=1.7mm]},bend right=0](0)edge node[left,inner sep=0.03cm,pos=0.5]{$x$}(3);
\end{tikzpicture}\hspace*{0.3cm}
\raisebox{0.5cm}{\begin{tikzpicture}[scale=0.8]
\node[circle,draw,minimum size=0.5cm](p1)at(0,1)[]{};
 \filldraw[black](0,1)circle(2pt);
\node[circle,draw,minimum size=0.5cm](p2)at(1,1)[]{};
 \filldraw[black](1,1)circle(2pt);
\node[draw,minimum size=0.4cm](a)at(-0.5,0){$c$};
\node[draw,minimum size=0.4cm](b)at(1.5,0){$b$};
\node[draw,minimum size=0.4cm](x)at(0.5,0){$x$};
\draw[-latex](p1)--(a);
\draw[-latex](p1)--(x);
\draw[-latex](p2)--(x);
\draw[-latex](p2)--(b);
\end{tikzpicture}}
\end{center}
\caption{Asymmetric confusion (left-hand side) and symmetric confusion (right-hand side),
according to Carl Adam Petri \cite{petri-gnt-1976}.}
\label{conf.fig}
\end{figure}

A second line of heritage of asymmetric and
dissymmetric choice nets can be traced back to none other than C.A. Petri.
In several of his works, he was trying to sort out different ways in which concurrent agents can interact
in a distributed setting.
He identified various constellations such as ``contact'', ``conflict'', and ``confusion''.
By ``confusion'', he meant, broadly speaking, interactions characterised by a mixture
of concurrency and conflict.
In \cite{petri-gnt-1976}, for instance,
we find the two prototypical confusion patterns shown 
in Figure \ref{conf.fig}. 

In asymmetric confusion as in Figure \ref{conf.fig}, if $c$ occurs before $b$,
there is a transient conflict between $a$ and $b$,
but if $b$ occurs before $c$, there is no such conflict,
and yet, $b$ and $c$ are completely independent.
Petri argues that since in a truly concurrent and distributed setting, it is ``physically'',
so to speak, not possible to determine a specific order between
$b$ and $c$, there is some fuzziness about the conflict between $a$ and $b$, which is
what the term ``confusion'' aims at expressing.

In symmetric confusion as in Figure \ref{conf.fig}, there is a related
kind of fuzziness: $b$ and $c$ are independent, and they could be physically very far away
from each other. The conflict between the two transitions $b$ (or $c$) and $x$ is resolved not just by one
of them occurring, but also by a likely uncontrollable occurrence of $c$ (resp. $b$).
This indicates that it could be quite hard to implement $b$ and $c$ in a distributed manner
such that the conflicts between $c,x$ and $x,b$ are respected as specified by the
``M'' shaped Petri net on the right-hand side of Figure \ref{conf.fig}, preferably not using
devices such as ``busy wait''.\footnote{There
have been thorough works, such as a paper by
van Glabbeek, Goltz, and Schicke-Uffmann
\cite{DBLP:journals/corr/GlabbeekGS13}, who aim at overcoming
the difficulties in truly distributing the concurrent actions $c$ and $b$ appearing
in the ``M'' shape.}

Actually, the dissymmetric choice restriction
can accommodate both of Petri's confusion patterns
shown in Figure \ref{conf.fig}. None of them violates the DC condition
\[(\pre{t}\cap\pre{t'}=\emptyset)\;\lor\;(\pre{t}\subseteq\pre{t'})\;\lor\;(\pre{t'}\subseteq\pre{t})
\]
and both sorts of confusion could occur in a DC net.
What DC does prohibit, though, is a particularly nasty form of dissymmetric choice
when the pattern shown on the left-hand side of Figure \ref{conf.fig}
is redoubled by adding a $d$ branch in parallel to the $c$ branch, leading into $b$.
This symmetricised version of asymmetric choice leads straight to
the net $\PNref{basic-exa.fig}$ in Figure~\ref{basic-exa.fig}.
As we have seen, the absence of this pattern in DC nets is one of the cruxes of our arguments
in Section \ref{main.sct}.
(And it is its presence which considerably complicates the proof of Ochma\'nski's conjecture
for plain, pure, and safe nets.)

Originally, Petri was aiming at finding rules for orderly means of communication,
and for some time, he was actually trying to argue that all kinds of confusion should be
avoided (if at all possible). This led to a lot of discussion because many real-life
use cases seem to involve confusion of some sort quite unavoidably.
If one follows Einar Smith in his function (if there is such a function)
as Petri's ``scientific executor'' \cite{DBLP:journals/tcs/Smith96},
then it should be advisable to come to terms with confused
situations, rather than trying to avoid them from the outset.
In particular, Smith argues, there are some pieces of hardware such as an arbiter that will necessarily contain
the ``nasty'' symmetric version of asymmetric confusion sketched in the
last paragraph and depicted in Figure \ref{basic-exa.fig}.\footnote{To
appreciate this, compare Figure 8 of \cite{DBLP:journals/tcs/Smith96} with Figure \ref{basic-exa.fig}.}
So, even from a philosophical point of view, it is hardly possible
to argue that we should not be interested in the
DC class of Petri nets and in solving Ochma\'nski's original conjecture one way or the other.

It is possible to prohibit asymmetric confusion by a structural condition
which is akin to the conditions defining AC and DC nets.
To this end, let us call a Petri net $N=(P,T,F)$ $\widetilde{\text{DC}}$ if
\begin{equation}\label{dcd.eq}
\forall p,p'\in P\colon \post{p}\cap\post{p'}\neq\es\impl(\post{p}\subseteq\post{p'})\lor(\pre{p'}\subseteq\pre{p})
\end{equation}
As it was noted in \cite{bestdesel-pnnewsletter},
(\ref{dcd.eq}) excludes Petri's asymmetric confusion (Figure \ref{conf.fig}),
and it was in fact designed to do so.
This indicates that (\ref{dcd.eq}) has some intuitive appeal, but not much seems
to have resulted in terms of any closer investigation of $\widetilde{\text{DC}}$ nets.
Besides, (\ref{dcd.eq}) is neither dual nor reverse dual to the AC or DC conditions,
so that we feel justified in proposing a more fitting change of names.

Like AC and DC, the $\widetilde{\text{DC}}$ property (\ref{dcd.eq})
generalises the FC property, but unlike DC,
it does not come close to characterising the pattern in Figure \ref{basic-exa.fig}
whose absence is a relevant ingredient of our main result.
For the sake of completeness, we show
a deadlock-free, safe, plain, pure,
non-free-choice net
which satisfies all of the properties AC, DC, and $\widetilde{\text{DC}}$.
A deadlocking and impure one is $\PNref{dia-lem.pn}$ shown in Figure~\ref{dia-lem.fig}.
These examples indicate
that it could be interesting, in future work, to
investigate the class of Petri nets which simultaneously satisfy AC, DC, and $\widetilde{\text{DC}}$
(but not necessarily FC),
and are thus confusion-free in the sense of Petri, without necessarily already being free-choice.

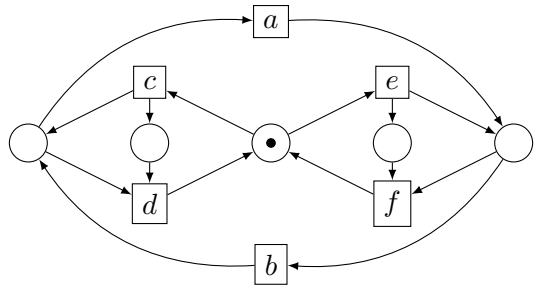
\begin{figure}[htb]
\begin{center}
\begin{tikzpicture}[scale=0.8]
\node[circle,draw,minimum size=0.5cm](1)at(0,2)[]{};
\node[circle,draw,minimum size=0.5cm](2)at(2,2)[]{};
\node[circle,draw,minimum size=0.5cm](3)at(4,2)[]{};
 \filldraw[black](4,2)circle(2pt);
\node[circle,draw,minimum size=0.5cm](4)at(6,2)[]{};
\node[circle,draw,minimum size=0.5cm](5)at(8,2)[]{};
\node[draw,minimum size=0.4cm](a)at(4,4){$a$};
\node[draw,minimum size=0.4cm](b)at(4,0){$b$};
\node[draw,minimum size=0.4cm](c)at(2,3){$c$};
\node[draw,minimum size=0.4cm](d)at(2,1){$d$};
\node[draw,minimum size=0.4cm](e)at(6,3){$e$};
\node[draw,minimum size=0.4cm](f)at(6,1){$f$};
\draw[-latex](1)edge[bend left=30](a);
\draw[-latex](a)edge[bend left=30](5);
\draw[-latex](5)edge[bend left=30](b);
\draw[-latex](b)edge[bend left=30](1);
\draw[-latex](c)--(1);\draw[-latex](1)--(d);
\draw[-latex](c)--(2);\draw[-latex](2)--(d);
\draw[-latex](d)--(3);\draw[-latex](3)--(c);
\draw[-latex](e)--(5);\draw[-latex](5)--(f);
\draw[-latex](e)--(4);\draw[-latex](4)--(f);
\draw[-latex](f)--(3);\draw[-latex](3)--(e);
\end{tikzpicture}
\end{center}
\caption{A net, taken from \cite{bestdesel-pnnewsletter}, which is AC, DC, and $\widetilde{\text{DC}}$, but not FC.
In \cite{bestdesel-pnnewsletter}, it is argued informally that no FC net can simulate it in any strong sense.
The argument is that after firing $c$, $a$ is enabled in conflict with $d$,
whereas after firing $eb$, $a$ is the only enabled transition, which is very much unlike the behaviour of an FC net.}
\label{acdcd.fig}
\end{figure}

\end{document}